\documentclass[a4paper,fleqn,usenatbib]{mnras}

\usepackage{newtxtext,newtxmath}

\usepackage[T1]{fontenc}

\usepackage{graphicx}	
\usepackage{amsmath}	
\usepackage{amssymb}	
\usepackage{listings}
\usepackage{color}
\definecolor{dkgreen}{rgb}{0,0.5,0}
\definecolor{gray}{rgb}{0.5,0.5,0.5}
\definecolor{mauve}{rgb}{0.58,0,0.82}
\definecolor{lightgray}{rgb}{0.95,0.95,0.95}
\newcommand{\Mpc}{{\rm Mpc}}
\newcommand{\km}{{\rm km}}

\newcommand{\Msun}{{\rm M}_\odot}
\newcommand{\Mstel}{M_\ast}
\newcommand{\logM}{\log\Mstel/\Msun}

\newcommand{\resp}{respectively}

\newcommand{\sfr}{{\rm SFR}}

\newcommand{\beq}{\begin{equation}}
\newcommand{\eeq}{\end{equation}}
\newcommand{\bitem}{\begin{itemize}}
\newcommand{\eitem}{\end{itemize}}
\newcommand{\benum}{\begin{enumerate}}
\newcommand{\eenum}{\end{enumerate}}

\mathchardef\mhyphen="2D

\newcommand{\grizli}{{\tt GRIZLI}}
\newcommand{\beams}{{\tt .beams.}}
\newcommand{\stack}{{\tt .stack.}}
\newcommand{\full}{{\tt .full.}}
\newcommand{\mastercat}{glass-acs-2019.05.10.fits}

\newcommand{\ntot}{22755} 
\newcommand{\nGold}{4636} 
\newcommand{\nLowRisk}{2319} 
\newcommand{\nHiRisk}{20436}
\newcommand{\midz}{0.60}

\newcommand{\facilities}{{\it Facilities:}}
\newcommand{\software}{{\it Software:}}

\title[The GLASS ACS data release]{The Grism Lens-Amplified Survey from Space (GLASS). XIII. 
	G800L optical spectra from the parallel fields}

\author[L.E.~Abramson et al.]{L.E.~Abramson$^{1, 2, 3,}$\thanks{E-mail: \href{mailto:labramson@carnegiescience.edu}{labramson@carnegiescience.edu}}, 
G.B.~Brammer$^{4, 5}$,
K.B.~Schmidt$^{6}$,
T.~Treu$^{1}$,
T.~Morishita$^{4}$,\newauthor 
X.~Wang$^{1}$,
B.~Vulcani$^{7},$
A.~Henry$^{4}$
\\
\\
$^1$	UCLA, 430 Portola Plaza, Los Angeles, CA 90095, USA\\
$^2$	Carnegie Observatories, 813 Santa Barbara Street, Pasadena, CA 91101, USA\\
$^3$	Princeton University, 4 Ivy Lane, Princeton, NJ 08544, USA\\
$^4$	Space Telescope Science Institute, 3700 San Martin Drive, Baltimore, MD 21218, USA\\
$^5$	Cosmic Dawn Centre, Niels Bohr Institute, University of Copenhagen, Lyngbyvej 2, DK-2100, Copenhagen, Denmark\\
$^6$ Leibniz-Institut f\"{u}r Astrophysik Potsdam (AIP), An der Sternwarte 16, 14482, Potsdam, Germany\\
$^7$ INAF -- Osservatorio Astronomico di Padova, Vicolo Osservatorio 5, IT-35122, Padova, Italy
}

\date{Submitted to {\it MNRAS} 31 May 2019}

\pubyear{2019}

\begin{document}
\label{firstpage}
\pagerange{\pageref{firstpage}--\pageref{lastpage}}
\maketitle

\begin{abstract}
	We present a catalogue of \ntot\ objects with slitless, optical, {\it Hubble Space
	Telescope} (HST) spectroscopy from the {\it Grism Lens-Amplified Survey from Space} 
	(GLASS). The data cover $\sim$220~sq.~arcmin to 7-orbit ($\sim$10~ks) depth in
	20 parallel pointings of the Advanced Camera for Survey's G800L grism. The fields 
	are located 6$'$ away from 10 massive galaxy clusters in the HFF and CLASH footprints. 
	Thirteen of the fields have
	ancillary HST imaging from these or other programs to facilitate a large number of applications, 
	from studying metal distributions at $z\sim0.5$, to quasars at $z\sim4$, to the star formation 
	histories of hundreds of galaxies in between. The spectroscopic catalogue has a median redshift 
	of $\langle z\rangle=\midz$ with a median uncertainty of $\Delta z / (1+z)\lesssim2\%$ at 
	$\rm F814W\lesssim23$~AB. Robust continuum detections reach a magnitude fainter.
	The 5\,$\sigma$ limiting line flux is $f_{\rm lim}\approx5\times10^{-17}\rm~erg~s^{-1}~cm^{-2}$  
	and half of all sources have 50\% of pixels contaminated at $\lesssim$1\%. All sources 
	have 1- and 2-D spectra, line fluxes/uncertainties and identifications, redshift probability 
	distributions, spectral models, and derived narrow-band emission line {\it maps} from the 
	Grism Redshift and Line Analysis tool (\grizli). We provide other basic sample 
	characterisations, show data examples, and describe sources and potential investigations 
	of interest. All data and products will be available online along with software to facilitate their use.
\end{abstract}




\section{Introduction}
\label{sec:intro}

The {\it Hubble Space Telescope} (HST) provides some of the best ultraviolet (UV) 
to near-infrared (NIR) imaging available. These data underpin most of our knowledge of  
the distant universe. Ground-based spectra---from, e.g., MOSDEF \citep{Kriek15} and LEGA-C 
\citep{vanDerWel16}---are critical, but the atmosphere limits continuum measurements 
to the most massive galaxies and blurs all spatial information. Using space-based data 
reduces these restrictions and yields a clearer picture of galaxy evolution.

HST's slitless grisms have played key roles here. They produce maps of every source at every 
wavelength at spatial scales inaccessible from Earth without adaptive 
optics. The two NIR grisms---G102 and G141 on WFC3\footnote{Wide Field Camera 3}---have proven 
especially effective, with, e.g., the 3D-HST survey \citep{Momcheva16} and our own Grism 
Lens-Amplified Survey from Space \citep[GLASS;][]{Schmidt14, Treu15} providing rest-optical 
$z>1$ spectroscopy over scores to hundreds of sq.~arcmin without the need for potentially 
biasing photometric preselection or slit masks. Paired with HST imaging, these data have 
extended our knowledge of the ages and star formation histories of galaxies at early cosmic 
times in ways comparable to ground based results at $z\lesssim1$ 
\citep[e.g.,][]{Whitaker12, Newman14, Nelson16, Wang17, Wang18, Abramson18b, 
Morishita18b, Morishita18}.


\begin{table*}
	\centering
	\caption{Basic survey information for the 20 GLASS ACS parallels. 
Field names reflect GLASS central pointing/IR grism (cluster) IDs and \texttt{ROOT} is the 
corresponding catalog keyword (Appendix \ref{sec:catCols}). PAs are ``{\tt PA\_V3}.''\newline
$^{\rm a}$ All objects with ${\rm c}|z - z_{\rm cl }|(1+z_{\rm cl})^{-1}<2000~\rm km\,s^{-1}$ irrespective of $z_{Q}$ where $z_{\rm cl}$ is the GLASS cluster redshift from Table 1 of \citet{Treu15}.\newline
$^{\rm b}$ Covering band(s) other than F814W to support SED fitting.\newline
$^{\rm c}$ CLASH -- \citet{Postman12}; \url{http://archive.stsci.edu/proposal_search.php?mission=hst&id=12067}; HFF -- \citet{Lotz17}; \url{http://archive.stsci.edu/proposal_search.php?mission=hst&id=13498}; Ebeling -- GO10420; \url{http://archive.stsci.edu/proposal_search.php?mission=hst&id=10420}; Riess -- GO13063; \url{http://archive.stsci.edu/proposal_search.php?mission=hst&id=13063}; Rodney -- GO13386; \url{http://archive.stsci.edu/proposal_search.php?mission=hst&id=13386}; Siana -- GO13389; \url{http://archive.stsci.edu/proposal_search.php?mission=hst&id=13389}.\newline 
$^{\rm d}$ One band.\newline
$^{\rm e}$ Grism data incorporated from GO12099 (PI Riess).}
	\label{tbl:fieldInfo}
	\begin{tabular}{ccccccccc} 
		\hline
Field & \texttt{ROOT} & RA [J2000] & DEC [J2000] & PA [deg] & $N_{\rm srcs}$ & $N_{\rm infall}$$^{\rm a}$ & G800L Exptime [s] & HST Imaging [program or PI]$^{\rm b,c}$\\
		\hline
Abell 2744 & j0014m3023 & 0:13:53.9 & -30:22:51 & 323 & 976 & 26 & 10076 & HFF, Siana, Rodney \\
Abell 2744 & j0014m3030 & 0:14:20.9 & -30:29:46 & 225 & 1066 & 16 & 10076 & HFF$^{\rm d}$, Rodney$^{\rm d}$ \\
Abell 370 & j0240m0132 & 2:39:31.6 & -1:31:42 & 343 & 1028 & 31 & 10076 & $\cdots$ \\
Abell 370 & j0240m0140 & 2:39:44.6 & -1:40:08 & 245 & 965 & 32 & 10344 & $\cdots$ \\
MACS0416 & j0416m2402 & 4:15:44.7 & -24:02:04 & 337 & 808 & 15 & 10076 & CLASH, Rodney \\
MACS0416 & j0416m2410 & 4:15:56.0 & -24:09:42 & 254 & 887 & 24 & 10186 & $\cdots$ \\
MACS0717 & j0717p3750 & 7:17:16.3 & 37:49:53 & 10 & 1123 & 46 & 9927 & HFF, CLASH, Siana, Ebeling \\
MACS0717 & j0718p3747 & 7:18:01.6 & 37:47:26 & 110 & 1186 & 37 & 10061 & $\cdots$ \\
MACS0744 & j0745p3922 & 7:45:08.6 & 39:22:16 & 194 & 853 & 37 & 9562 & CLASH \\
MACS0744 & j0745p3930 & 7:45:20.6 & 39:30:10 & 109 & 934 & 35 & 10061 & $\cdots$ \\
MACS1149 & j1150p2218 & 11:49:40.3 & 22:18:02 & 215 & 1117 & 37 & 10084 & HFF, CLASH$^{\rm d}$, Siana \\
MACS1149 & j1150p2225 & 11:50:00.9 & 22:25:25 & 122 & 1090 & 47 & 9854 & CLASH \\
RXJ1347 & j1347m1140 & 13:47:19.7 & -11:40:11 & 13 & 1049 & 31 & 10076 & CLASH \\
RXJ1347 & j1347m1147 & 13:47:09.4 & -11:47:36 & 293 & 954 & 20 & 10344 & $\cdots$ \\
MACS1423 & j1424p2401 & 14:24:06.8 & 24:00:38 & 178 & 813 & 40 & 10076 & CLASH \\
MACS1423 & j1424p2408 & 14:24:09.4 & 24:08:19 & 98 & 939 & 33 & 10344 & $\cdots$ \\
MACS2129 & j2130m0736 & 21:29:31.5 & -7:35:40 & 58 & 1131 & 39 & 10061 & Rodney \\
MACS2129$^{\rm e}$ & j2130m0742 & 21:29:54.4 & -7:41:45 & 136 & 3910 & 127 & 18789 & CLASH, Riess \\
RXJ2248 & j2249m4433 & 22:49:18.0 & -44:32:37 & 143 & 906 & 16 & 9372 & HFF, CLASH, Riess \\
RXJ2248 & j2249m4438 & 22:48:46.0 & -44:37:39 & 223 & 1020 & 27 & 10344 & CLASH$^{\rm d}$\\
		\hline
	\end{tabular}
\end{table*}


Beyond their scientific utility, HST's grisms play a pathfinding role: the astronomical 
community has decided that space-based, wide-field, slitless spectroscopy will be an increasingly 
large part of its portfolio with the forthcoming operation of JWST\footnote{\it James Webb Space Telescope.}, Euclid, and WFIRST\footnote{\it Wide-Field Infrared Survey Telescope.}.
Hence, building intuition, applications, and tools to handle these data is prudent. 

HST's less widely used optical disperser---the ACS\footnote{Advanced Camera for Surveys} G800L 
grism---is powerful in this context. Covering $\lambda=0.5$--1.0\,\micron\ with $\sim$2.5$\times$ 
WFC3's footprint, G800L extends space-based grism surveys to lower redshifts and more than 
doubles their area. The instrument's capability has been exploited by the GRAPES 
\citep[GO9793, PI Malhotra;][]{Pirzkal04} and PEARS 
\citep[GO10530, PI Malhotra;][]{Straughn08, Straughn09} surveys, and 3D-HST, with 
all results obtained prior to 3D-HST compiled by \citet{Kummel11}. 

Here, we describe GLASS' use of the G800L grism in parallel observing mode and present a 
source catalogue covering the 20 fields this survey comprises to 7-orbit depth (cf.\ 2 orbits for 
3D-HST). These data may be useful to anyone interested in a range of scientific 
questions---from infalling cluster galaxies to $z\gtrsim4$ Lyman-$\alpha$ emitters---and 
illustrate what could be done with higher signal-to-noise ratio ($S/N$) or spectral resolution 
data from future facilities.

Below, Section \ref{sec:data} describes data acquisition, field locations, reduction procedures,
and output products. Section \ref{sec:sampChar} describes the catalogue's redshift, magnitude, and 
contamination distributions, and highlights some notable sources for further study. Section 
\ref{sec:discussion} proposes some uses of the data. Section \ref{sec:summary} summarises. We 
provide file descriptions and some useful pieces of code in the Appendix. 
Throughout, we take $(H_{0}~\mathrm{km^{-1}~s}~\Mpc, \Omega_{m}, \Omega_{\Lambda}) 
= (70,0.3,0.7)$ and quote AB magnitudes \citep{Oke74}. All data/products are available upon 
request and will be made public.


\section{Data}
\label{sec:data}

\subsection{Acquisition}
\label{sec:acquisition}

GLASS covers ten $z\sim0.4$--0.6 cluster sightlines: six from HFF\footnote{Hubble Frontier Fields.} 
\citep[][]{Lotz17} and four more from CLASH\footnote{Cluster Lensing and Supernova Survey with Hubble. 
Due to CLASH/HFF overlap, eight GLASS sightlines have CLASH coverage.} \citep[][]{Postman12}. 
WFC3 was placed on the cluster centre at each sightline where 
gravitational lensing maximises the chances of detecting high-redshift Lyman-$\alpha$ emitters.
G102 and G141 were exposed there in series for 5 + 2 orbits, \resp. This process was repeated in two visits at 
two roughly orthogonal position angles (PAs) to facilitate source deblending. A $2\times$2 dither 
pattern with sub-pixel offsets was used to support image interlacing \citep{Brammer12}, the format of the 
GLASS IR coadds.

In parallel with the NIR grism observations, G800L was exposed to a region $\sim$6$'$ 
($\sim$2\,Mpc) away from the cluster centres. Given HST's rotation between visits, this strategy 
yielded 20, non-overlapping G800L survey fields; i.e., a spectroscopic database covering 
$\sim$220 sq.~arcmin to 7-orbit ($\sim$10 ks) depth. 
Data were acquired between 24 December 2013 and 18 January 2015. Table \ref{tbl:fieldInfo} provides basic information.

\begin{figure}
	\centering
	\includegraphics[width = 0.9\linewidth, trim = 0cm 0cm 0cm 0cm]{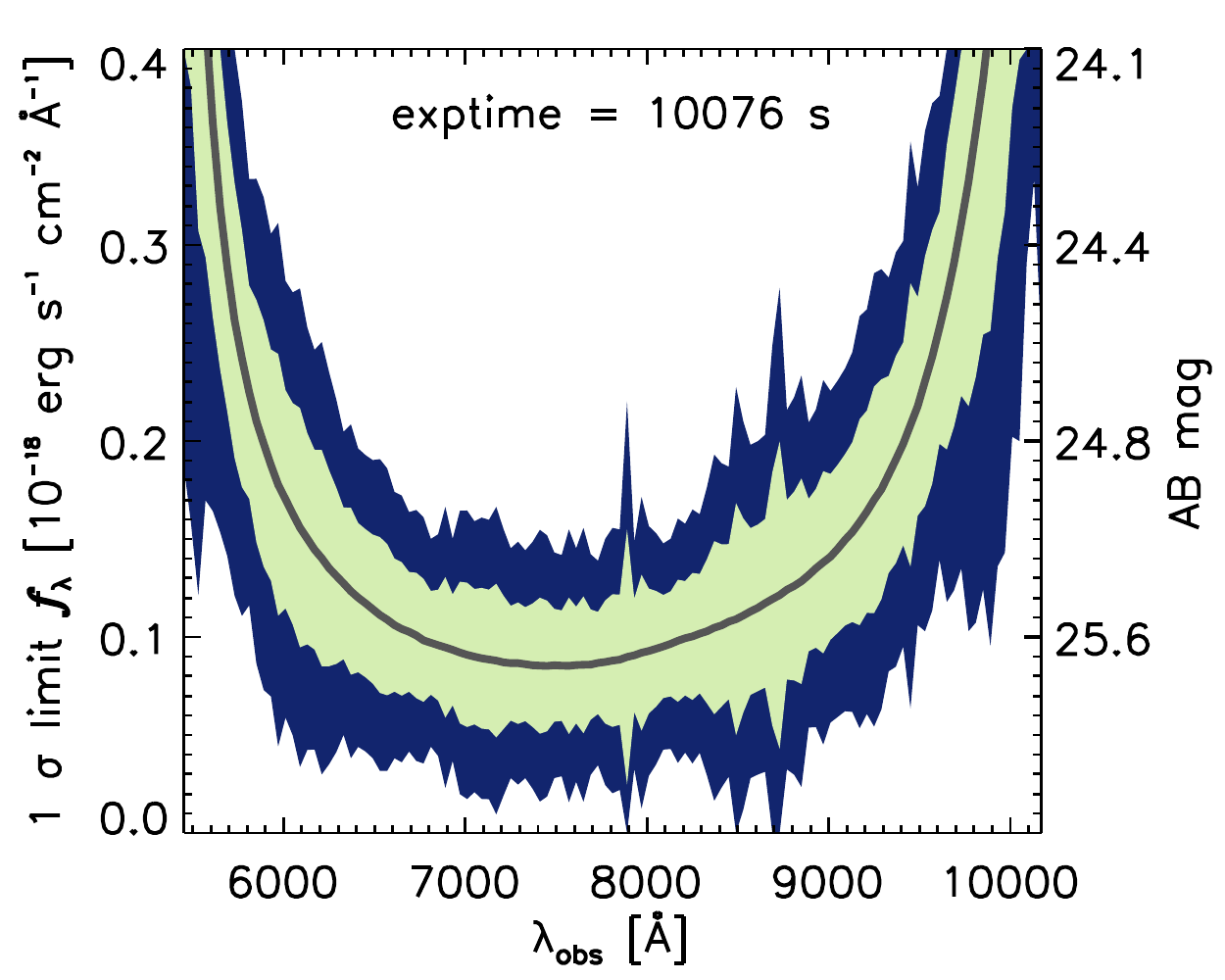}
	\caption{GLASS G800L sensitivity based on a $1''$ spatial extraction aperture. 
			This plot describes Abell 2744 (at PA 323) but is typical of all pointings.
			Thresholds are the median 1\,$\sigma$ $f_{\lambda}$ limit (grey) and 
			that value's 1\,$\sigma$ and 2\,$\sigma$ scatter (shaded). The median 5\,$\sigma$ 
			limiting line flux is $f_{\rm lim}\approx5\times10^{-17}\,{\rm erg~s^{-1}~cm^{-2}}$
			($\log f_{\rm lim}=-16.29\pm0.36$).}
	\label{fig:sens}
\end{figure}

A F814W pre-image was taken with each grism exposure (2 per orbit per PA). These data are 
necessary for source detection, spectral ID assignment, and wavelength calibration. Where possible, 
one PA was aligned with planned or actual ancillary imaging. This was achieved in 13 fields, where 
GLASS' F814W pre-images complement or supplement other HST photometry (Table 
\ref{tbl:fieldInfo}). GLASS pre-images are the only HST images in the remainder (e.g., MACS0717 
\texttt{j0718p3747}). We provide 
example spectrophotometry in Section \ref{sec:joint} and catalogue matching instructions in 
Appendix \ref{sec:algorithms}, but all analyses here {\it including redshift estimation} use 
no additional photometry unless stated.

\subsection{Reduction}
\label{sec:reduction}

Whereas the GLASS NIR data \citep{Schmidt14,Treu15} were reduced using a modified version 
of the 3D-HST pipeline \citep{Brammer12, Momcheva16}, the ACS parallel data were extracted 
using a python package with improved capabilities: \grizli---the Grism Redshift and Line Analysis 
tool \citep[see][Brammer et al.\ in preparation]{Wang17, 
Wang18}.\footnote{Available at \url{https://github.com/gbrammer/grizli}.} As with the original 
pipeline, the basic ingredients are two full-field ACS frames: one direct
image, and one containing the dispersed 2D spectra of all sources therein. The former 
serves as the reference frame to which the latter is anchored. 

\grizli\ first identifies cosmic rays in the F814W and G800L exposures using standard 
settings with the AstroDrizzle software \citep{Gonzaga12}, and fits and removes a two dimensional 
master sky image from the grism exposures.\footnote{We use the G800L 
configuration files provided at \url{http://www.stsci.edu/hst/acs/analysis/STECF/wfc_g800l.html}} 
We refine the astrometric alignment including both a fine relative alignment between exposures 
and a global alignment to an absolute reference frame, here defined by the GAIA DR2 
catalogue \citep{GaiaDR2}. As there are no telescope offsets between the paired F814W direct 
and G800L grism exposures, the
alignment of the former is applied directly to the latter.  Finally, we combine the 
F814W direct exposures into a rectified mosaic with AstroDrizzle and generate a
source catalog and segmentation map from this mosaic using the \texttt{sep} software 
package.\footnote{\texttt{sep} is a Python implementation of the \texttt{SExtractor} software \citep{BertinSEX} designed to exactly replicate its functionality; \url{https://github.com/kbarbary/sep/}} 

\begin{figure}
	\centering
	\includegraphics[width = 0.9\linewidth]{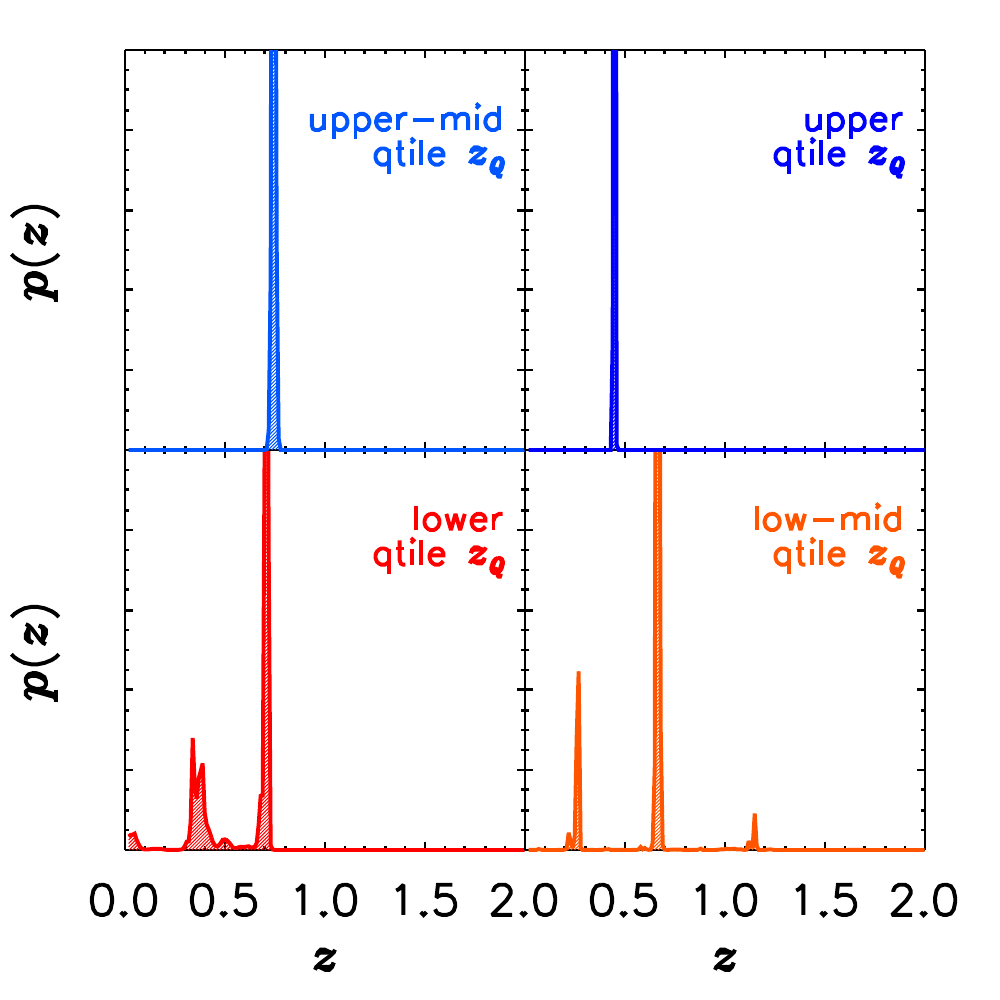}
	\caption{Random examples of redshift estimation quality in the 4 quartiles of 
			the redshift quality parameter, $z_{Q}$ based on the  $m_{814}<24~{\tt or}~S/N\geq5$ 
			in any optical strong line quality cut. Variation in the PDFs is noticeable.}
	\label{fig:zqual}
\end{figure}

The global contamination model for a given field is constructed in two passes. In the first, we 
assume that each object has a spectrum flat in units of $f_\lambda$ flux density with a normalisation 
set by its observed flux in the F814W direct image (integrated over the object's segmentation 
polygon). In the second pass, we step through objects in the catalogue sorted by increasing magnitude 
and compute a third-order polynomial fit to each spectrum after subtracting the contamination 
model from any neighbours (i.e., either the flat spectra of fainter sources or the polynomial models 
of brighter sources). This refinement is iterated three times.

In both passes, the \textit{spatial} template of the dispersed 2D model spectra is taken from the 
observed F814W cutout within an object's segmentation polygon. Note that this approach neglects 
both morphological variation across the G800L bandpass (e.g., the PSF, continuum colour gradients,
and morphological differences between line- and continuum-emitting regions). However, it 
provides a spatial model with much higher fidelity than any simple parametrised representation 
(e.g., Gaussian or S\'{e}rsic approximations). 

\begin{figure*}
	\includegraphics[width = 0.95\linewidth, trim = 0cm 0.5cm 0cm 0cm]{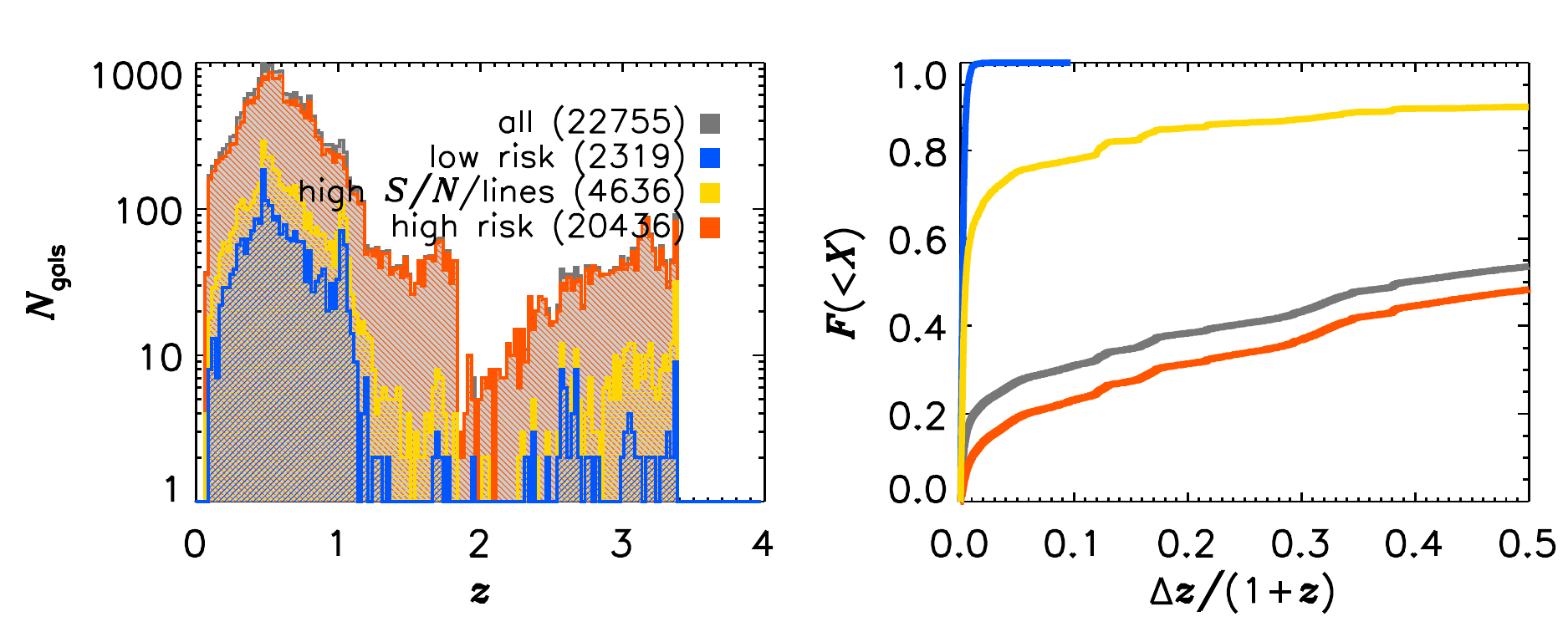}
	\caption{The redshift (left) and redshift uncertainty (right) distributions for
			the GLASS G800L sample. Here and in all following plots, the ``high'' 
			and ``low risk'' samples are plotted in red and blue, \resp. These
			are defined based on the median $z_{Q}$ of all sources with 
			$m_{814}<24~{\tt or}~S/N\geq5$, whose distributions are plotted in
			gold. The median redshift for those sources is $\langle z\rangle\simeq\midz$, with the
			low risk sample at $\langle z\rangle\simeq0.58$. Median formal uncertainties 
			rise from $<1\%$ for low risk to 
			nearly 50\% for high risk objects. Half of the full sample has 
			$\Delta z / (1+z)\lesssim0.35$. These uncertainties would be reduced
			by the addition of archival photometry where available, especially
			at $m_{814}\gtrsim23$ (Table \ref{tbl:fieldInfo}, Figure \ref{fig:magZerr}).}
	\label{fig:zDist}
\end{figure*}

\begin{figure*}
	\centering
	\includegraphics[width = 0.4\linewidth, trim = 0cm 0cm 0cm 0cm]{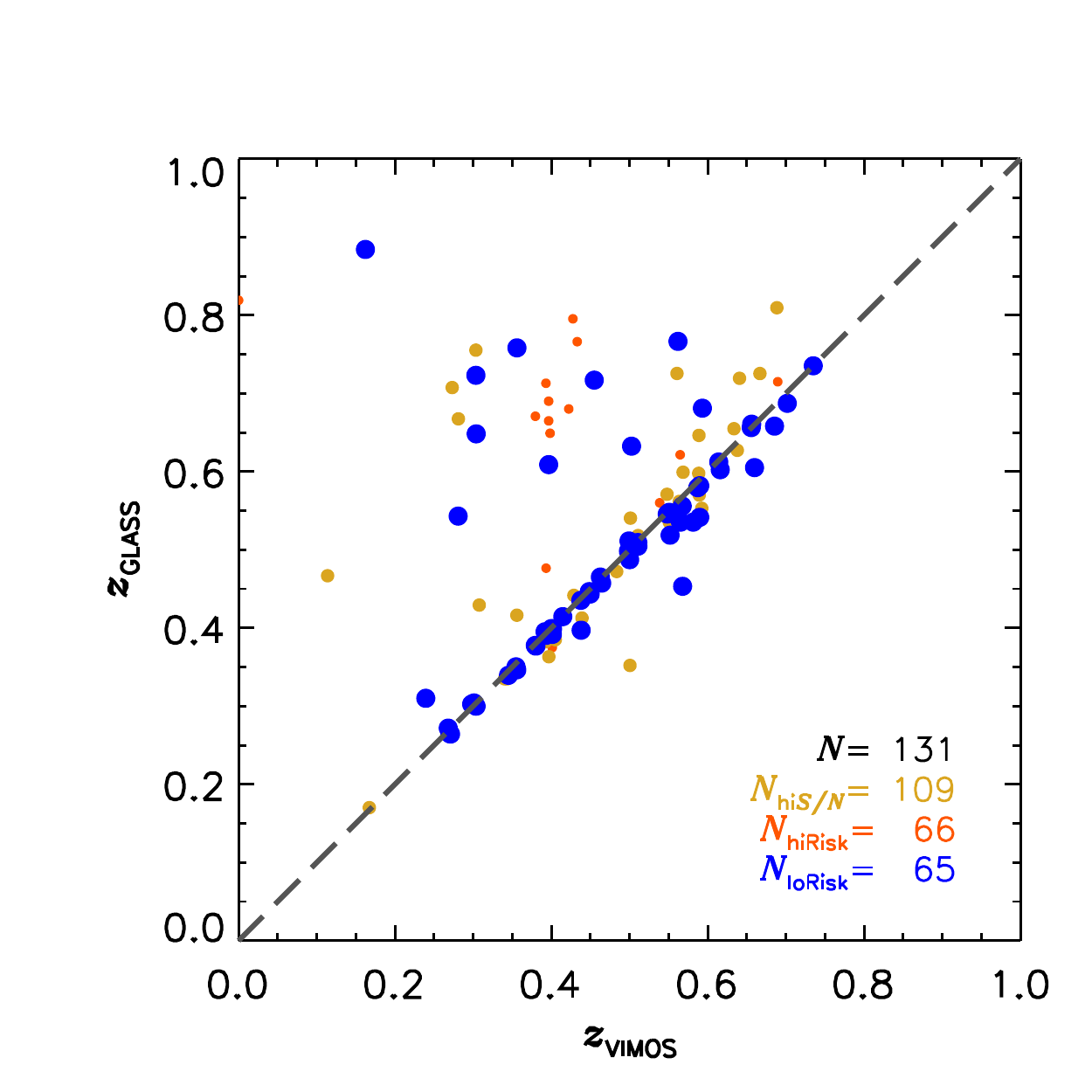}
	\includegraphics[width = 0.45\linewidth, trim = 0cm 0cm 0cm 0cm]{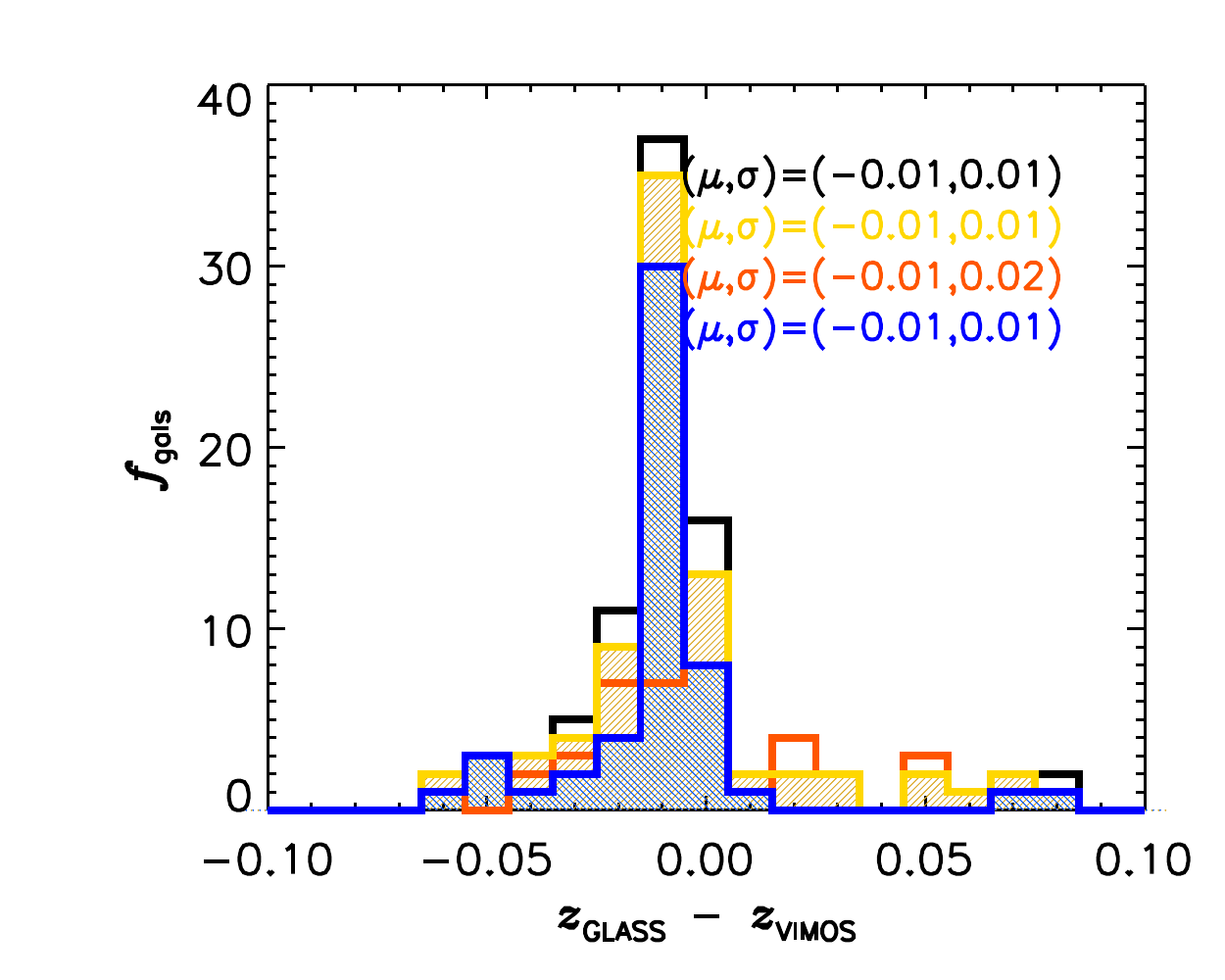}
	\caption{Top: GLASS G800L \grizli\ redshifts compared to those from VLT VIMOS as
			taken by various authors (see text). Agreement is quite good at $z<1$, with the global
			sample biased to lower redshifts by just $\Delta z =-0.01$ and with a scatter
			of the same order (right panel). A similar comparison with HFF photo-$z$s is presented
			in the Appendix.}
	\label{fig:specSpec}
\end{figure*}

The above process automatically yields a source and contamination model for each galaxy detected in the
reference image. These can be used to ``decontaminate'' source spectra---i.e., remove overlapping light
from neighbouring sources---to provide more accurate 
line and continuum characterisations. This was critical in GLASS' crowded cluster G102 and G141 pointings, 
but the ACS parallels are sparse enough that source contamination is often negligible
(Section \ref{sec:sampChar}). Note that no spectra are modelled or extracted for objects not detected in the 
reference frame. As such, the catalogue presented here is truncated at a F814W flux of $m_{814}=26$
with various subsamples defined by brighter flux limits (Section \ref{sec:sampChar}).

There are two main differences between \grizli\ and the previous GLASS pipeline. 
First, \grizli\ automatically handles HST grism spectra taken at one or more telescope 
roll angles that translate into different spectral dispersion position angles. This is implemented 
in that the 2D spectral models are computed in the pixel space of each separate grism exposure.  
This is the space in which the field-dependent grism dispersion configuration is measured and 
defined \citep[e.g.,][]{Kummel09}. This approach allows for evaluating model fits in the 
space of the original detector pixels with a well-calibrated noise model. Recent works have 
demonstrated the strengths of this approach \citep[see][]{Wang18, Morishita18b,Morishita18}, 
though, note that the GLASS G800L data described here are observed at a single position 
angle. This implementation causes the ACS data format to differ from the WFC3IR data \citep{Schmidt14,Treu15} in that each field's 14 subframes are drizzle- and not interlace-combined. 
All weight maps account for this fact and contain the appropriate pixel-level covariances.

The second difference is that \grizli\ provides automatic redshift estimates of each source, 
along with its uncertainty, and ``risk factor'' (\citealt{Tanaka18}; proportional to $1/\chi^{2}$; Section 
\ref{sec:sampChar}). This is done by evaluating fits of galaxy templates similar to those used by 
the {\tt EAZY} code \citep{Brammer08}. At the best redshift (typically where $\chi^2(z)$ is 
minimised but with an option to include a separate redshift prior), \grizli\ produces 
measurements and uncertainties of emission line fluxes and equivalent widths, as well as 
continuum-subtracted, 2D narrow-band maps around the emission line species. We do not use 
all of the \grizli\ outputs here, but they may be of value to other investigations.

{\it Note that all of the above processing is based exclusively on the G800L spectra}; no ancillary 
photometry is employed. This caveat should be kept in mind for analyses based solely on
our catalog entries as it affects any quantity that may depend substantially on 
information outside the grism bandpass (e.g., stellar mass).

\subsection{Products}
\label{sec:products}

The GLASS ACS \grizli\ reductions yield a suite of derived data products beyond the reference
and spectral data frames:
\benum
	\item A master FITS catalogue containing IDs and summary metrics for all extracted sources.
	\item A ``\beams'' FITS file containing 2D cutouts straight from each grism exposure 
		covering a given object, and the metadata necessary for using them to generate spectral 
		models with \grizli. 
	\item A ``\stack'' FITS table for each object containing its 2D spectrum, error map, 
		contamination and source models, and F814W direct image. All are rectified such that
	 	the spectral dispersion aligns with the data's $x$-axis. Unlike in the \beams\
		files, this rectification (and drizzling) necessitates a resampling of the original detector 
		pixels. This file's header contains a wavelength solution, exposure time, source 
		location, and ancillary information.
	\item A ``\full'' FITS table for each object containing redshift solutions, the
		covariance matrix and $\chi^2(z)$ of the template SED fits, the best-fit high-resolution 
		template SED with and without emission lines, the rectified, drizzled F814W direct 
		image and error map, and rectified, drizzled narrow-band images at wavelengths
		surrounding possible emissions lines (Section \ref{sec:joint}). These stamps comprise 
		a line map, 	continuum estimate, contamination, and error array for each line listed 
		in the FITS header. The header also contains a wavelength solution, exposure time, 
		source location, line fluxes, and ancillary information.
	\item  A ``{\tt .1D.}'' FITS table for each object comprising the wavelength array, 
		contamination-subtracted source spectrum ($\rm e^{-}\,s^{-1}$), sensitivity 
		curve [$\rm e^{-}\,s^{-1} / (erg\,s^{-1}\,cm^{-2}$\,\AA$^{-1}$)], RMS error
		($\rm e^{-}\,s^{-1}$), and line emission and continuum models.  
		These extractions use an optimal weighting \citep{Horne86} based on
		an object's spatial profile derived from the direct image itself as described above.
\eenum

\begin{figure}
	\centering
	\includegraphics[width = 0.9\linewidth, trim = 0.5cm 0.5cm 0cm 0cm]{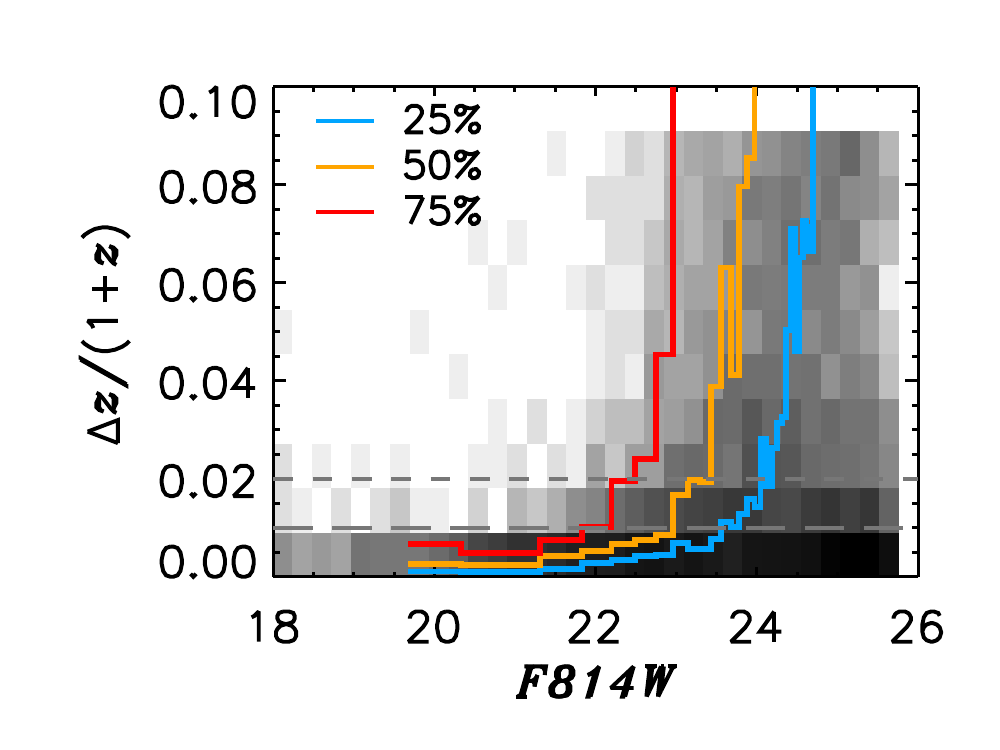}
	\caption{GLASS G800L \grizli\ redshifts are precise to $<1\%$ at
			$m_{814}\lesssim22$ for most sources but degrade rapidly by $m_{814}\sim24$, 
			where ancillary SED coverage
			or strong emission lines are needed to ensure fidelity. Coloured lines show 
			redshift uncertainty quartiles as a function of flux.}
	\label{fig:magZerr}
\end{figure}

All of the above files are detailed in Appendix \ref{sec:fileContents}, are available upon request,
and will be published on GLASS' MAST 
website.\footnote{\url{https://archive.stsci.edu/prepds/glass/}.} Useful software tools for 
selecting sources and matching to external catalogues will also be released (Appendix \ref{sec:algorithms}).


\section{Sample Characteristics}
\label{sec:sampChar}

\subsection{Counts, quality levels, and depth}
\label{sec:cts}

GLASS' G800L spectral database contains \ntot\ objects extracted to a limiting magnitude of 
$m_{814}=26$---roughly 1000 sources per pointing. For the purposes of this characterisation,
we split these into three, somewhat overlapping, quality-based subsets defined by continuum and 
line $S/N$. These are:
\benum
	\item All non-point sources with $m_{814} \leq 24~{\tt or}~S/N_{\rm line} \geq 5$ 
	(``high-$S/N$+lines''; $N=\nGold$);
	\item The subset of such galaxies below that subsample's median \grizli\ $z_{Q}$ (``low risk''; $N=\nLowRisk$);
	\item Everything not in (ii) (``high risk''; $N=\nHiRisk$).
\eenum		
Above, $S/N_{\rm line}$ refers to the $S/N$ in any of the following features as identified
by \grizli: H$\alpha$, H$\beta$, [\ion{O}{ii}], [\ion{O}{iii}], [\ion{S}{ii}], [\ion{Mg}{ii}], or Ly$\alpha$.
The magnitude cut corresponds roughly to the inferred $5\,\sigma$ continuum sensitivity---$\sim$$5\times
10^{-19}\,{\rm erg\,s^{-1}\,cm^{-2}}$\,\AA$^{-1}$ (Figure \ref{fig:sens})---as measured in a $1''$ spatial 
aperture. ``$z_{Q}$'' is the \grizli-characterised redshift quality/risk (lower is better), which correlates
well with redshift errors estimated as $\Delta z\equiv\,$84th$-$16th $P(z)$ percentile. Figure \ref{fig:zqual} 
shows $P(z)$ PDFs for each quartile of the $z_{Q}$ distribution.

\begin{figure*}
	\centering
	\includegraphics[width = 0.8\linewidth, trim = 1cm 0cm 0cm 0cm]{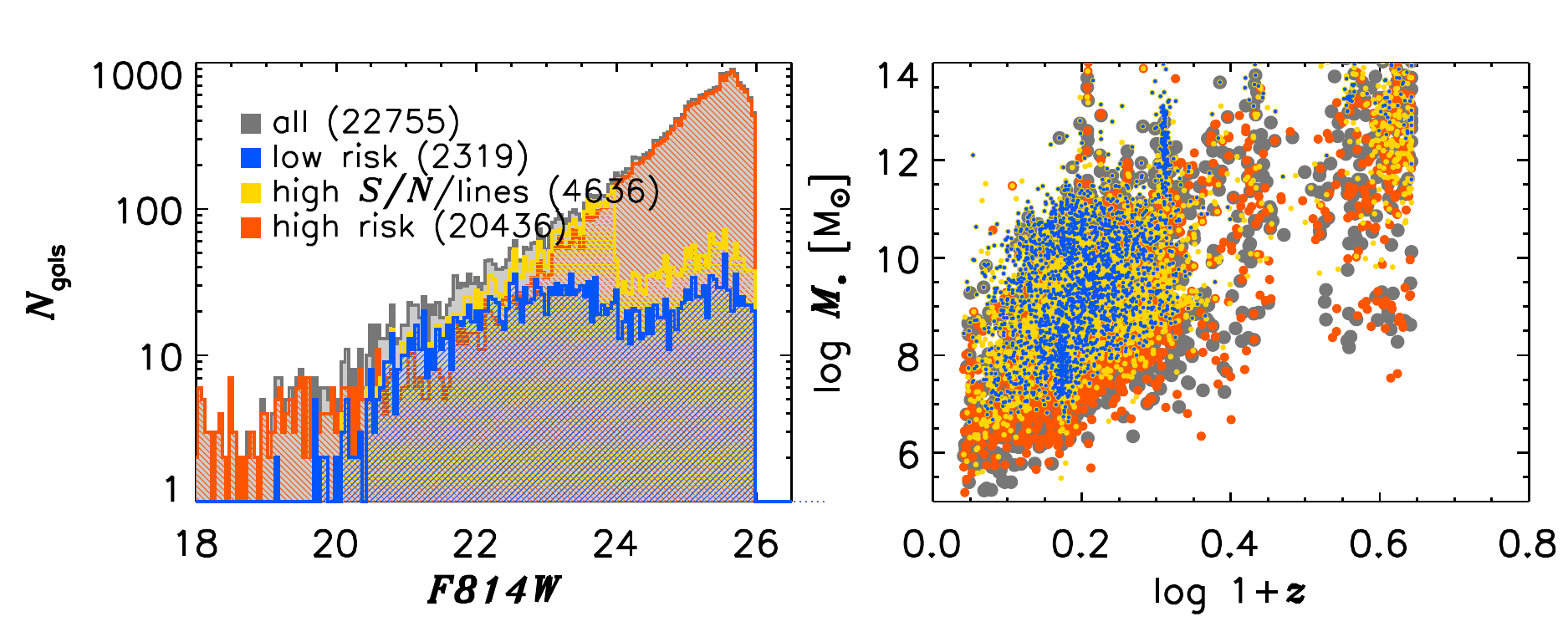}
	\caption{F814W magnitude distributions (left) and estimated stellar mass coverage as a 
	function of redshift (right) for the full and quality-defined samples. Color coding is as in 
	Figure \ref{fig:zDist}. Sources in the low risk sample fainter than $m_{814}=22$ tend
	to be high-EW line emitters. Lines therefore form the basis of redshifts in that regime, 
	with the continuum shape playing a larger role for brighter objects. High quality sources 
	reach $\logM\sim7$ and $\sim$9 at $z=\midz$ and 1.0, \resp, though
	these numbers are based on the relatively small G800L bandpass.}
	\label{fig:magDist}
\end{figure*}

\begin{figure}
	\centering
	\includegraphics[width = 0.9\linewidth, trim = 0.5cm 0cm 0cm 0cm]{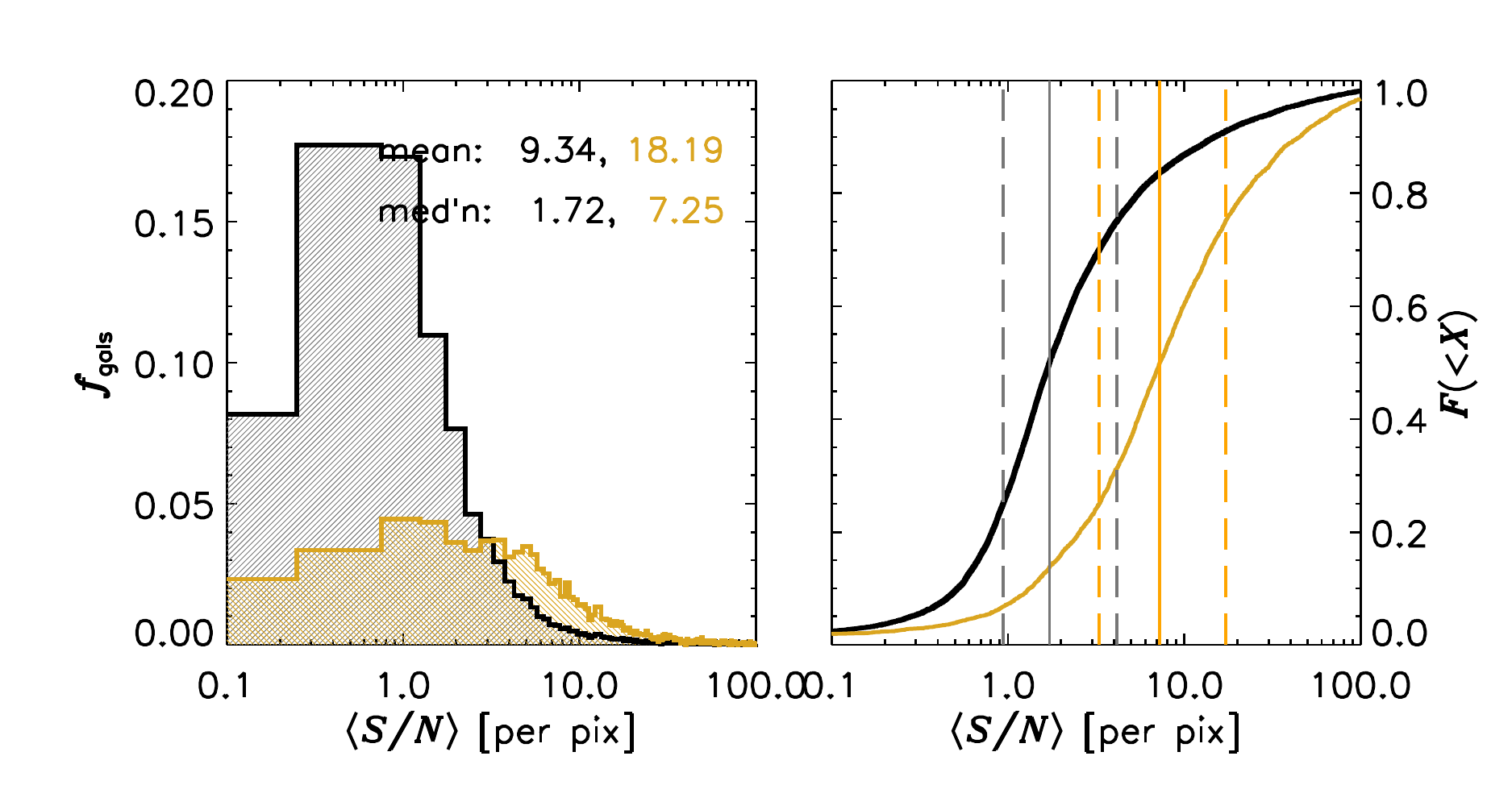}
	\includegraphics[width = 0.9\linewidth, trim = 0.5cm 0cm 0cm 0cm]{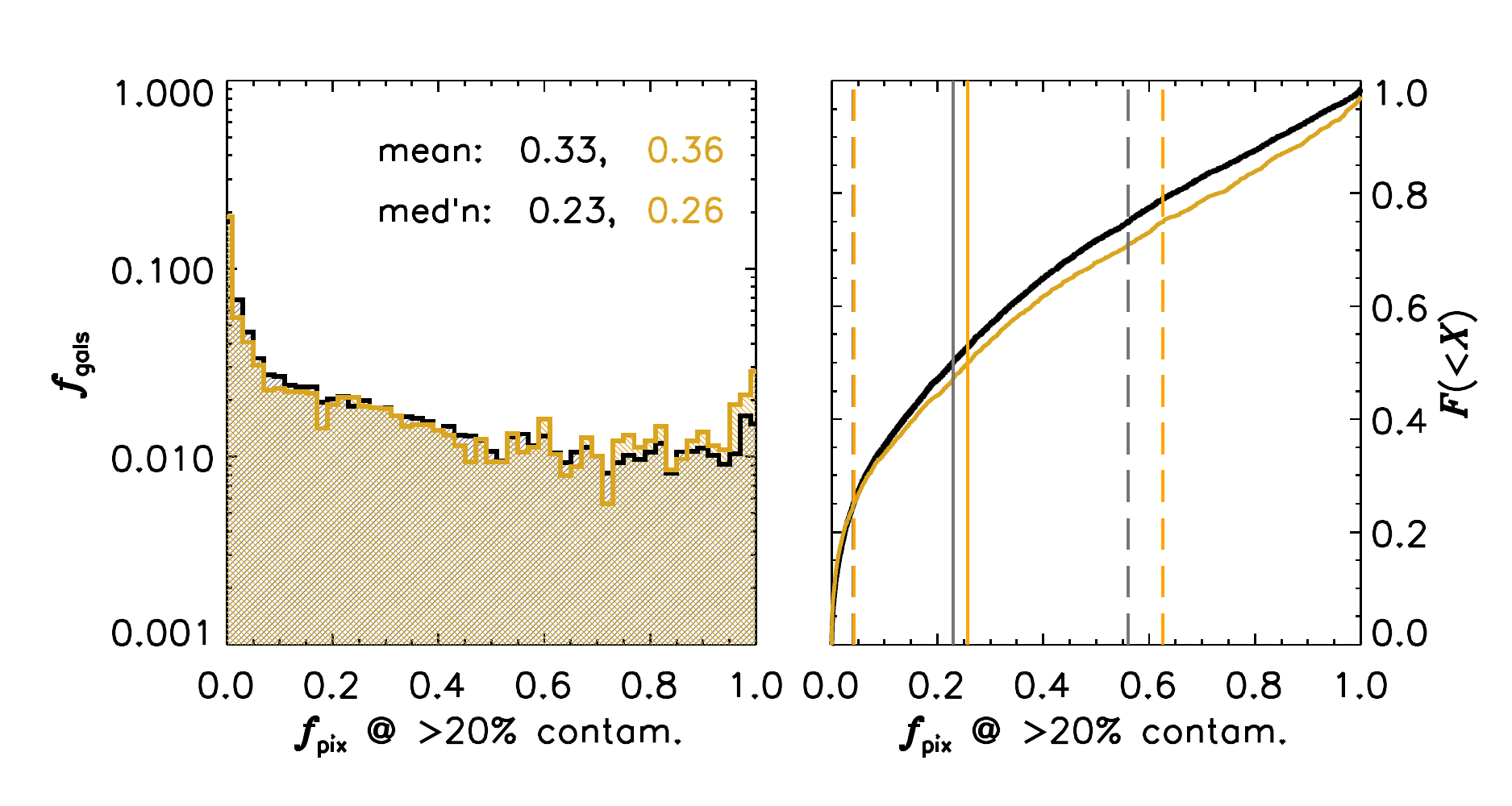}
	\includegraphics[width = 0.9\linewidth, trim = 0.5cm 0cm 0cm 0cm]{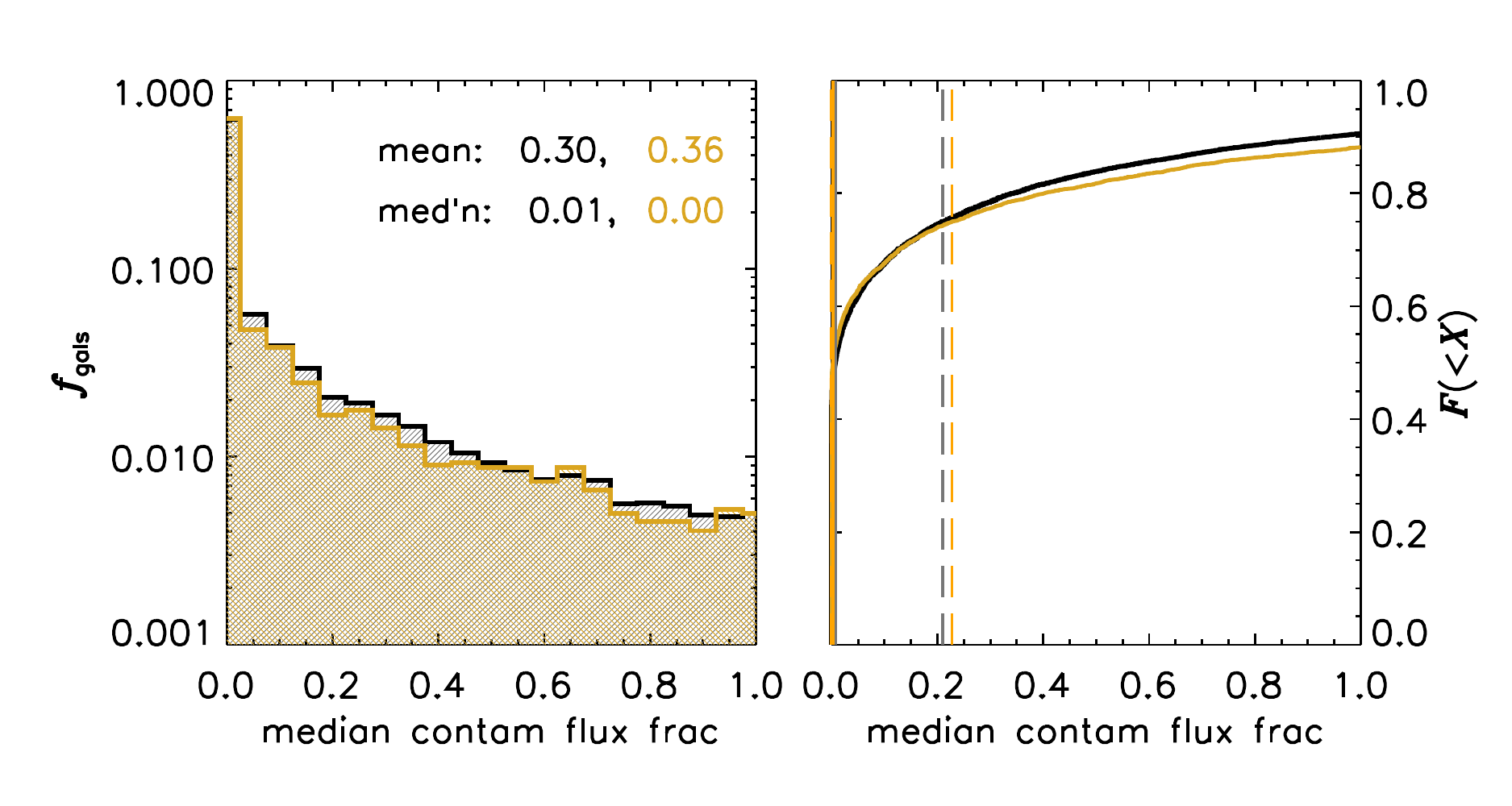}
	\caption{Top: median $S/N$ per spectral pixel from the optimal 1D 
			extractions	for all galaxies (grey) and those in the high-$S/N$+lines 
			subsample (gold). The latter has about 3$\times$ the $S/N$ as the former.
			Middle: the fraction of source 2D spectral pixels whose flux is contaminated
			by overlapping traces to at least 20\%. Unlike in the WFC3 data, fully half
			of G800L GLASS spectra have fewer than 20\% substantially contaminated pixels.
			Bottom: median contaminating flux per pixel per source; another metric of 
			contamination. Cumulative distributions for all quantities are shown at right: 
			75\% of objects have a median per-pixel contamination 
			of $<$20\%; 75\% of high-$S/N$+lines objects have typical 
			continuum $S/N>4~\rm pix^{-1}$}		
	\label{fig:sn1}
\end{figure}

\begin{figure}
	\includegraphics[width = \linewidth, trim = 0.5cm 0.5cm 0cm 0cm]{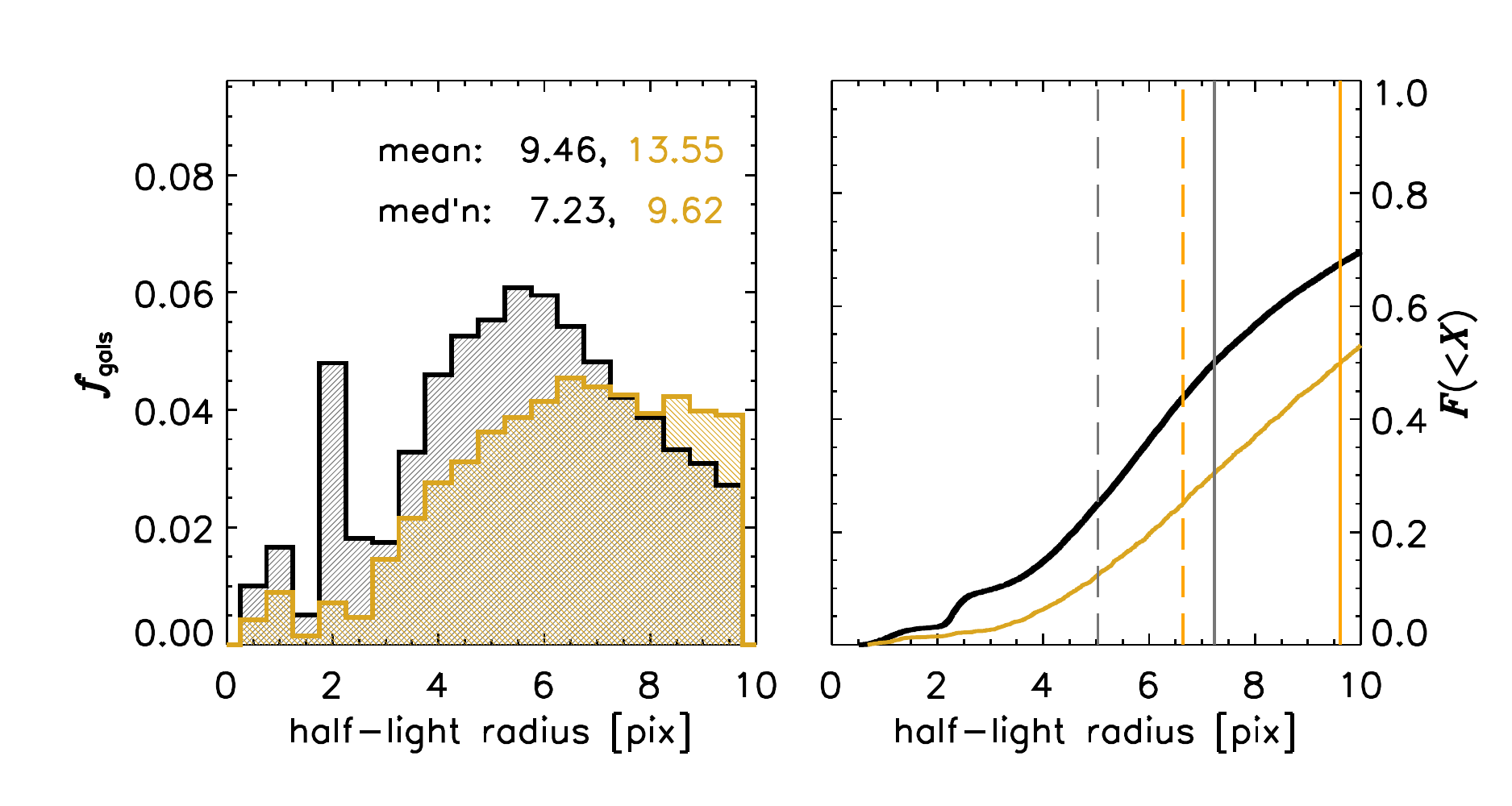}
	\caption{Sample size distribution. The high-$S/N$+lines sample comprises larger
			sources on average, contributing to their slightly higher contamination levels.}
	\label{fig:sn2}
\end{figure}

Figure \ref{fig:zDist} shows the sources' inferred redshift distribution (left) and redshift error 
distribution (right; $\equiv\Delta z / (1+z)$). Note that, due to the inclusion of template lines spanning
[\ion{O}{ii}]$\,\lambda3727$ and redward, there is an artifact in the redshift distribution
at $z\sim1.7$ where that feature drops out of the G800L bandpass. The sample's median redshift 
is $z\simeq\midz$, irrespective of quality cuts.  Compared to the high risk subsample, the enhanced quality of the 
other objects' redshifts is clear, with the $\sim$2200 low risk systems having estimates formally 
precise to $<$1\%. Visual inspection of high quality objects suggests these redshifts are systematically
accurate to within the 2\,$\sigma$ formal uncertainties. The $R\sim40$ spectral resolution 
makes more quantitative crosschecks difficult, but comparisons to much higher resolution
VLT VIMOS spectra for 131 common objects suggest agreement is at the $\Delta z=0.01$ level
with a scatter of $\sigma_{z}=0.01$ below $z=1$ (Figure \ref{fig:specSpec}; data from \citealt{Grillo16,Caminha17,Karman17, Monna17}).\footnote{Available at \url{https://sites.google.com/site/vltclashpublic/data-release}.} 
Figure \ref{fig:magZerr} illustrates that this quality is 
typical at $m_{814}\lesssim22$ ($S/N\sim20$ per spectral pixel), and true for at least 50\% of objects 
up to a magnitude fainter.

Figure \ref{fig:magDist}, left, shows the F814W apparent magnitude distribution for 
the above samples. Half of the full sample is brighter than $m_{814} \approx 25$, though it is 
dominated by insecure redshifts at that limit. The distribution of ``high-$S/N$+lines'' objects cuts off 
sharply at $m_{814}=24$ by construction (see definition above), though counts in the low risk subsample remain relatively
flat beyond this point. Comparing these histograms amplifies Figure \ref{fig:magZerr}'s results,
suggesting that robust automated redshift estimations probably require ancillary SED information 
or spectral lines beyond $m_{814}\sim22$ in these data, so catalogue values should be used cautiously. 
Inverted, this statement provides a selection criterion: GLASS sources fainter than that limit with 
tight redshifts distributions are probably high-EW line emitters.

Figure \ref{fig:magDist}, right, presents the samples' rough stellar mass ($\Mstel$) limit. For the low
risk objects, this extends from $\logM\sim7$ at $z\sim0.6$ to $\logM\sim9$ at 
$z\sim1$---similar to the main GLASS catalogue limits \citep{Morishita17}. We
caution, however, that these estimates are based only on the information in the G800L spectrum and 
therefore may change significantly when ancillary photometry is included in the inference 
(Section \ref{sec:joint}).

Of course, all of the above depends fundamentally on the grism $S/N$. Figure \ref{fig:sn1}, top,  
shows the median per-pixel $S/N$ for the full sample (grey) and the baseline quality cut ($m_{814}\leq 24$
or any line detected at $>$5\,$\sigma$). In the case of the former, half of the spectra have 
$\langle S/N\rangle>2\,\rm pix^{-1}$, and 25\% have at least twice that.
The high-confidence sample has a median and upper quartile of $\sim$7 and $\sim$20, \resp, more
than triple the $S/N$ of the general population. We therefore recommend basing most general 
analyses on at least our ``high-$S/N$ + lines'' quality cut, and only proceeding beyond this
after more rigourous exploration of the data. Fortunately, Figure \ref{fig:sn1}'s middle and 
bottom rows reveal that contamination is often not an impediment: half the sample has 
$>$50\% of pixels free of any contamination. Alternatively, half of sources have 
$\lesssim$20\% of pixels contaminated at $>$20\%. Contamination (by either metric) is slightly 
higher for the higher quality data subset. This is likely due in part to their typically $\sim$30\% 
larger half-light radii (Figure \ref{fig:sn2}), which increases the probability of spectral collisions.

\subsection{Higher level outputs: spectral line maps}
\label{sec:lineMaps}

\begin{figure}
	\centering
	\includegraphics[width = \linewidth, trim = 0.5cm 1cm 0cm 0cm]{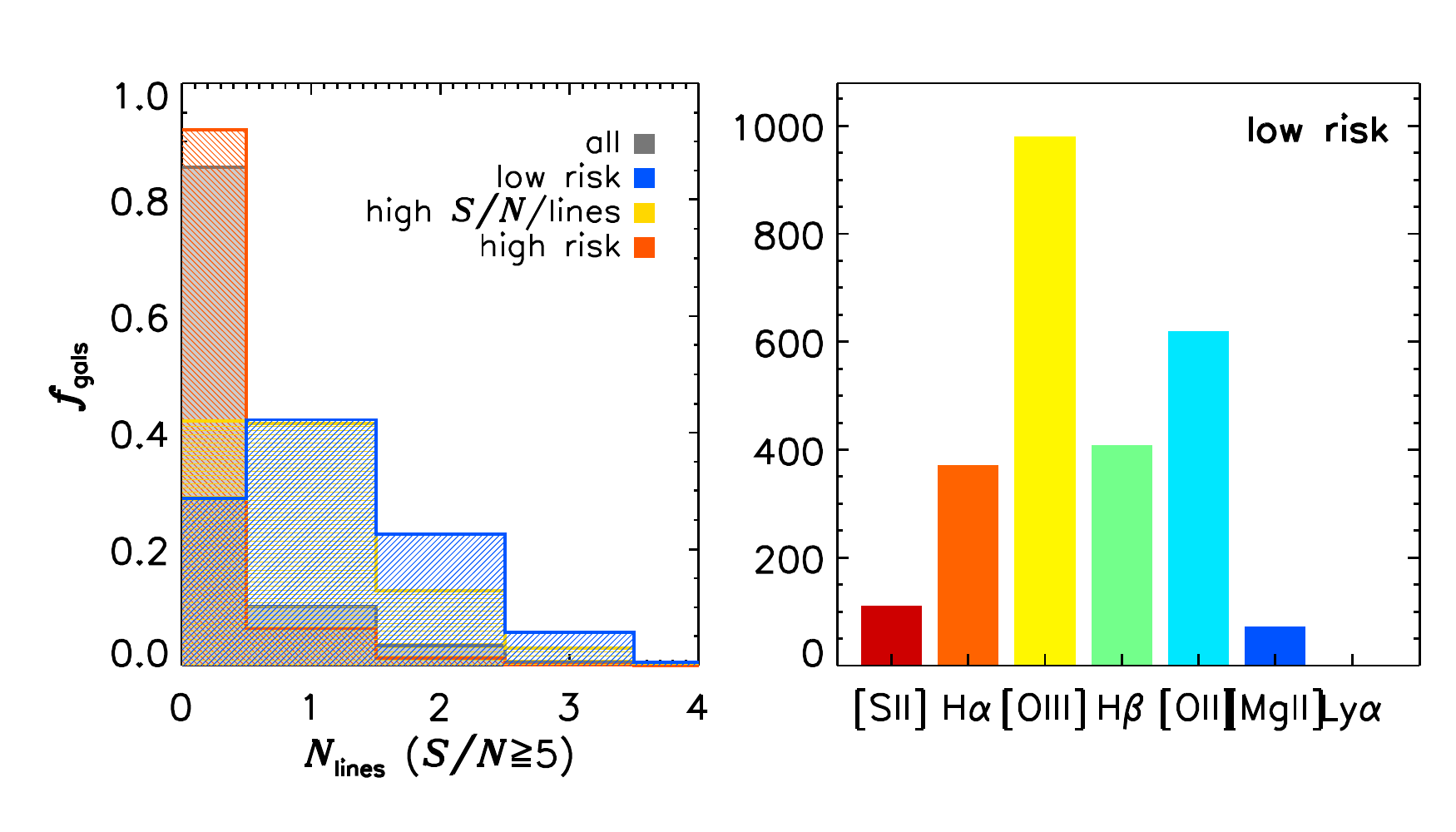}
	\caption{Line detection statistics for all galaxies and quality subsamples split as in
		Figure \ref{fig:zDist}. Line IDs from the low risk sample cut (blue histogram at
		left) are shown in the right hand panel. All lines have integrated $S/N\ge5$. Unsurprisingly,
		higher quality subsamples tend to have more high-confidence line detections, with 
		the mode in the full and high-risk samples being zero lines. Of detections, the most
		common line is [\ion{O}{iii}]. {\it Note:} since the line is not explicitly modelled, 
		sources identified as [\ion{Mg}{ii}] emitters typically reflect reduction errors or noise spikes.}
	\label{fig:lines}
\end{figure}

The GLASS G800L catalogue contains automated spectral line identifications. Figure \ref{fig:lines}, left, 
shows the distribution of galaxy fractional counts as a function of the number of lines detected at
$S/N_{\rm line}\geq5$ and quality cut. The corresponding line IDs
 for the low risk sample are shown at right. As expected, the higher quality samples typically 
 have 1--2 well detected lines while the general population has zero. Of detected lines, the most 
 common are [\ion{O}{iii}], [\ion{O}{ii}], and the strong Balmer lines. 
 However, even at moderate $S/N_{\rm line}$, these IDs should be treated with care: visual inspection 
 suggests $\sim$20--40\% of sources at fixed $N_{\rm lines}\in[1,3]$ may reflect spurious 
 characterisations. Furthermore, because the line is not explicitly modelled, objects identified as
 [\ion{Mg}{ii}] emitters nearly entirely reflect reduction errors or noise spikes and so should
 not be taken {\it prima facie} as, e.g., AGN or Lyman continuum leaker candidates \citep{Finley17,Henry18}. 
Appendix \ref{sec:algorithms} contains example code to extract high confidence line emitters.

\begin{figure*}
	\includegraphics[width=0.475\linewidth]{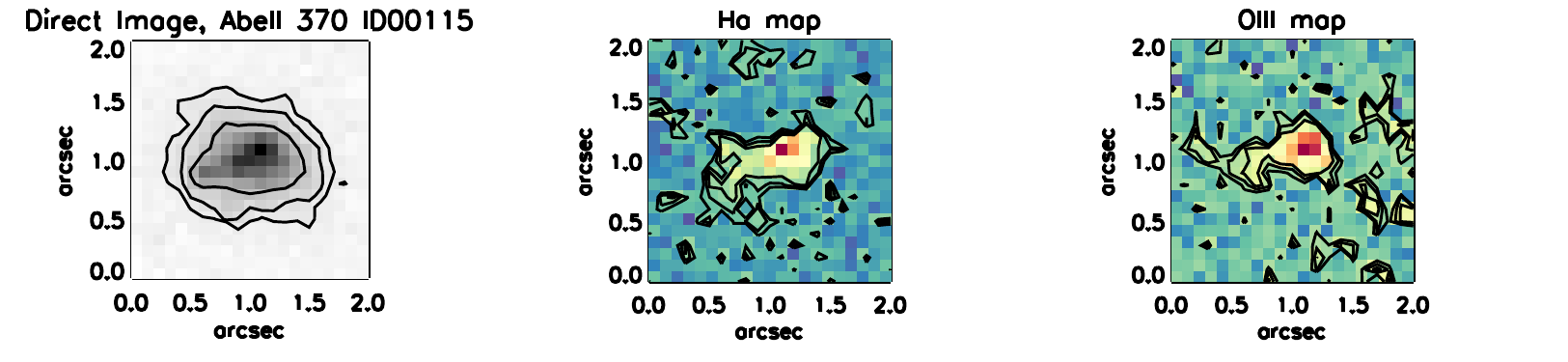}
	\includegraphics[width=0.475\linewidth]{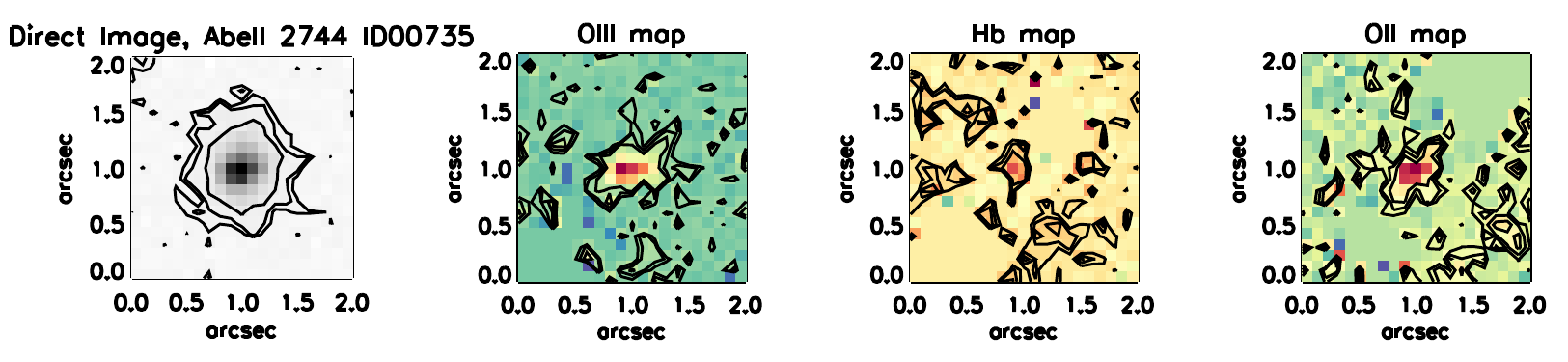}
	\includegraphics[width=0.475\linewidth]{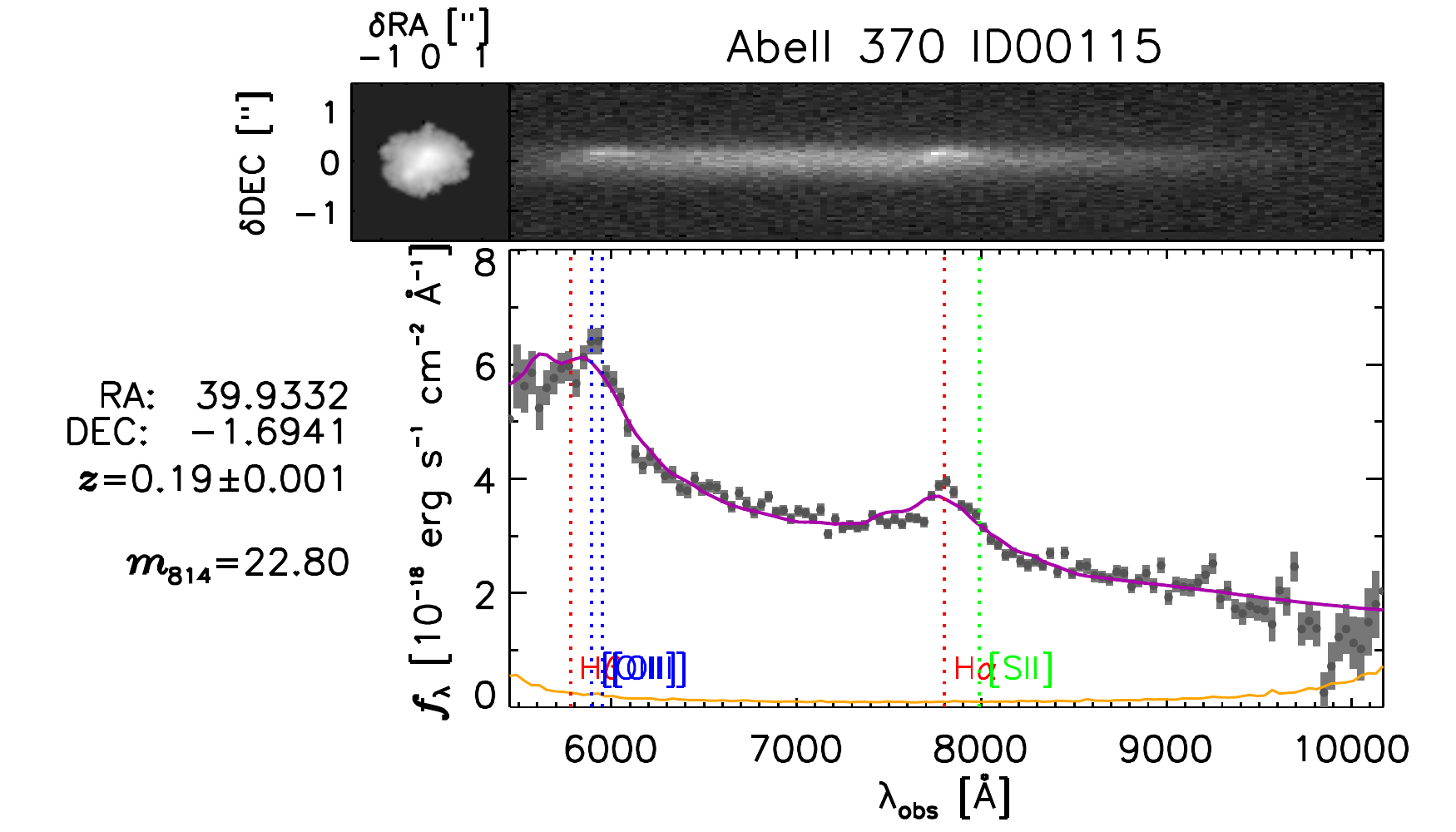}    
	\includegraphics[width=0.475\linewidth]{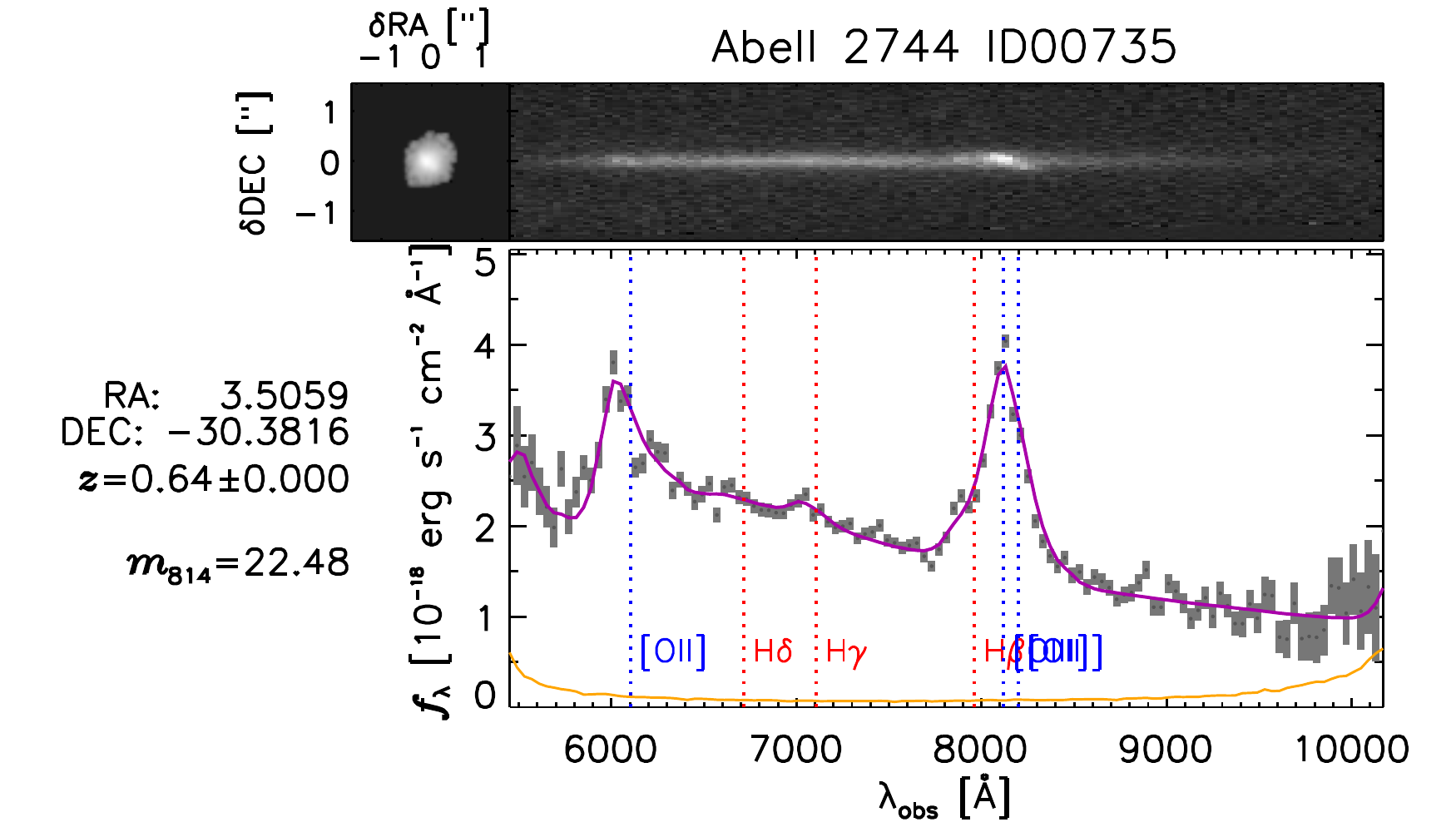}
	\caption{Top: line maps for two low risk subsample sources. Contours show 70, 80, 90\% flux contours.
			Bottom: the sources' direct images and 2D spectra above their 1D optimal extractions
			with \grizli\ model (purple) and line IDs overlaid (blue = oxygen, red = hydrogen, 
			green = sulphur). The
			source on the right shows a typical redshift systematic offset for these sources. Ancillary
			source details are printed to the left of both spectra (coordinates, redshift, flux).}
	\label{fig:lineMaps}
\end{figure*}

In addition to the total fluxes and errors for every line in every source, \grizli\ outputs
2D spectral cutouts in the relevant wavelength regions. Figure \ref{fig:lineMaps} shows examples
of these line maps for two high-quality sources with their direct images and full 1- and 2-D
spectra. The source on the left is a typical $z\sim0.2$ starforming galaxy with prominent and extended
H$\alpha$. The source on the right is a high-EW [\ion{O}{iii}] emitter at $z\sim0.65$ with quite
different oxygen and H$\beta$ morphologies, perhaps suggesting the presence of an active galactic
nucleus (though H$\beta$ is poorly resolved). We discuss these maps further in Section \ref{sec:lineMapping}.

\subsection{Notable sources}
\label{sec:specialSources}

\begin{figure*}
	\includegraphics[width=0.475\linewidth]{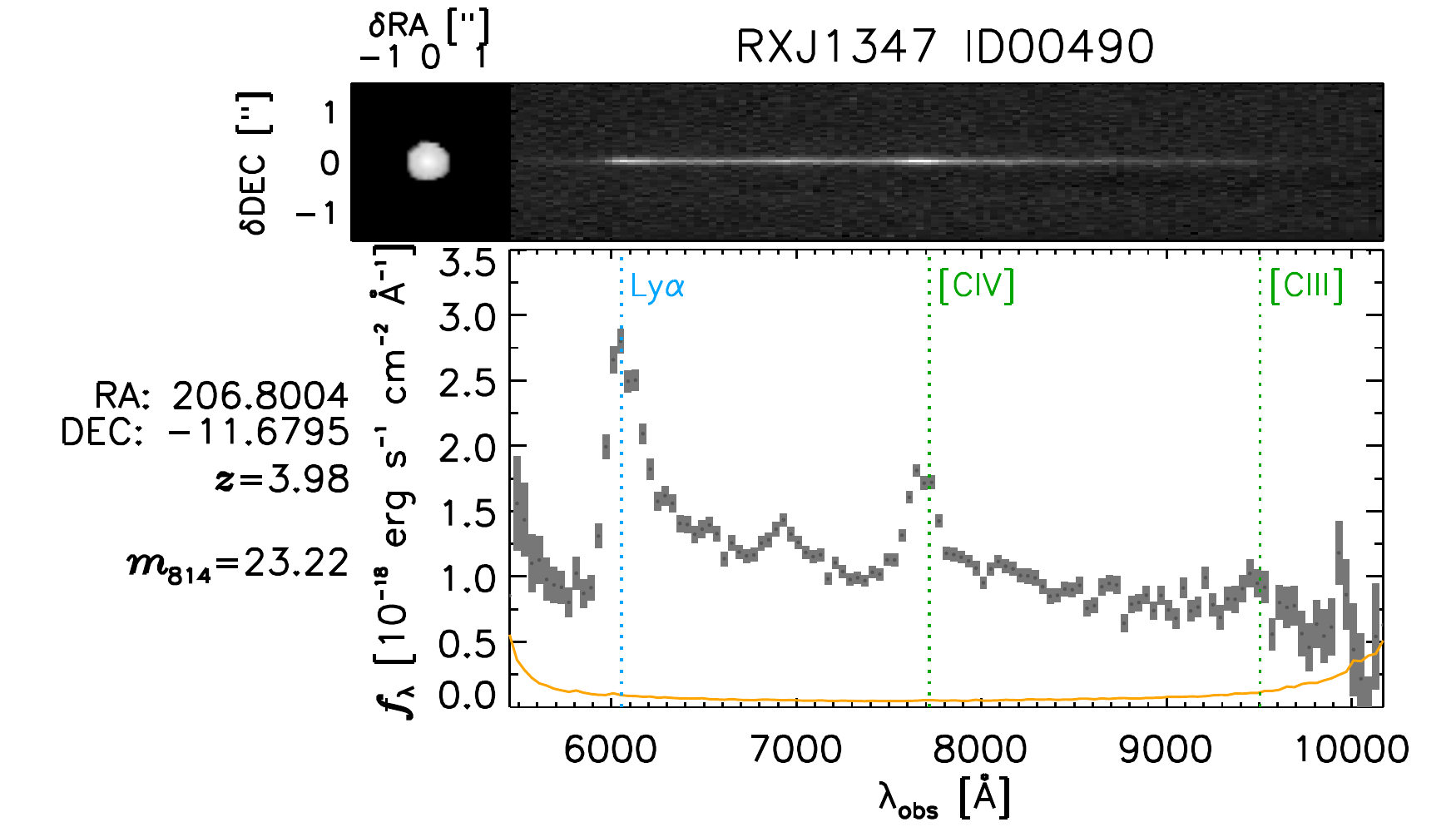}
	\includegraphics[width=0.475\linewidth]{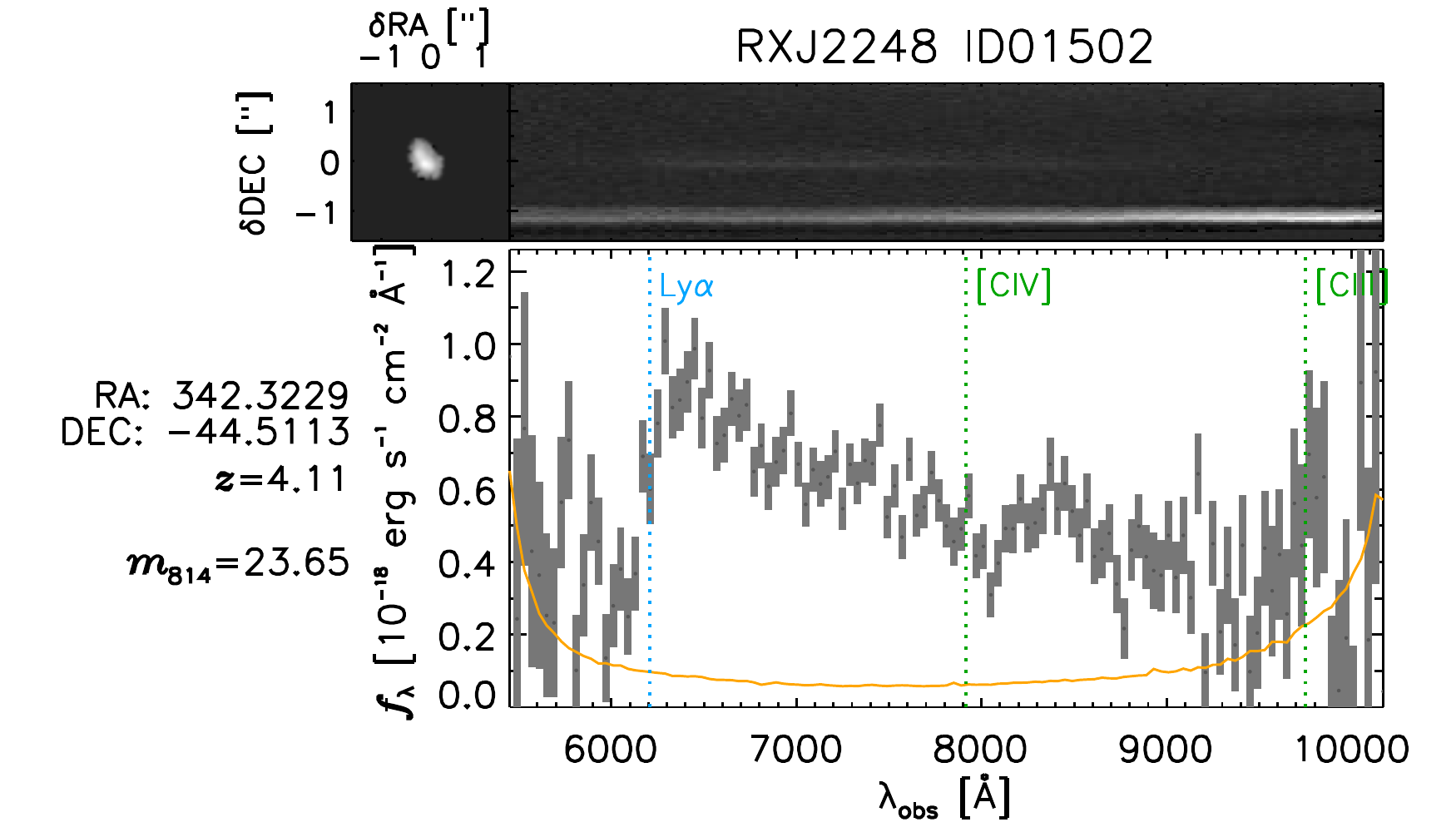}
	\caption{Examples of two known redshift misclassifications. Both
		the QSO (left) and LBG (right) are at $z\sim4$ but were initially placed
		at $z=0.89$ and $z=0.70$, \resp. More examples of the $\sim$300
		bright, $S/N\geq5$ [\ion{O}{iii}] emitters in the GLASS G800L database are
		shown in Appendix \ref{sec:moreExamples}.}
	\label{fig:notable}
\end{figure*}

Slitless spectroscopy avoids the need for photometric preselection.
As such, it maximises the chance of serendipitously capturing interesting sources. Figure \ref{fig:notable}
shows two of these in the GLASS G800L footprint that are also known redshift misclassifications (due to 
the modelling of lines only at and blueward of [\ion{O}{ii}]$\,\lambda3727$ in the template spectra; 
see Section \ref{sec:cts}). As revealed by its prominent Ly$\alpha$,
carbon lines, and pointlike morphology, the source at left is a $z\sim4$ quasar. Of course, GLASS data 
simultaneously provide a redshift survey along this QSO's line of sight, and thus immediately suggest 61 
(164) foreground candidates with a $\rho\leq150$ (300)\,kpc impact parameter from the quasar
suitable for characterizing \ion{H}{i} and low-ionization species such as \ion{Mg}{ii} \citep{Chen10, Prochaska11, 
Rudie12, Johnson15}. As such,
this source could support higher resolution spectroscopic follow-up to learn about the circumgalactic 
media (CGM) of scores of galaxies (though 30-m class facilities will be needed for $R\sim40,000$ studies). 
Combined with their ISM maps (Figure \ref{fig:lineMaps}), the CGM 
enrichment levels/temperatures/ionization states derived from such follow-up could provide powerful 
empirical constraints on metal transport, and therefore evolutionary models 
\citep[e.g.,][]{Dave11b,Dave17, Peeples14, Muratov15}. Both of these studies are enabled by slitless spectroscopy.

Figure \ref{fig:notable}, right, shows another $z\sim4$ source. As opposed to the QSO, this object's
lack of Ly$\alpha$ emission, strong Ly$\alpha$ break, and extended morphology reveal it to be a 
Lyman Break Galaxy (LBG). Large samples of such objects at these redshifts exist 
\citep[e.g.,][]{Bouwens15}, but spectroscopy remains difficult. The low backgrounds in 
space and avoidance of slit losses, however, make HST efficient at spectroscopically identifying LBGs. 
Indeed, at $m_{814}=23.8$, this object shows that GLASS' data reach continuum 
levels comparable to those from surveys on 10-m class ground-based telescopes for similar integration 
times \citep[albeit at lower spectral resolution;][]{Steidel99, Shapley03}. 

\begin{figure}
	\includegraphics[width = \linewidth]{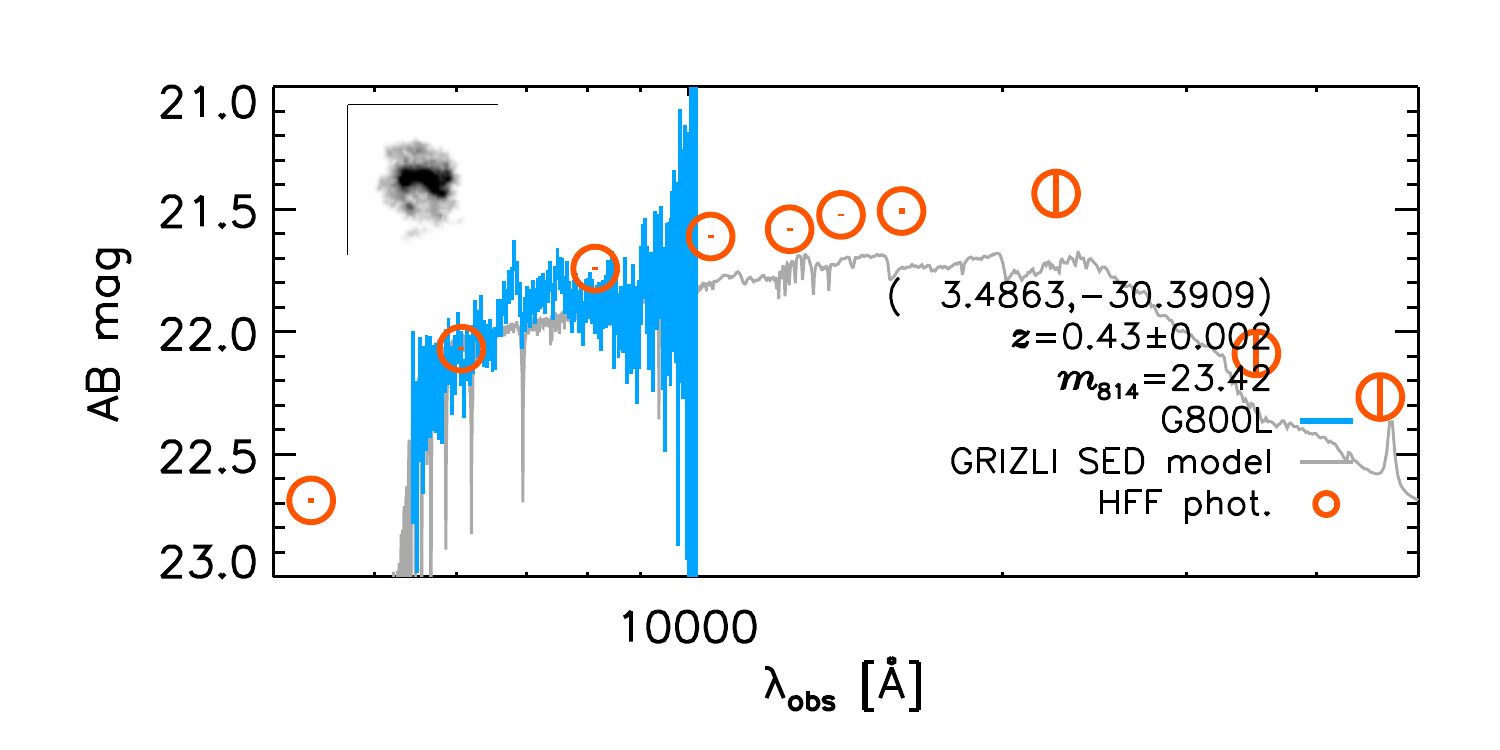}\\
	\includegraphics[width = \linewidth]{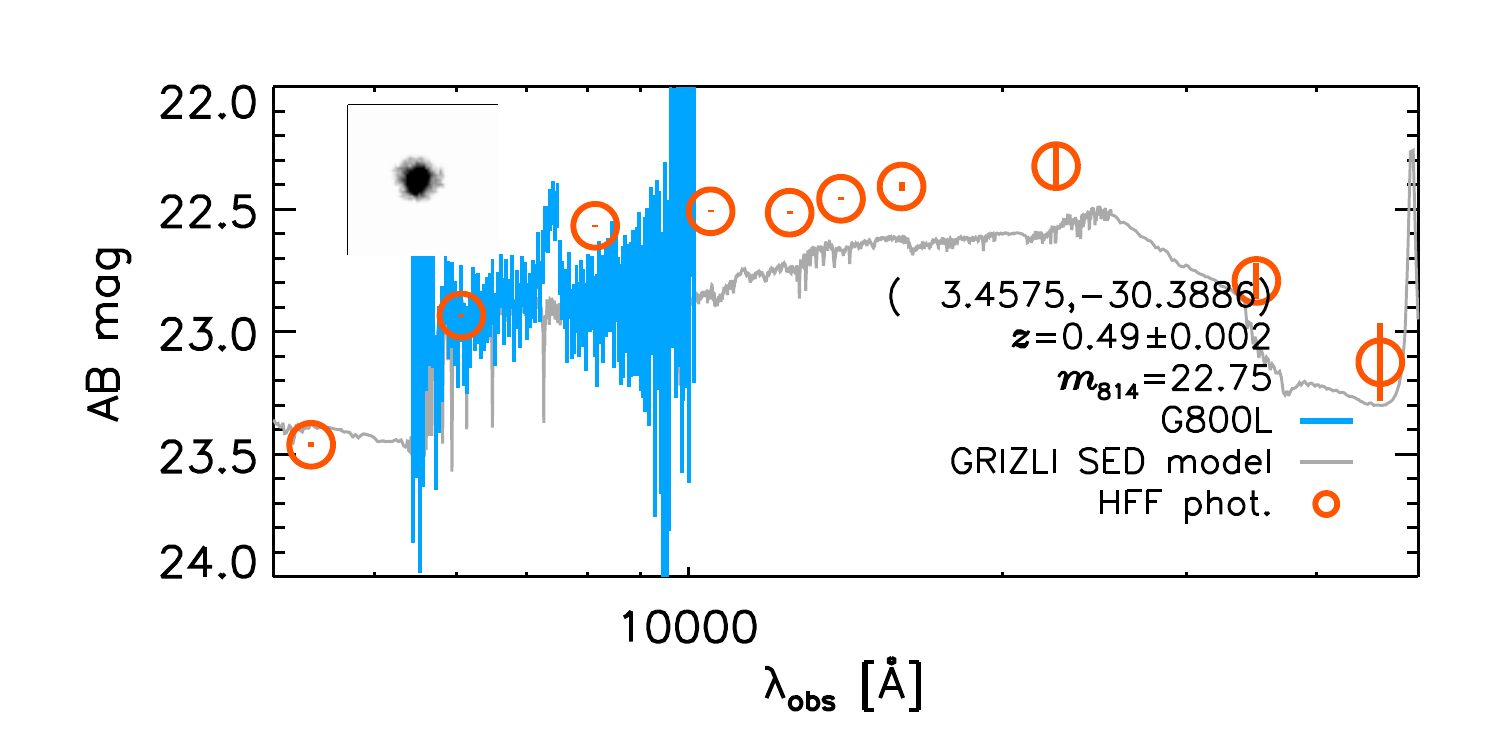}\\
	\includegraphics[width = \linewidth]{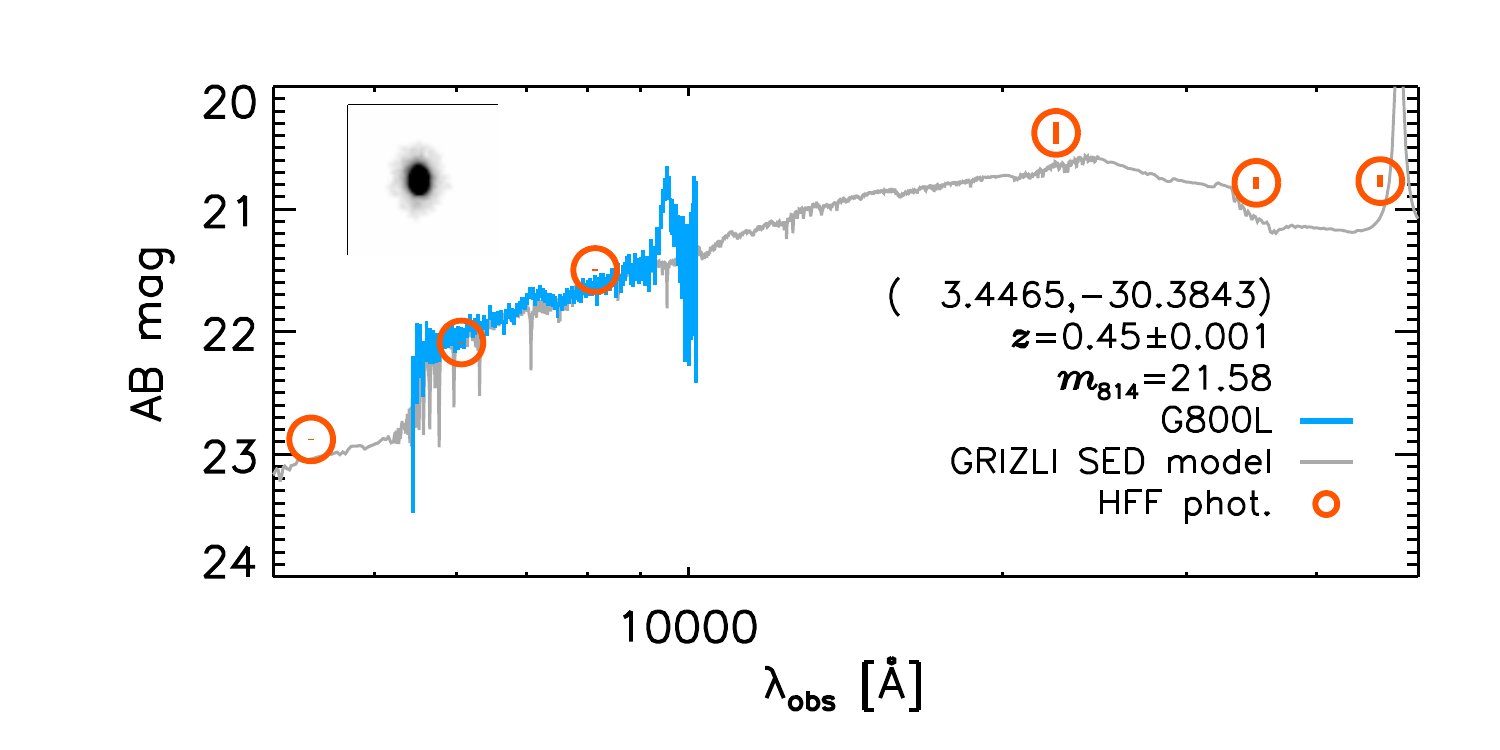}
	\caption{Examples of GLASS G800L spectra (blue) and \grizli-inferred continuum SEDs (grey)
			shown with matched source photometry from the HFF campaign 
			\citep[][orange circles]{Shipley18}. Some inferences agree remarkably well
			with photometry lying well outside the grism bandpass despite its 
			limited coverage; others are less consistent. Such
			offsets---typically at the $\sim$0.25\,mag level (Section \ref{sec:joint})---will 
			affect stellar mass and SFR estimates. However, the consistent
			location of observed and inferred features (e.g., the 1.6\,\micron\ bump) support
			the accuracy of these sources' grism-determined redshifts (see also Figure \ref{fig:photoz}). 
			These data are available to anyone wishing to perform joint spectrophotometric 
			SED fitting on the  the 13 fields where HST photometry exist (Table \ref{tbl:fieldInfo}). 
			More examples of low risk objects in the HFF footprints are shown in Appendix \ref{sec:moreExamples}.}
	\label{fig:specPhotExamples}
\end{figure}


\section{Discussion and Further Applications}
\label{sec:discussion}

The GLASS G800L database will support a range of investigations. We note three 
potentially fruitful avenues below beyond the QSO and LBG follow-up studies just 
discussed.

\subsection{Infalling cluster members}
\label{sec:infall}

As shown in Table \ref{tbl:fieldInfo}, each pointing in the catalogue contains $\sim$20--40
objects within $2000~\km\,\rm s^{-1}$ of the HFF or CLASH cluster redshifts. These are
likely galaxies either falling into the cluster potential for the first time, or ``splashing back''
after one or more crossings through the clusters \citep{More15,Baxter17}. Both cases present 
opportunities to study the forces affecting galaxies in the densest environments in the Universe.
We encourage anyone interested in ram pressure stripping \citep{GG72} or preprocessing 
\citep{ZabludoffMulchaey98} to explore these
objects. Analogous analyses to \citet[][see Section \ref{sec:lineMapping}]{Vulcani15,Vulcani16,Vulcani17} but
based on [\ion{O}{iii}] may, for example, prove powerful.

\subsection{Joint analyses with HST photometry}
\label{sec:joint}

Over half of the sources in the database lie in regions covered by CLASH, HFF, or other
HST imaging. Given the limited bandpass of the G800L grism, incorporating these data
into any further SED fitting is valuable, especially for analyses that rely on accurate stellar 
mass or SFR inferences. Figure \ref{fig:specPhotExamples} shows the GLASS spectra 
overlaid on HFF photometry for three of the 859 common GLASS/\citet{Shipley18} sources 
in the Abell 2744 parallel field. For low risk sources, the relative fluxing between the grism 
data and HFF photometry is good to $\Delta m_{814}=0.1$ mag (such that HFF fluxes are brighter) 
with a scatter of 0.25\,mag about that offset.\footnote{This measurement accounts for zeropoint and
Milky Way extinction, but not aperture corrections, to the HFF data.} The grism-derived continuum 
SED model extrapolated well outside the G800L bandpass can be predictive to similar 
levels---low risk median $m_{\grizli}-m_{\rm HFF}= [0.19, 0.25, 0.25, 0.25]$ in [F105,125,140,160W] with
$>$0.5\,mag scatters around these offsets---but larger disagreements in the IR and UV can obviously occur. 

\begin{figure*}
	\centering
	\includegraphics[width = \linewidth]{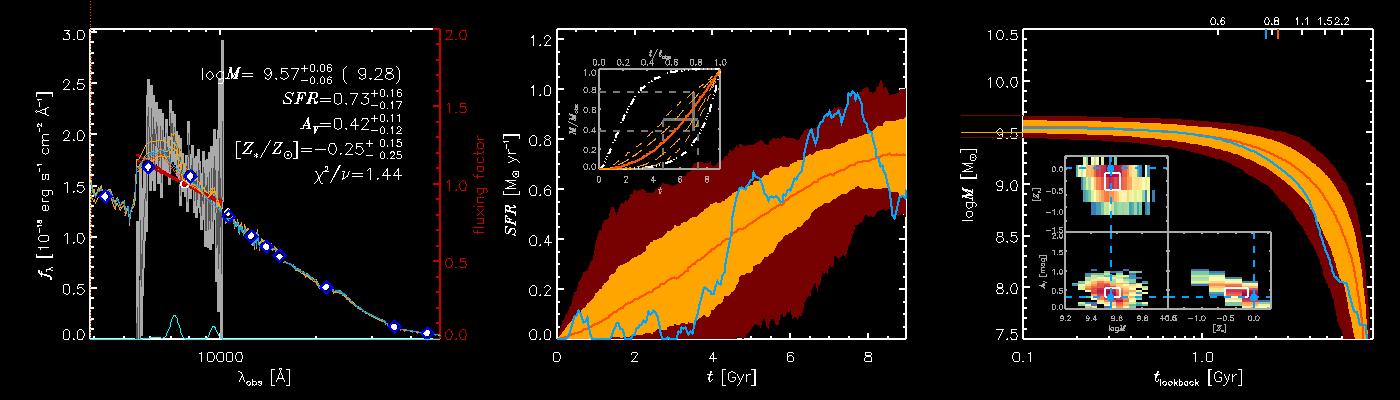}
	\caption{Left: a GLASS G800L spectrum of a starforming galaxy at $z = 0.44$ (grey bars showing 1\,$\sigma$ errors)
			with HFF + {\it Spitzer} photometry from \citet{Shipley18} (white circles). The maximum
			likelihood \citet{BC03} SED fit is overlaid in blue (photometry in diamonds) with the 5\%, 50\%, and
			95\% probability fits in orange. The inferred contribution from emission lines is plotted in cyan at bottom.
		Middle: the SFH ($\sfr(t)$) inferred from this SED using a library of 3000 $H=1$ 
			stochastic model SFHs
			from \citet{Kelson14}. The blue line is the maximum likelihood output, with the rust- and mustard-coloured 
			bands showing the 95\% and 68\% confidence envelopes, \resp, and the median plotted in orange. 
			The inset shows the fractional mass growth history (same intervals) with the 95\% envelope from the
			prior plotted in hatched white lines.
		Right: the inferred absolute mass growth history with $\Mstel$, $Z_{*}$, and $A_{V}$ covariances
			plotted in the inset. According to this fit---which marginalises over the choice of a Reddy, Calzetti, 
			SMC, and Milky Way dust law---this galaxy has an observed $\logM=9.3\pm0.06$ (\citealt{Chabrier03} IMF)
			with a half-mass redshift of $z\simeq0.8$ and has grown in a relatively linear fashion for at least 
			$0.5\,t_{\rm Hubble}$. The machinery that performed this fit will be described in a future paper
			(L.~Abramson, in preparation), but is similar to that described by \citet{Pacifici12}. The GLASS G800L
			data combined with extant photometry will support analyses of SFHs derived using these and 
			other tools.
			 }
	\label{fig:sfhRecons}
\end{figure*}

Moreover, the combination of low resolution grism spectroscopy covering, e.g., the Balmer
or 4000\,\AA\ breaks with broadband photometry to the red and blue is now being used to great 
effect in inferring the star formation {\it histories} (SFHs) of galaxies at $0.3 < z < 3$, not just observed 
quantities \citep[][]{Dressler16, Dressler18, Abramson18b, Morishita18}. Figure \ref{fig:specPhotExamples}
shows the GLASS data will support similar analyses at $z\sim0.4$--1.25, and Figure \ref{fig:sfhRecons}
shows an example estimate (L.~Abramson, in preparation). Critically, given their high
spatial resolution and low contamination, these data could support spatially resolved SED
analyses to constrain individual galaxies' joint mass and structural evolution over at least the past 
$\sim$1--2~Gyr. These empirical inferences can be compared to simulations to provide direct, 
longitudinal tests of numerical physical prescriptions, not just bulk predictions
for the galaxy population at large. \citet{Abramson18b} performed such an analysis, but were limited to 
just 4 systems due to the high contamination rates in GLASS' central WFC3 pointings. The G800L data 
do not suffer from this issue, so spatially resolved spectrophotometric SFH reconstructions based on
functional [e.g., by {\tt pyspecfit} \citep{Newman14}, or other means \citep{Iyer17}] or free-form inferential 
techniques \citep{Pacifici12, Kelson14a, Leja17, Morishita18} should yield a valuable database of hundreds of high 
quality mass, SFR, and structural histories over large ranges in {\it observed} mass, SFR, and structural
parameters.

\subsection{Emission line mapping}
\label{sec:lineMapping}

As mentioned in Sections \ref{sec:intro}, \ref{sec:lineMaps}, Figure \ref{fig:lineMaps} illustrates 
one of the key advantages of HST and future space-based slitless spectroscopy: the automatic production 
of spatially resolved spectral features mapped at the diffraction limit---600\,pc at $z = 0.6$ 
(${\rm FWHM}=0\farcs09$). These maps can be used, for example, to infer galaxy metallicity distributions
and even outflow patterns at sub-kpc scales to challenge numerical feedback and star formation models 
in new ways. For example, \citet{Vulcani15,Vulcani16,Vulcani17} used the G102 H$\alpha$ maps from GLASS' 
central pointings to study the connection between galaxy stellar and gas morphology, and associate this
with various features of the local and global environment in clusters. \citet{Jones15} and \citet{Wang17} 
used oxygen and
H$\beta$ to produce gas-phase metallicity maps and gradients for sources at $z\sim1$--3 and unprecedentedly
low stellar masses ($\log\Mstel\lesssim8$). \citet{Wang18} identified two such sources with steeply positive
gradients ($[{\rm O/H}] \propto r$) and used simple models to produce outflow maps and mass 
loading factors as a function of underlying stellar mass density, showing that individual systems are not 
described well by {\it only} energy or momentum driven wind models. The GLASS G800L data will 
support similar studies based on [\ion{O}{ii}] through H$\alpha$ lines
at $z\lesssim1$, both outside and in the infall regions around massive clusters. Further two-dimensional 
explorations of the ISM will no doubt be fruitful.


\section{Summary}
\label{sec:summary}

We present a catalogue of \ntot\ objects with ACS G800L slitless spectroscopy from the 
20 GLASS parallel fields (Table \ref{tbl:fieldInfo}). The catalogue extends to $m_{814}=26$ 
with uniform 7-orbit ($\sim$10\,ks) coverage over $\sim$220 sq.\ arcmin. Sources have 
a median redshift of $z=\midz$ with median uncertainties of $\Delta z / (1+z)\lesssim0.02$ 
at $m_{814}\leq23$ (Figures \ref{fig:zDist}, \ref{fig:magZerr}) and are typically
contaminated at the 0\%--20\% level (Figure \ref{fig:sn1}). About a quarter of the sample either 
has continuum flux detected at $\geq$5\,$\sigma$ ($m_{814}\lesssim24;~f_{\lambda}
\approx5\times10^{-19}~{\rm erg~s^{-1}~cm^{-2}~\AA^{-1}}$) or $S/N>5$ in at 
least one spectral line ($f_{\rm lim}\approx5\times10^{-17}~{\rm erg~s^{-1}~cm^{-2}}$) such that median 
redshift errors are $<$1\%. Incorporating photometry 
for the 13 fields that overlap with extant HST imaging from CLASH, HFF, or other programs will 
allow redshifts to be obtained to much greater depth and support a rich variety of spectrophotometric 
studies into galaxy star formation histories at $z\sim0.5$--1.5 (Section \ref{sec:joint}).

The full catalogue also contains full 2D spectra, contamination, and source models for each object,
along with automated line identifications, fluxes, uncertainties, and 2D spatially resolved
maps produced by \grizli. Optimal 1D extractions are also provided, as are full UV--sub-mm 
best-fit SEDs and redshift PDFs. All data and derived products are available on request, and will
be published on MAST along with code for performing various basic operations---which we also give 
here in Appendix \ref{sec:algorithms}.

These data and products will support a wide range of investigations---from mapping
galaxy outflows at $z<0.8$ (Sections \ref{sec:lineMaps}, \ref{sec:lineMapping}) to identifying 
QSOs and LBGs at $z\sim4$ (Section \ref{sec:specialSources})---and serve as useful 
intuition-builders as the field prepares for the increasing ubiquity of slitless spectroscopy
in the eras of JWST and WFIRST.\\

\noindent\facilities\ HST ACS.

\noindent\software\ IDL (Coyote libraries; \url{http://www.idlcoyote.com/}), python (\grizli).


\section*{Acknowledgements}

L.E.A.~thanks Camilla Pacifici and Dan Kelson for inspiring and aiding in the 
writing of the SFH inference code debuted here. GLASS (HST-GO-13459) is supported 
by NASA through a grant from STScI 
operated by AURA under contract NAS 5-26555. This work uses data and catalogue 
products from HFF-DeepSpace, funded by the National Science Foundation and STScI.
This work also uses catalogues derived from data from VLT programmes 186.A-0798, 
094.A-0115(B), 094.A-0525(A), 60.A-9345(A), and 095.A-0653(A).




\bibliographystyle{mnras}
\bibliography{/Users/labramson/lit} 




\appendix
\label{sec:appendix}

\section{File Names, Formats, and Contents}
\label{sec:fileContents}

The core component of the GLASS G800L database is the catalogue master FITS table:
\benum
	\item[] {\tt \mastercat}.
\eenum
This table contains field names and IDs for all extracted sources---from which their
data filenames can be constructed---along with a wealth of summary metrics described 
at the end of this Appendix.

The master FITS file is designed for the database's organization into 20 folders---one for each 
ACS pointing. These are named by their field centroid coordinates in the format:
\benum
	\item[]{\tt jRARA$^{\rm p}_{\rm m}$DEDE}, 
\eenum
such that `{\tt j0014m3023}' corresponds to the Abell 2744 parallel at PA 323.\footnote{A program
to translate these coordinates to a cluster field name is available on request and will be posted to the 
MAST website.}
These correspond to the {\tt ROOT} column in the master catalogue. Within each folder, 
all objects are identified by a source number, corresponding to the {\tt ID} column. 
As such, the master catalogue can be used to point to the {\tt ii}-th object's datafiles by writing:
\begin{lstlisting}[backgroundcolor=\color{lightgray}]
	folder = mastercat[ii].ROOT
	sourceID = string(mastercat[ii].ID, f = '(FI05)')
	sourceFileBase = folder+'/'+folder+'_'+sourceID
\end{lstlisting}
which will construct the base filename from which a source's data files are built (Appendix \ref{sec:algorithms}).

\subsection{Individual source files}

As mentioned in Section \ref{sec:sampChar}, each source is associated with three files. These
are defined by the following suffixes appended to their base ID {\tt jRARA$^{\rm p}_{\rm m}$DEDE\_0IDID}:

\benum
	\item ``{\tt .stack.fits}'' -- A 9 layer 2D FITS table containing:
	\benum
		\item[{\tt [1]}] {\tt SCI} -- A source's rectified, drizzled 2D spectral cutout;
		\item[{\tt [2]}] {\tt WHT} -- inverse variance array;
		\item[{\tt [3]}] {\tt CONTAM} -- contamination model;
		\item[{\tt [4]}] {\tt MODEL} -- source model;
		\item[{\tt [5]}] {\tt KERNEL} -- F814W direct image used to convolve the 
								2D trace to a common spatial sampling. 
	\eenum
	Layers 1, 2, 4, and 5 are then repeated with the source resampled to a common PA.
	Since each GLASS source is observed only once, this PA is set to the PA of observation, 
	such that the layers are identical (to $<$1\%).  This file's header also contains the 
	wavelength solution, exposure time, RA/DEC, and other ancillary information.
	\item ``{\tt .full.fits}'' -- An $N\geq5$ layer 2D FITS table containing:
	\benum
		\item[{\tt [1]}] {\tt ZFIT\_STACK} -- A 6-column table containing redshift
					estimation information. This includes the {\tt ZGRID} redshift
					array, $P(z)$ distribution ({\tt PDF}), redshift {\tt RISK} 
					and {\tt CHI2} goodness of fit distributions,
					and SED template/redshift covariance matrix ({\tt COVAR});
		\item[{\tt [2]}] {\tt COVAR} -- Covariance matrix of the template fit coefficients at the 
		best-fit redshift (taken to be where $P(z)$ is maximised from the previous extension).
		\item[{\tt [3]}] {\tt TEMPL} -- A 3-column table containing the best fit SED ({\tt FULL}),
					continuum-only model ({\tt CONTINUUM}), and wavelength array
					spanning $0.02<\lambda/\micron<2\times10^{4}$ ({\tt WAVE});
		\item[{\tt [4]}] {\tt DSCI} -- F814W source direct image;
		\item[{\tt [5]}] {\tt DWHT} -- F814W image weight map;
		\item[{\tt ([6])}] {\tt LINE001} -- Map of {\tt LINE001} in the file header;
		\item[{\tt ([7])}] {\tt CONTINUUM} -- Continuum near {\tt LINE001} in the file header;
		\item[{\tt ([8])}] {\tt CONTAM} -- Contamination near {\tt LINE001} in the file header;
		\item[{\tt ([9])}] {\tt LINEWHT} -- Weight map  near {\tt LINE001} in the file header;
		\item[{\tt (...)}] {\tt LINE002} -- As above, but for {\tt LINE002} in the file header;
	\eenum
	Layers 6--$N$ will not be present in an object with no entries in the master catalogue's {\tt HASLINES}
	column (see below) or no lines in this file's header. Otherwise, e.g., H$\alpha\beta\gamma\delta$, 
	[\ion{O}{ii}] and [\ion{O}{iii}], [\ion{S}{ii}] have names ``Ha, Hb, Hg, Hd, OII, OIII, SII,'' etc.
	There is one set of 4 layers for each line. The header also contains the wavelength solution, 
	exposure time, RA/DEC, line fluxes, and other ancillary information.
	\item ``{\tt .1D.fits}'' -- A 6 column 1D FITS table containing:
	\benum
		\item[{\tt [1]}]  {\tt WAVE} -- Wavelength array spanning only the G800L bandpass
					($5450 < \lambda/$\AA\,$< 10170$);
		\item[{\tt [2]}] {\tt FLUX} -- Optimally extracted, contamination subtracted 1D source
					spectrum ($\rm e^{-}\,s^{-1}$);
		\item[{\tt [3]}] {\tt ERR} -- 1\,$\sigma$ noise on {\tt FLUX} ($\rm e^{-}\,s^{-1}$);
		\item[{\tt [4]}] {\tt FLAT} -- Sensitivity curve needed to transform the above into $f_{\lambda}$
							($\rm e^{-}\,s^{-1}/erg\,s^{-1}\,cm^{-2}\,\AA^{-1}$);
		\item[{\tt [5]}] {\tt LINE} -- Best fit galaxy SED template, including emission lines.
		\item[{\tt [6]}] {\tt CONT} -- Best fit galaxy SED template, continuum only.
	\eenum
\eenum

The {\tt getfname} function in Appendix \ref{sec:algorithms} will return the filenames for all of the above FITS 
tables given a source's {\tt ROOT} and {\tt ID} entry in the GLASS master catalogue.

\subsection{Master catalogue column definitions}
\label{sec:catCols}

The following entries are found in the GLASS G800L master catalogue FITS table:

\bitem
\setlength\itemsep{0ex}
	\item {\tt  ROOT} -- ACS field (string)
	\item {\tt  ID} -- Source ID in {\tt ROOT} (int)
	\item {\tt  RA} -- Source RA (J2000, decimal deg.; double)
	\item {\tt  DEC} -- Source DEC (J2000, decimal deg.; double)
	\item {\tt  NINPUT} -- $N$ frames in source's G800L coadd (int; 1 --14)
	\item {\tt  REDSHIFT} -- Source redshift estimate ($=z_{\rm MAP}$; float)
	\item {\tt  NUMLINES} -- Number of potential lines in source spectrum (int; 0 --10)
	\item {\tt  HASLINES} -- Emission line IDs (string; space-separated)
	\item {\tt  CHI2POLY} -- $\chi^2$ of a third-order polynomial fit to the spectrum (float)
	\item {\tt  DOF} -- Approximate spectral degrees of freedom given by the total number 
	of unmasked pixels in the ``beams'' spectra extracted from the grism exposures (float)
	\item {\tt  CHIMIN} -- Minimum $\chi^2$ of the SED template fit on a redshift grid (float)
	\item {\tt  CHIMAX} -- Maximum $\chi^2$ of the SED template fit on a redshift grid (float)
	\item {\tt  BIC\_POLY} -- ``Bayesian Information Criterion'' of the polynomial fit; 
	$\ln(\mathtt{DOF})\cdot k + (\mathtt{CHI2POLY-CHIMIN})$, where $k=4+N_b$ and 
	$N_b$ the number of spectra in the \texttt{.beams.} file, since an additive background component 
	is fit for each.
	\item {\tt  BIC\_TEMP} -- Bayesian Information Criterion of the template fit, where 
	$k=N_c + N_b$ and $N_c$ is the number of non-zero template fit coefficients.
	\item {\tt  BIC\_DIFF} -- Difference $\mathtt{BIC\_POLY}-\mathtt{BIC\_TEMP}$.  
	Large values generally correspond to cases where the galaxy SED templates much better 
	explain the observed spectrum than the trivial polynomial fit.
	\item {\tt  Z02} -- 2\,$\sigma$ redshift lower limit (float)
	\item {\tt  Z16} -- 1\,$\sigma$ redshift lower limit (float)
	\item {\tt  Z50} -- 50\% confidence redshift estimate (float)
	\item {\tt  Z84} -- 1\,$\sigma$ redshift upper limit (float) 
	\item {\tt  Z97} -- 2\,$\sigma$ redshift upper limit (float) 
	\item {\tt  ZWIDTH1} -- 1\,$\sigma$ $\Delta z$ ($84-16$; float)
	\item {\tt  ZWIDTH2} -- 2\,$\sigma$ $\Delta z$ ($97-2$; float)
	\item {\tt  Z\_MAP} -- Maximum likelihood redshift (same as {\tt REDSHIFT}; float)
	\item {\tt  Z\_RISK} -- Redshift where the ``risk'' is minimised, where risk is defined 
	as in \citet{Tanaka18} (float)
	\item {\tt  MIN\_RISK} -- Risk estimate at {\tt Z\_RISK} (float)
	\item {\tt  FLUX\_\{LINE\}} -- Line flux (erg sec$^{-1}$ cm$^{-2}$ \AA$^{-1}$; float)
	\item {\tt  ERR\_\{LINE\}} -- Line error (erg sec$^{-1}$ cm$^{-2}$ \AA$^{-1}$; float) 
	\item {\tt  EW50\_\{LINE\}} -- Median line equivalent width drawn from the 
	line-plus-continuum covariance matrix (observed-frame \AA; float)
	\item {\tt  EWHW\_\{LINE\}} -- Half-width between the 16 and 84th percentiles of the 
	EW estimates for a given line (observed-frame \AA; float)
	\item {\tt  SN\_\{LINE\}} -- Line $S/N=\tt FLUX/ERR$ (float)
	\item {\tt  CHINU} -- Spectral model fit reduced $\chi^{2}$ ($\tt CHIMIN / DOF$; float)
	\item {\tt  ZQ} -- {\tt EAZY} redshift risk assessment (float)
	\item {\tt  IDX} -- Vizier search query URL for this source (string)
	\item {\tt  ELLIPTICITY} -- {\tt SExtractor ELLIPTICITY} measure (float)
	\item {\tt  MAG\_AUTO} -- {\tt SExtractor MAG\_AUTO} (F814W; float)
	\item {\tt  FLUX\_RADIUS} -- {\tt SExtractor} half-light radius (F814W pix; float)
	\item {\tt  A\_IMAGE} -- {\tt SExtractor} semimajor axis length (F814W pix; float)
	\item {\tt  T\_G800L} -- Total G800L exposure time (seconds; float)
	\item {\tt  IS\_POINT} -- Point source flag (bit)
\eitem
where ``{\tt \{LINE\}}'' refers to each of the following:
\bitem
\setlength\itemsep{0ex}
	\item[\textbullet] \ion{He}{i} $\lambda1083$, \ion{He}{ii} $\lambda1640$; 
	\item[\textbullet] Lyman $\alpha$;
	\item[\textbullet] \ion{N}{v} $\lambda1240$, \ion{N}{iv} $\lambda1487$, \ion{N}{iii}] $\lambda1750$; 
	\item[\textbullet] \ion{C}{iv} $\lambda1549$, \ion{C}{iii}] $\lambda1908$;
	\item[\textbullet] [\ion{O}{iii}] $\lambda1663$, [\ion{O}{ii}] $\lambda3727$, 
				[\ion{O}{iii}] $\lambda4363$, [\ion{O}{iii}] $\lambda\lambda4959,5007$, 
				[\ion{O}{i}] $\lambda6302$;
	\item[\textbullet] [\ion{Mg}{ii}];
	\item[\textbullet] [\ion{Ne}{v}] $\lambda3346$, [\ion{Ne}{vi}] $\lambda3426$, [\ion{Ne}{iii}] $\lambda3867$;
	\item[\textbullet] H$\delta$, H$\gamma$, H$\beta$, H$\alpha$+[\ion{N}{ii}];
	\item[\textbullet] [\ion{S}{ii}] $\lambda\lambda6716,6731$, [\ion{S}{iii}] $\lambda\lambda9069,9532$.
\eitem

Various quantities inferred from the template SED fits ($M/L$, $\Mstel$, sSFR, and implied 
fluxes in various filters) are also given, but---as noted in the main text---these estimates are
not typically reliable and so caution against their use (Section \ref{sec:cts}).

\section{Useful Algorithms}
\label{sec:algorithms}

Below are some useful IDL routines to perform some basic data operations mentioned
in the main text.

\subsection{Return the high $S/N$+lines and low/high risk subsamples:}

\begin{lstlisting}[backgroundcolor = \color{lightgray}]
function getGoldSample, mastercat, $
                        MAGLIM = maglim, $
                        SNLIM = snlim

  if NOT keyword_set(MAGLIM) then maglim = 24.
  if NOT keyword_set(SNLIM) then snlim  = 5.
  
  ;; Define `high S/N + lines' objects
  hiQ = where(mastercat.MAG_AUTO le maglim OR $
               (mastercat.SN_HA ge snlim  OR $
                mastercat.SN_HB ge snlim  OR $
                mastercat.SN_OII ge snlim  OR $
                mastercat.SN_OIII ge snlim  OR $
                mastercat.SN_SII ge snlim  OR $
                mastercat.SN_MGII ge snlim  OR $
                mastercat.SN_LYA ge snlim) AND NOT $
                mastercat.IS_POINT, compl = nhiQ)
                
  ;; Use highQ to define `low' and `high risk'
  ;; after nulling the not-hiQ qualities
  mastercat[nhiQ].ZQ = 0 
  medrisk = median(mastercat[highQ].ZQ)
  loRisk = where(mastercat.ZQ le medrisk, $
                               compl = hiRisk)
  indices = {HIQ: hiQ, NOTHIQ: nhiQ, $
              LORISK: loRisk, HIRISK: hiRisk}

  RETURN, indices
end

\end{lstlisting}

\subsection{Match to HFF photometry from \citet{Shipley18}:}

\begin{lstlisting}[backgroundcolor = \color{lightgray}]
function loadHffPhot, hffPhotFile

  ;; Restore the Shipley et al. .save catlaog
  restore, hffPhotFile
  ninds = n_elements(ra)
  
  ;; Find the NIR intrument
  if n_elements(f_KS_HAWKI) gt 0 then begin
     fks = f_KS_HAWKI
     eks = e_KS_HAWKI
     flgks = redflag_KS_HAWKI
  endif else begin
     fks = f_KS_MOSFIRE
     eks = e_KS_MOSFIRE
     flgks = redflag_KS_MOSFIRE
  endelse

  ;; Create the SED, error, and flag arrays 
  ;; (F435W--Spitzer 4.5 um)
  fluxes = transpose([[f_F435W], [f_F606W], [f_F814W], [f_F105W], [f_F125W], [f_F140W], [f_F160W], [fks], [f_CH1], [f_CH2]])
  errs = transpose([[e_F435W], [e_F606W], [e_F814W], [e_F105W], [e_F125W], [e_F140W], [e_F160W], [eks], [e_CH1], [e_CH2]])
  flags = transpose([[redflag_F435W], [redflag_F606W], [redflag_F814W], [redflag_F105W], [redflag_F125W], [redflag_F140W], [redflag_F160W], [flgks], [redflag_CH1], [redflag_CH2]])

  ;; Add the MW dust and zeropoint corrections
  ;; (should they be necessary; key names differ 
  ;; slightly between catalogues)
  mwcorr = transpose([[mw_F435W], [mw_F606W], [mw_F814W], [mw_F105W], [mw_F125W], [mw_F140W], [mw_F160W], [1.0], [1.0], [1.0]])
  if N_ELEMENTS(zpcorr_f814w) gt 0 then $
     zpcor = transpose([[zpcorr_F435W], [zpcorr_F606W], [zpcorr_F814W], [zpcorr_F105W], [zpcorr_F125W], [zpcorr_F140W], [zpcorr_F160W], [1.0], [1.0], [1.0]]) $
  else $
     zpcor = transpose([[zpcor_F435W], [zpcor_F606W], [zpcor_F814W], [zpcor_F105W], [zpcor_F125W], [zpcor_F140W], [zpcor_F160W], [1.0], [1.0], [1.0]])

  ;; Set up the output
  nbands = n_elements(fluxes[*,0])
  results = {LAMBDA: [4350., 6060., 8140., $
                      10500., 12500., 14000., 16000., $
                      22500., 35000., 46000.], $
             SED   : fltarr(nbands), $
             ESED  : fltarr(nbands), $
             FLAGS: bytarr(nbands), $
             RA    : 0.d, $
             DEC   : 0.d, $
             MWCORR: fltarr(nbands), $
             ZPCORR: fltarr(nbands)}
  results = replicate(results, ninds)
  
  ;; Fill the output structure
  ;; get fluxes to erg/s/cm2/Hz
  ninds = n_elements(ra)
  for ii = 0, ninds - 1 do begin
     results[ii].SED = fluxes[*,ii]/10.^29.44
     results[ii].ESED = errs[*,ii]/10.^29.44
     results[ii].FLAGS = flags[*,ii]
     results[ii].RA = ra[ii]
     results[ii].DEC = dec[ii]
     results[ii].MWCORR = mwcorr
     results[ii].ZPCORR = zpcor
  endfor

  RETURN, results
end

;;
;;
;;

pro getSpecPhot, mastercat, field

  if field ne 'Abell 2744' AND $
     field ne 'MACS0717' AND $
     field ne 'MACS1149' AND $
     field ne 'RXJ2248' then $
        print, 'NO HFF ANCILLARY PHOTOMETRY AVAILABLE' $
  else begin
     
      ;; Load spectral database
     data = mrdfits(mastercat, 1)
     
     ;; Load photometry
     phot = loadhffphot(field)

     ;; Match with a 1'' radius
     spherematch, phot.RA, phot.DEC, $
     			data.RA, data.DEC, 1./3600, $
				minds, sinds, len
                  
     ;; Trim and align databases              
     data = data[sinds]
     phot = phot[minds]

  endelse      

  ... ;; Operations go here
end

\end{lstlisting}

\subsection{Get sources emitting $N_{\rm lines}$ lines at high confidence:}

\begin{lstlisting}[backgroundcolor = \color{lightgray}]
function getHighSnLines, mastercat, $
                        SNCUT = sncut

  if NOT keyword_set(SNCUT) then sncut = 5.
  
  ;; Set up the output storage
  ;; allowing for up to 10 high-sn lines
  output = {NUMLINES: 0, $
            IDS: strarr(10), $
            SNS: fltarr(10)}
  output = replicate(output, n_elements(mastercat))

  ;; Go through the lines and get their S/N
  for ii = 0, n_elements(mastercat) - 1 do begin

     lines = mastercat[ii].HASLINES
     lines = strsplit(lines, ' ', /extr)

     if mastercat[ii].NUMLINES gt 0 then begin

        tsn = fltarr(mastercat[ii].NUMLINES)
        
        for jj = 0, mastercat[ii].NUMLINES - 1 do begin
           tl = lines[jj]
           case tl of
              'Ha': tsn[jj] = mastercat[ii].SN_HA
              'Hb': tsn[jj] = mastercat[ii].SN_HB
              'OII': tsn[jj] = mastercat[ii].SN_OII
              'OIII': tsn[jj] = mastercat[ii].SN_OIII
              'SII': tsn[jj] = mastercat[ii].SN_SII
              'MgII': tsn[jj] = mastercat[ii].SN_MGII
              'Lya': tsn[jj] = mastercat[ii].SN_LYA               
              ... ;; More line names here.
           endcase
        endfor

        ;; Count and store lines about S/N threshold
        foo = where(tsn ge sncut, nlines)
        output[ii].NUMLINES = nlines
        if nlines gt 0 then begin
           output[ii].IDS[0:nlines-1] = lines[foo]
           output[ii].SNS[0:nlines-1]  = tsn[foo]
        endif
     endif
  endfor

  RETURN, answer
end

\end{lstlisting}

\subsection{Get all of the data file names for a given source:}

\begin{lstlisting}[backgroundcolor = \color{lightgray}]
function getfname, root, id

  fname = root+'/'+root+'_'+string(id, f = '(I05)')+'.XXX.fits'
  stack = repstr(fname, 'XXX', 'stack')
  full  = repstr(fname, 'XXX', 'full')
  oneD  = repstr(fname, 'XXX', '1D')
  srcName = root+'_'+string(id, f = '(I05)')
  
  fnames = {ONED: oneD, $
            FULL: full, $
            STACK: stack, $
            SRCNAME: srcName}
  
  RETURN, fnames
end
\end{lstlisting}

\section{Comparison to Extant Photometric Redshifts}
\label{sec:photoZcomp}

\begin{figure}
	\centering
	\includegraphics[width = 0.9\linewidth]{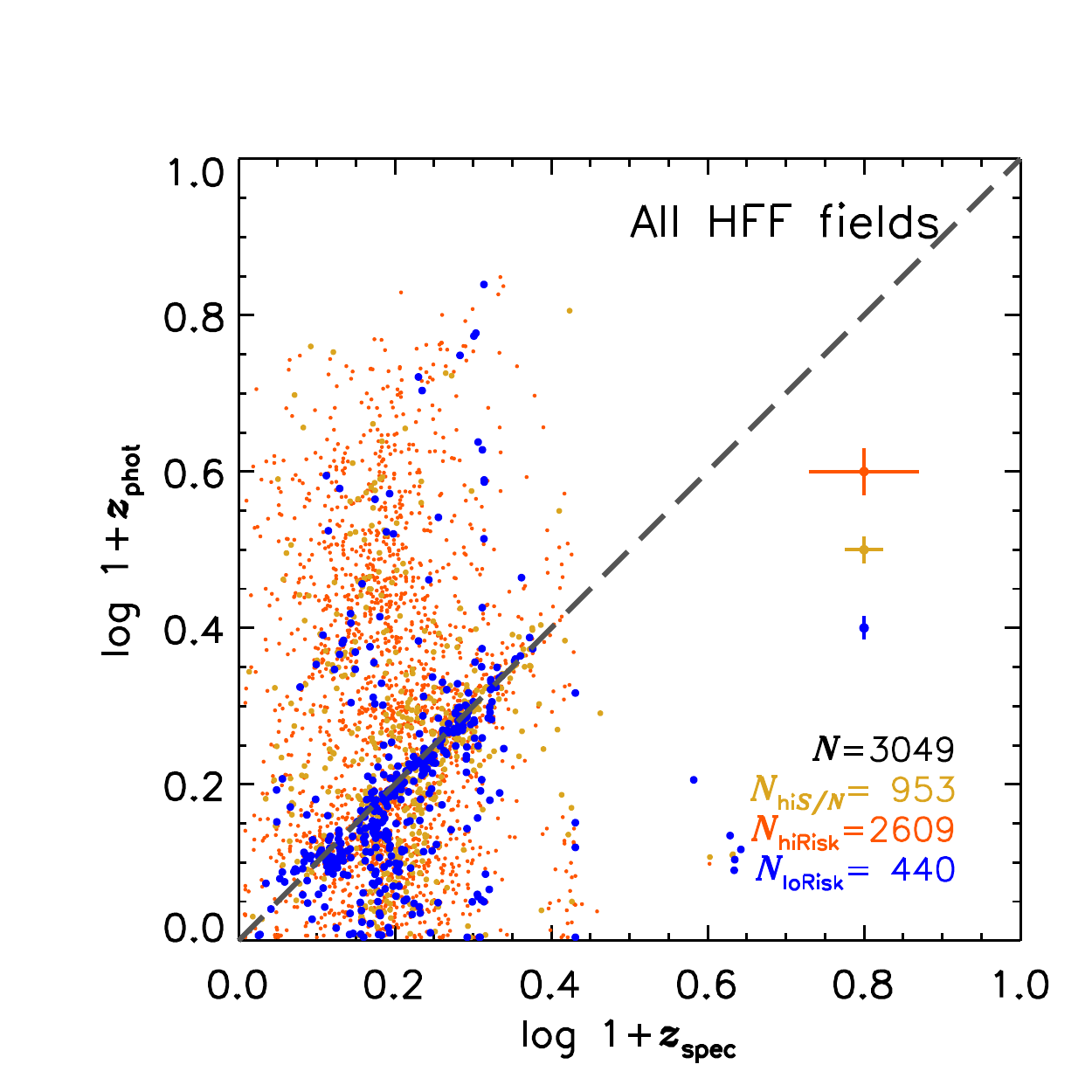}\\
	\includegraphics[width = 0.95\linewidth]{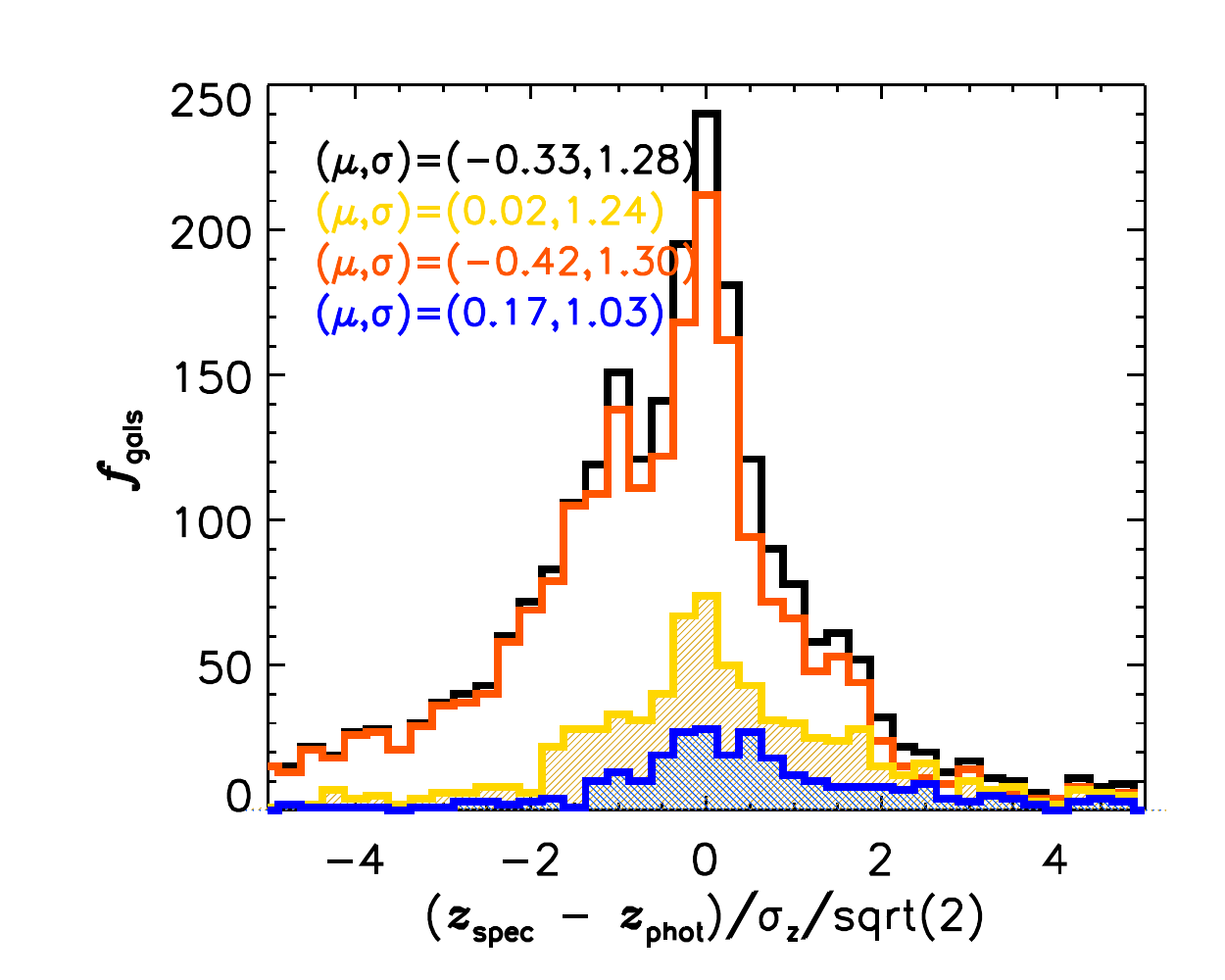}
	\caption{Similar to Figure \ref{fig:specSpec} but plotting HFF photometric redshifts
			from \citet{Shipley18} ($z_{\rm phot}$) against GLASS G800L redshifts 
			($z_{\rm spec}$). Distributions are much wider, but agreement is still 
			fair, with low risk sources having photo-$z$ errors consistent with
			gaussian noise. All samples show slight biases, but those for high-$S/N$+lines
			and low-risk sources are small in the mean.}
	\label{fig:photoz}
\end{figure}

\citet{Shipley18} provides photometric redshifts for GLASS G800L sources in the HFF 
footprint. Figure \ref{fig:photoz} compares these to the grism-only redshifts. Agreement is 
quite good for low risk sources, and unbiased for the full high-$S/N$+lines sample. The low
risk sample is consistent with gaussian errors, though errors may be slightly underestimated 
for the other samples ($\sigma/\sqrt{2}\langle{\rm err}\rangle > 1$). Nevertheless, this
analysis supports our conclusion that the $\sim$2200 low risk objects should be immediately 
useful to analyses ``straight out of the box.''

\section{Further Data Examples}
\label{sec:moreExamples}

\begin{figure*}
	\includegraphics[width=0.475\linewidth]{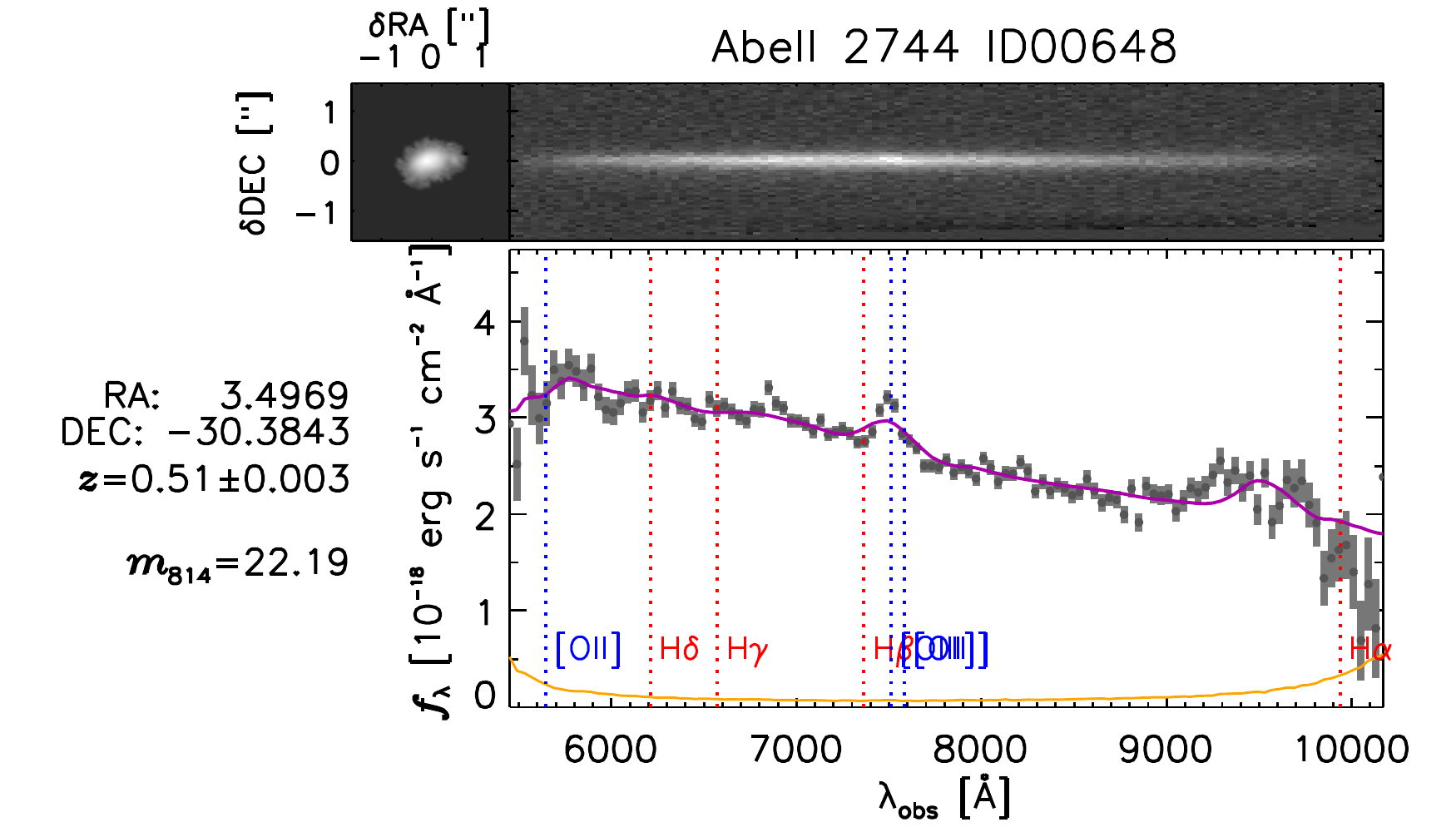}
	\includegraphics[width=0.475\linewidth]{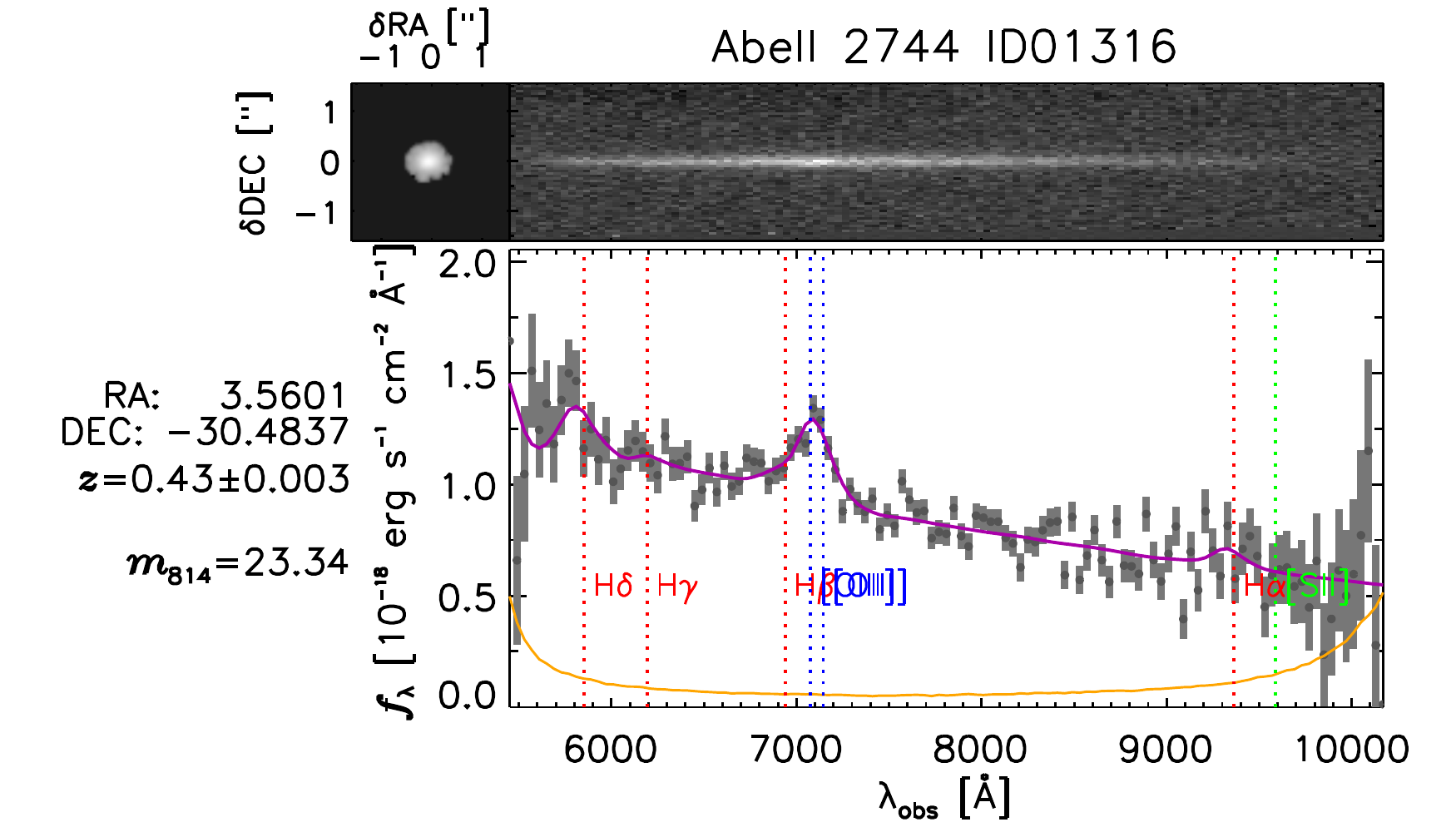}
	\includegraphics[width=0.475\linewidth]{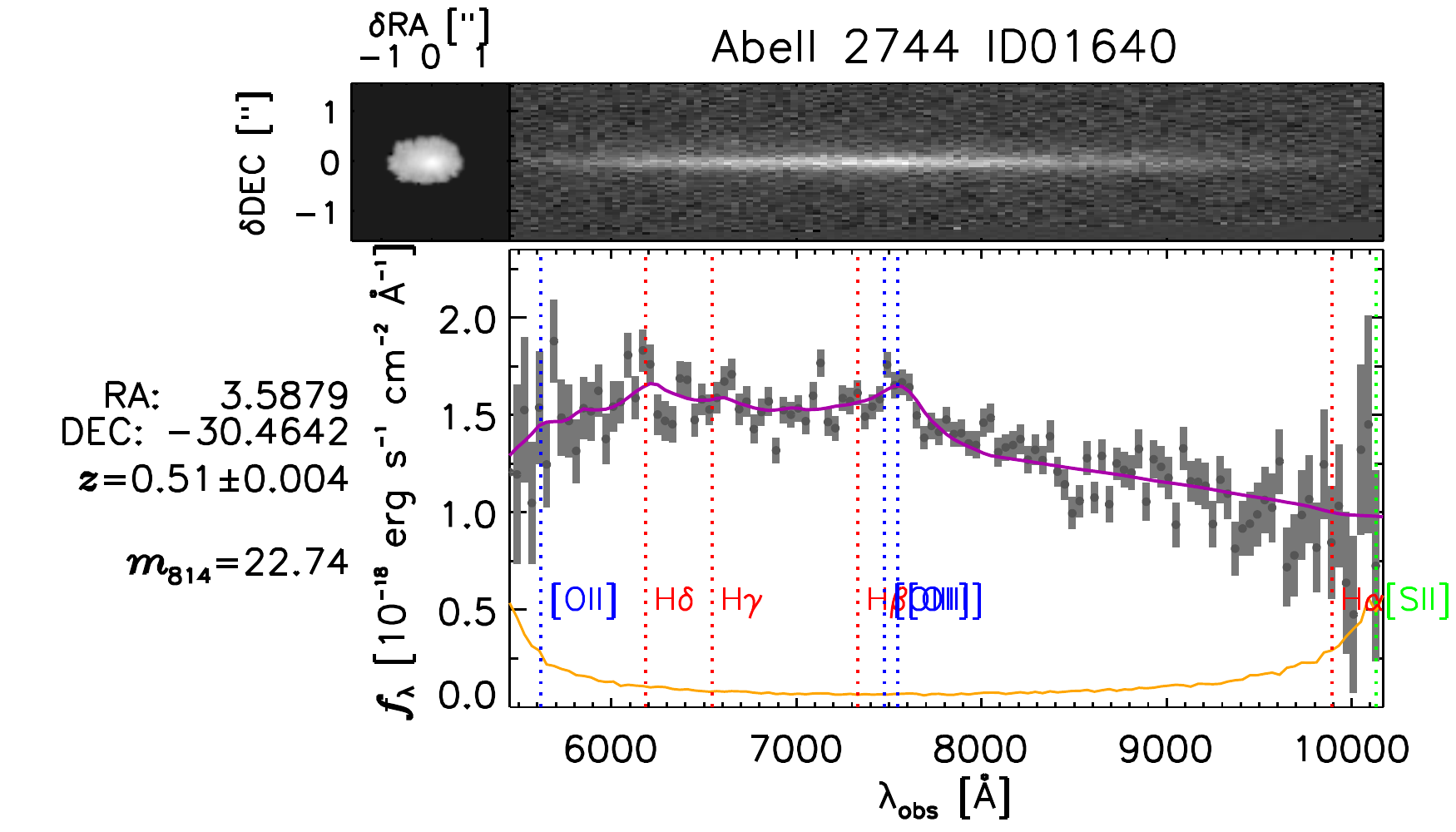}
	\includegraphics[width=0.475\linewidth]{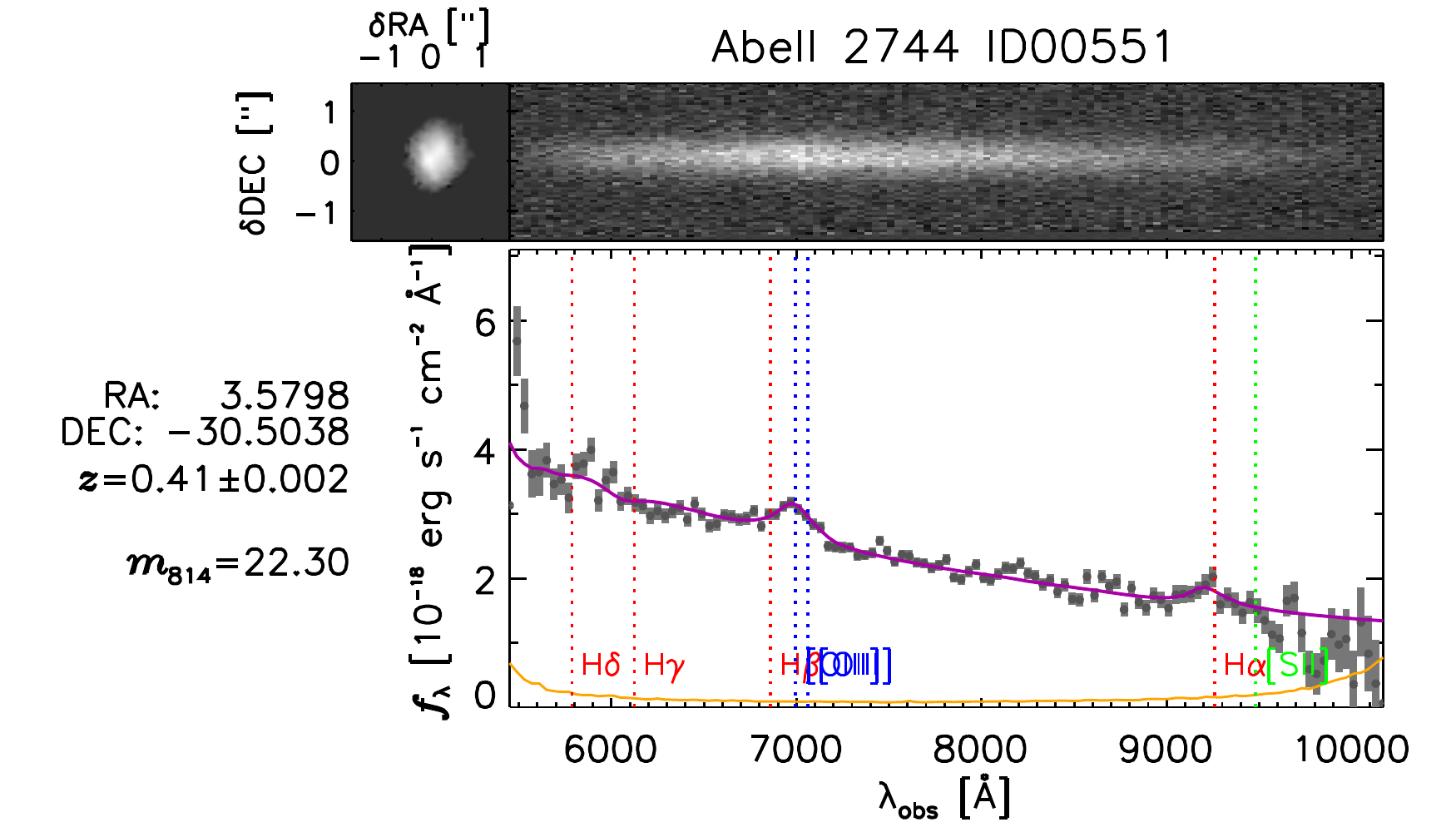}
	\includegraphics[width=0.475\linewidth]{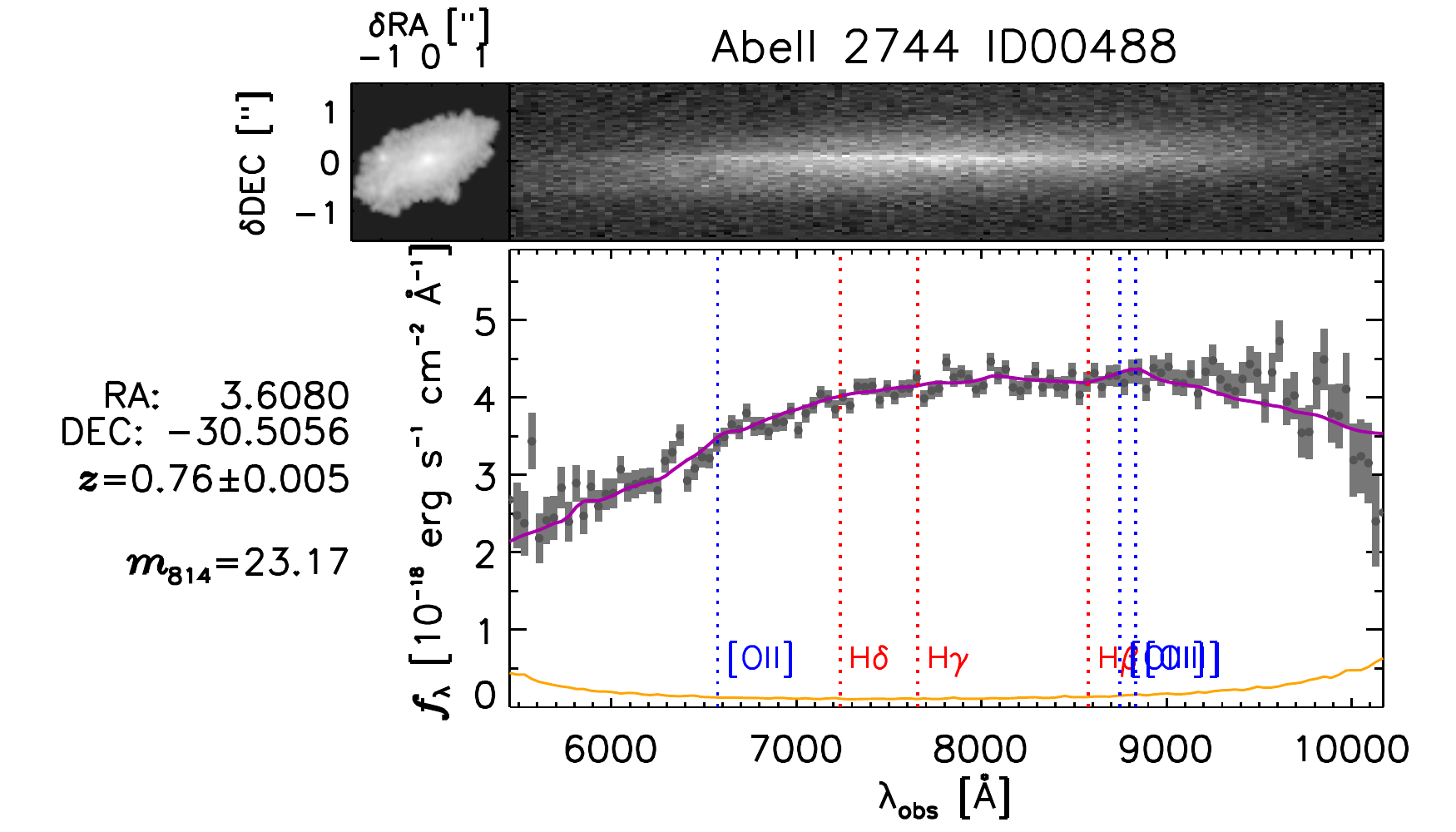}
	\includegraphics[width=0.475\linewidth]{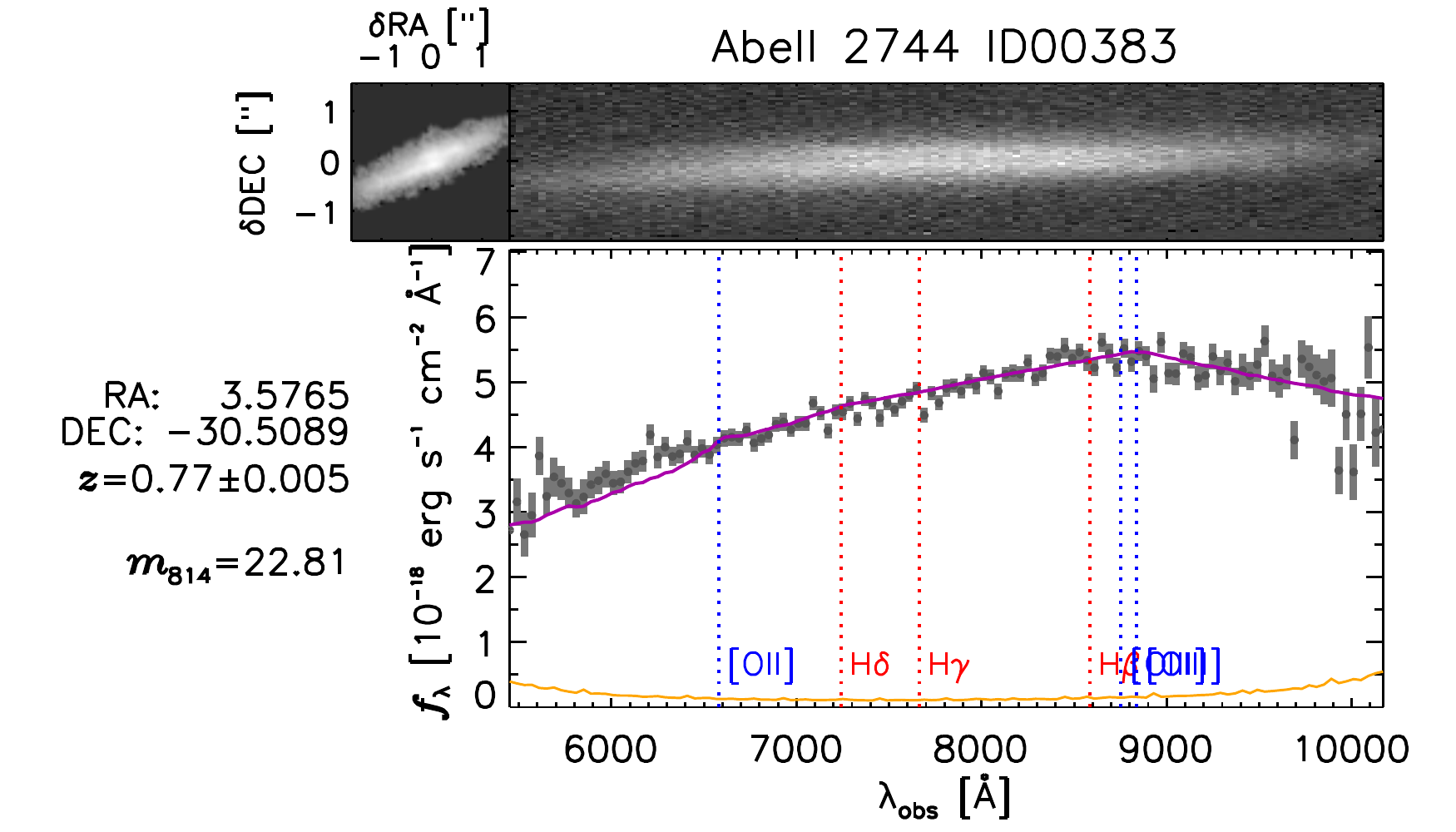}
	\includegraphics[width=0.475\linewidth]{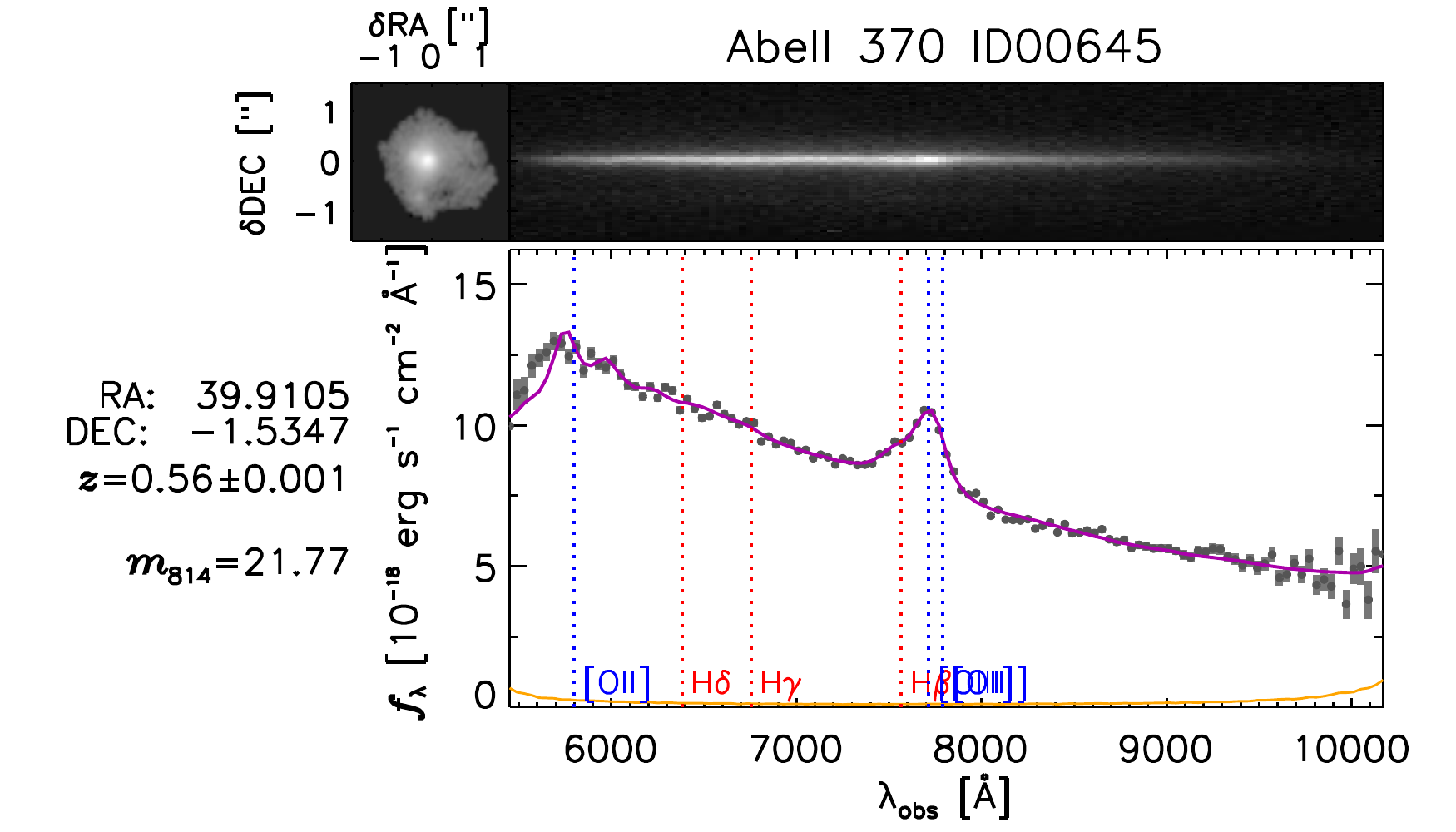}
	\includegraphics[width=0.475\linewidth]{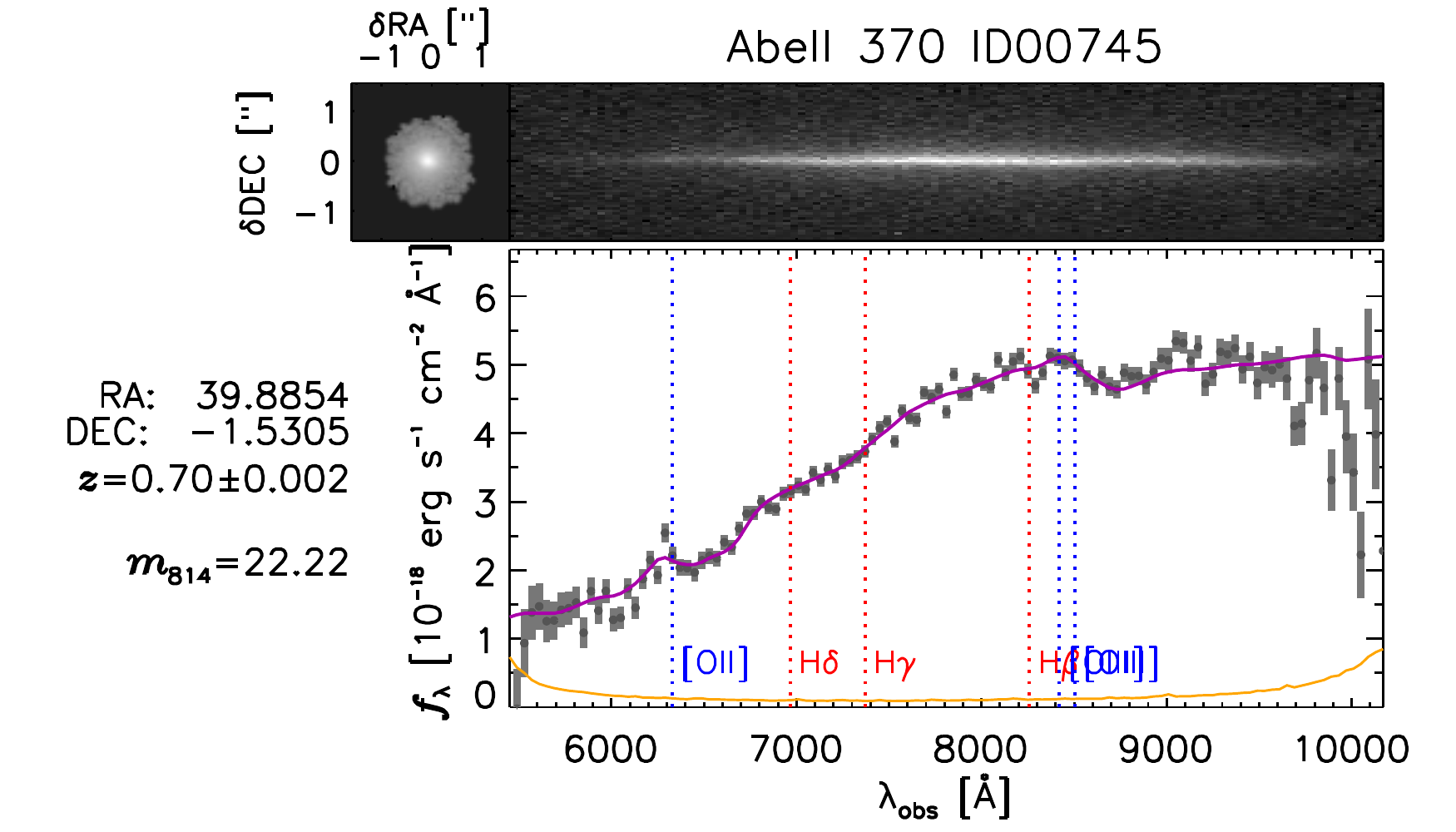}	   
	\caption{Examples of the $\sim$300 bright, $S/N\geq5$ [\ion{O}{iii}] 
		emitters in the glass database. This figure continues on
		the following page.}
	\label{fig:specExamples}
\end{figure*}

Below are more examples of GLASS G800L spectra and combined spectrophotometry
for select sources in the HFF footprint. All galaxies shown Figure \ref{fig:specExamples} 
have $S/N_{\rm[OIII]}\geq5$ and are above the median brightness in the low risk sample. 
There are 304 such sources in the GLASS ACS database. All galaxies in 
Figure \ref{fig:specPhotExamples2} are in the low risk sample with matching HFF photometry.
There are 383 such objects in the GLASS ACS database (3411 in total with HFF overlap).

\addtocounter{figure}{-1}

\begin{figure*}
	\includegraphics[width=0.475\linewidth]{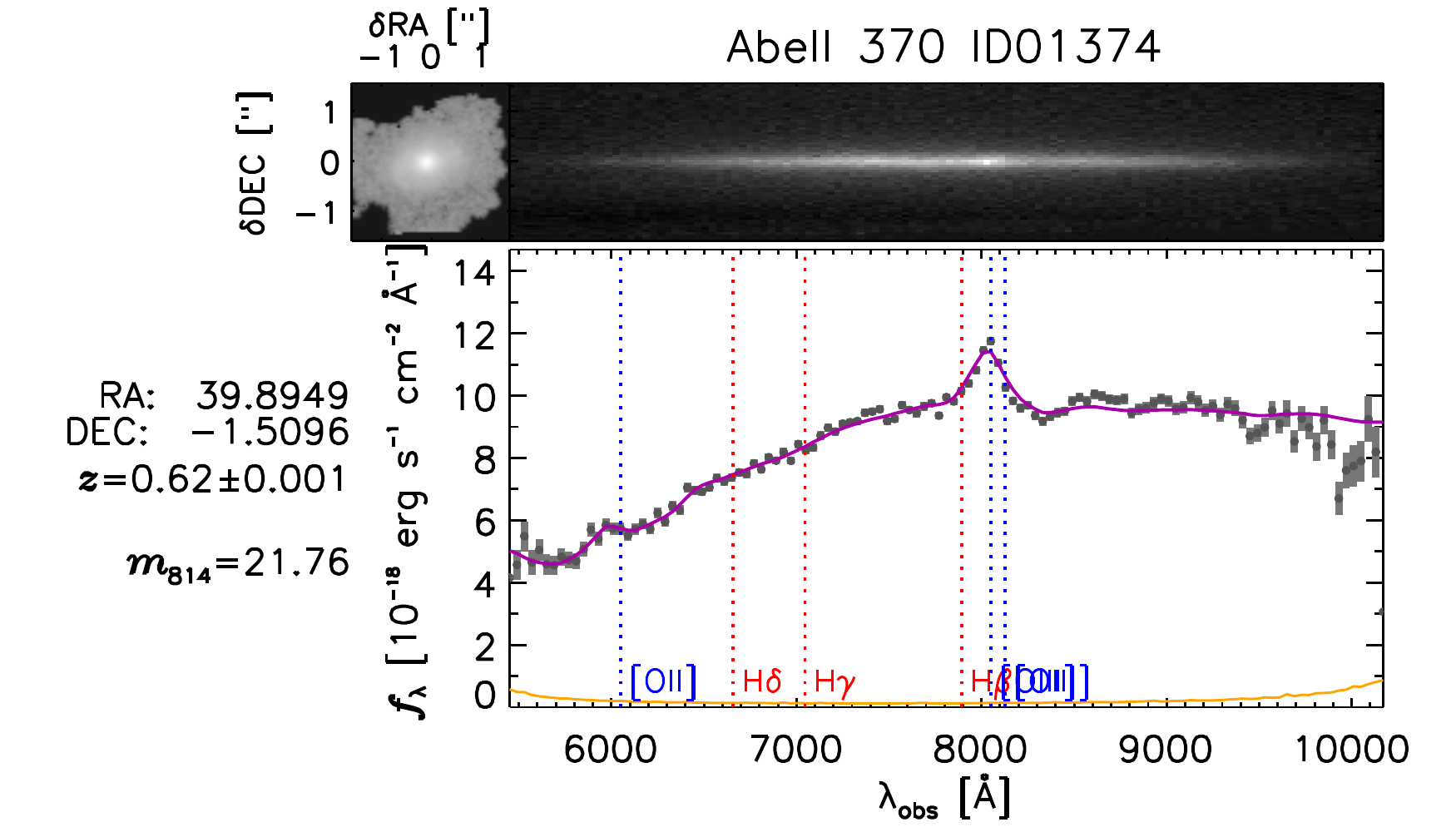}    
	\includegraphics[width=0.475\linewidth]{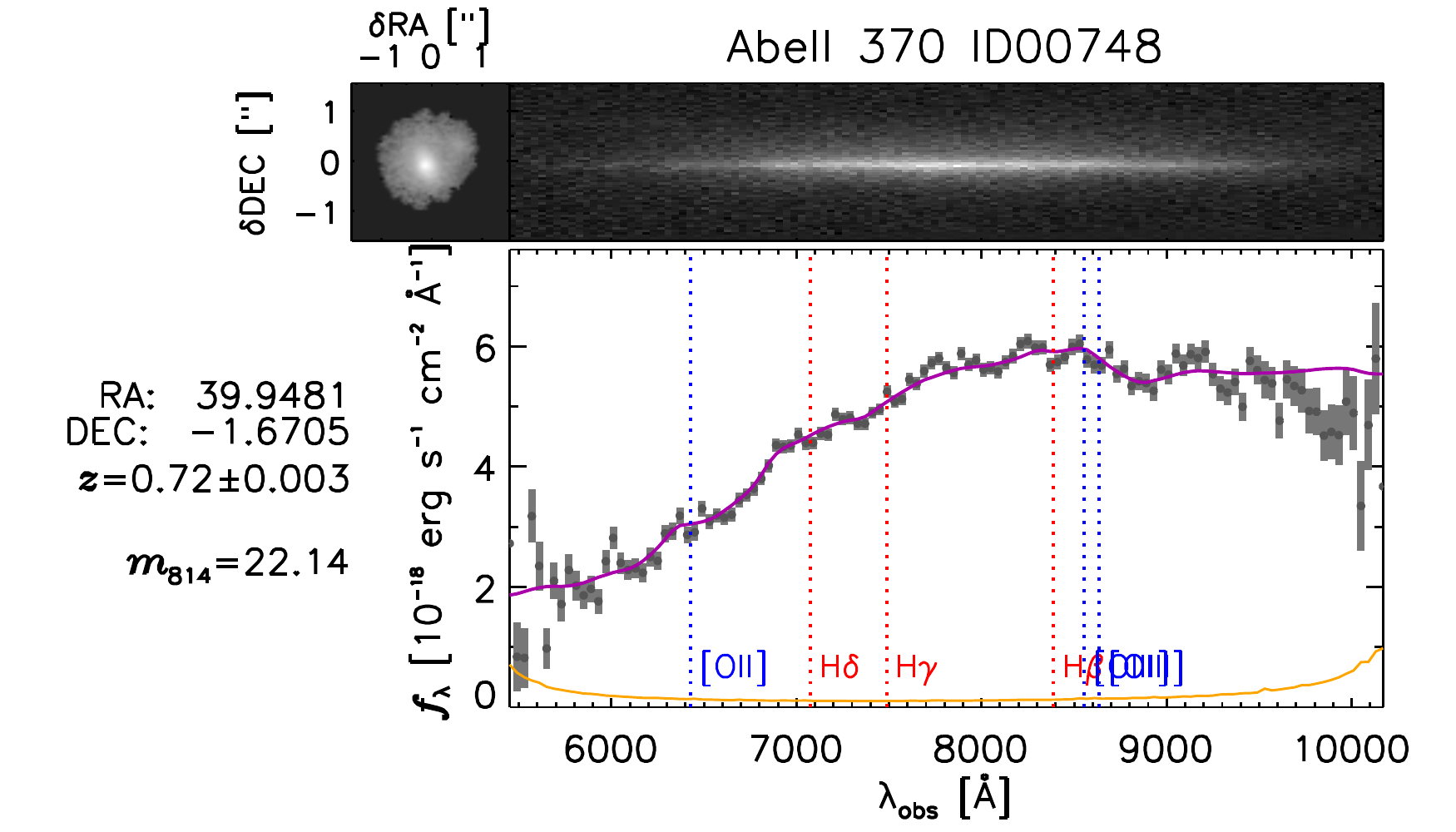}    
	\includegraphics[width=0.475\linewidth]{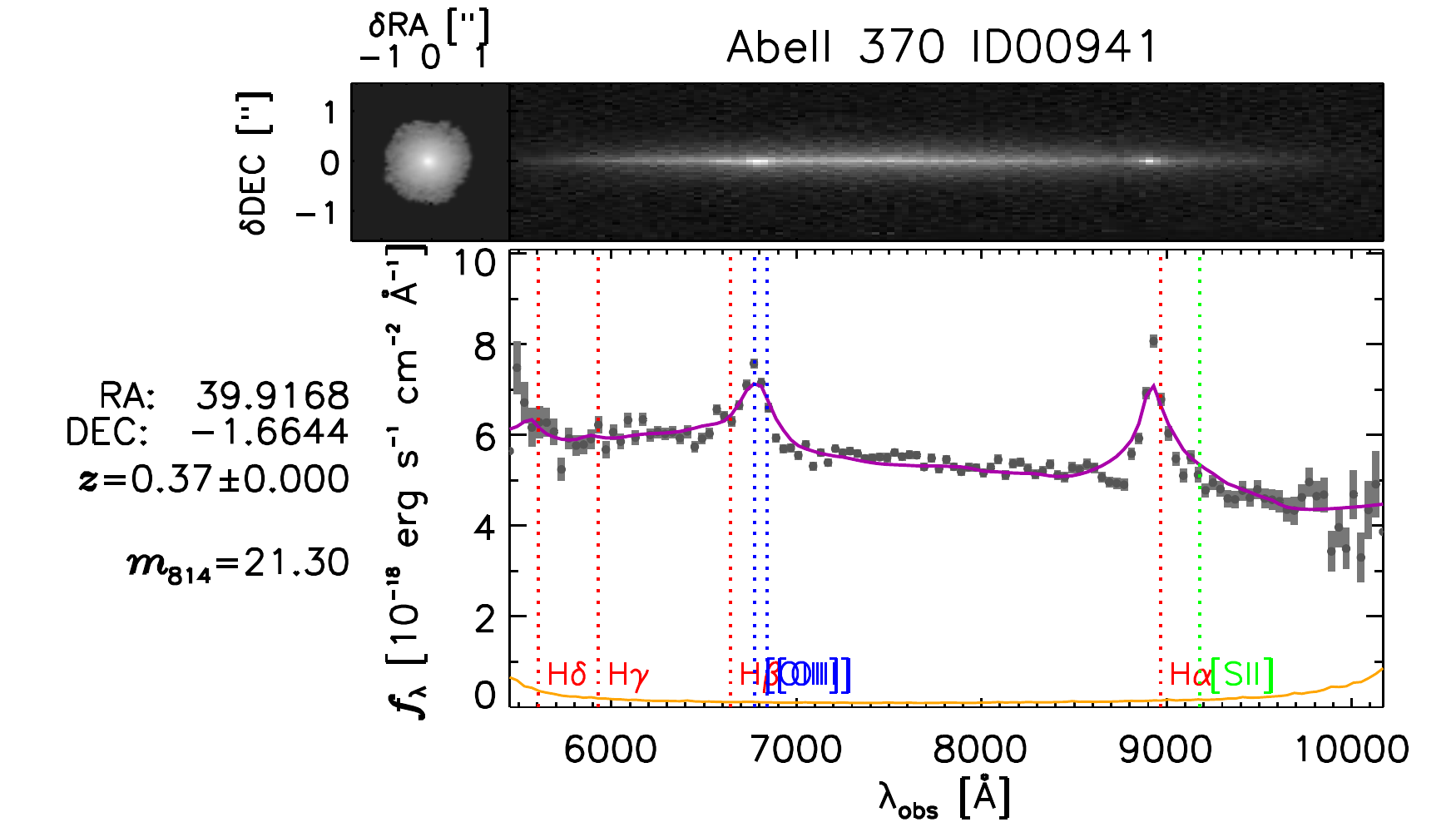}    
	\includegraphics[width=0.475\linewidth]{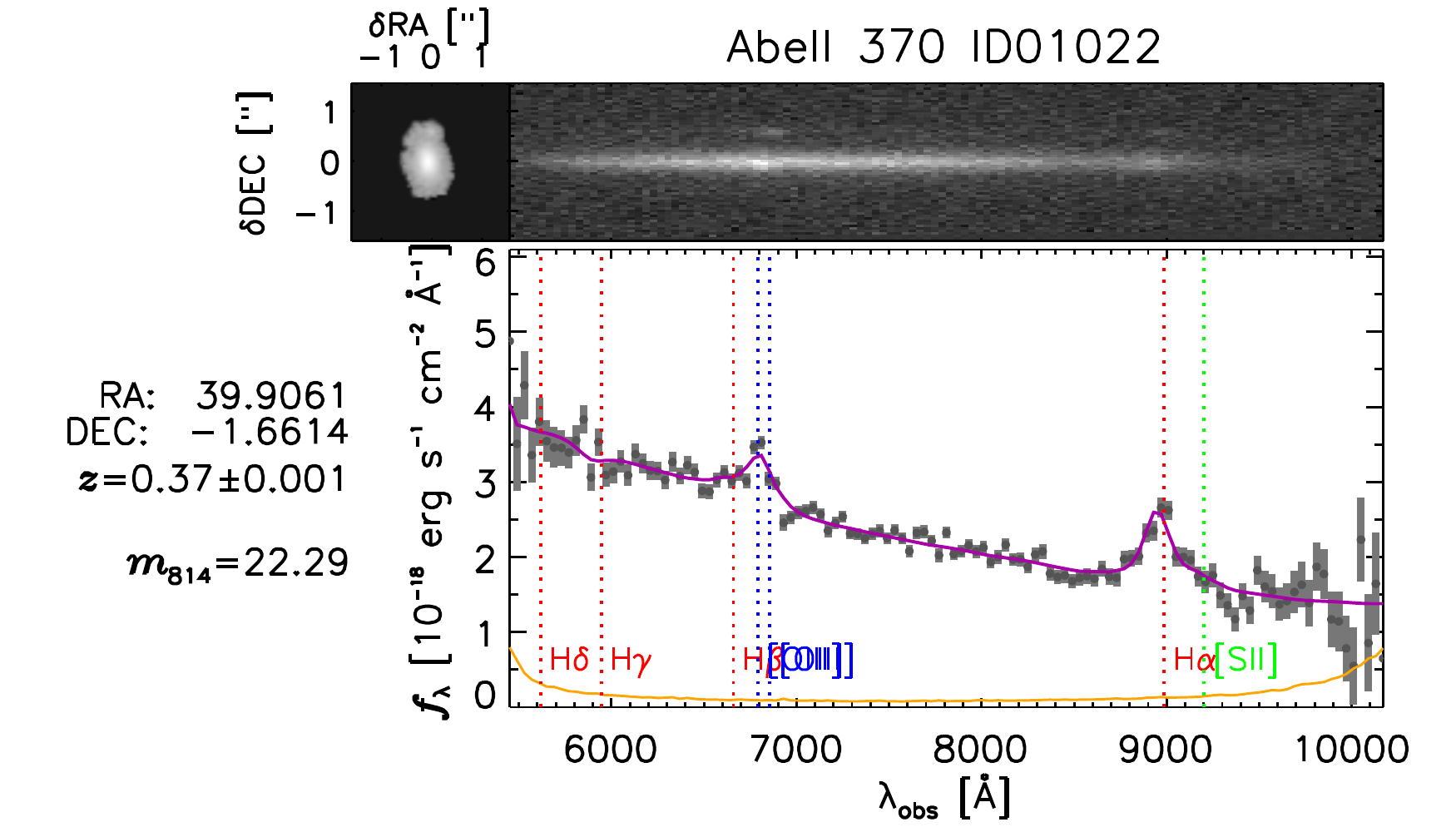}    
	\includegraphics[width=0.475\linewidth]{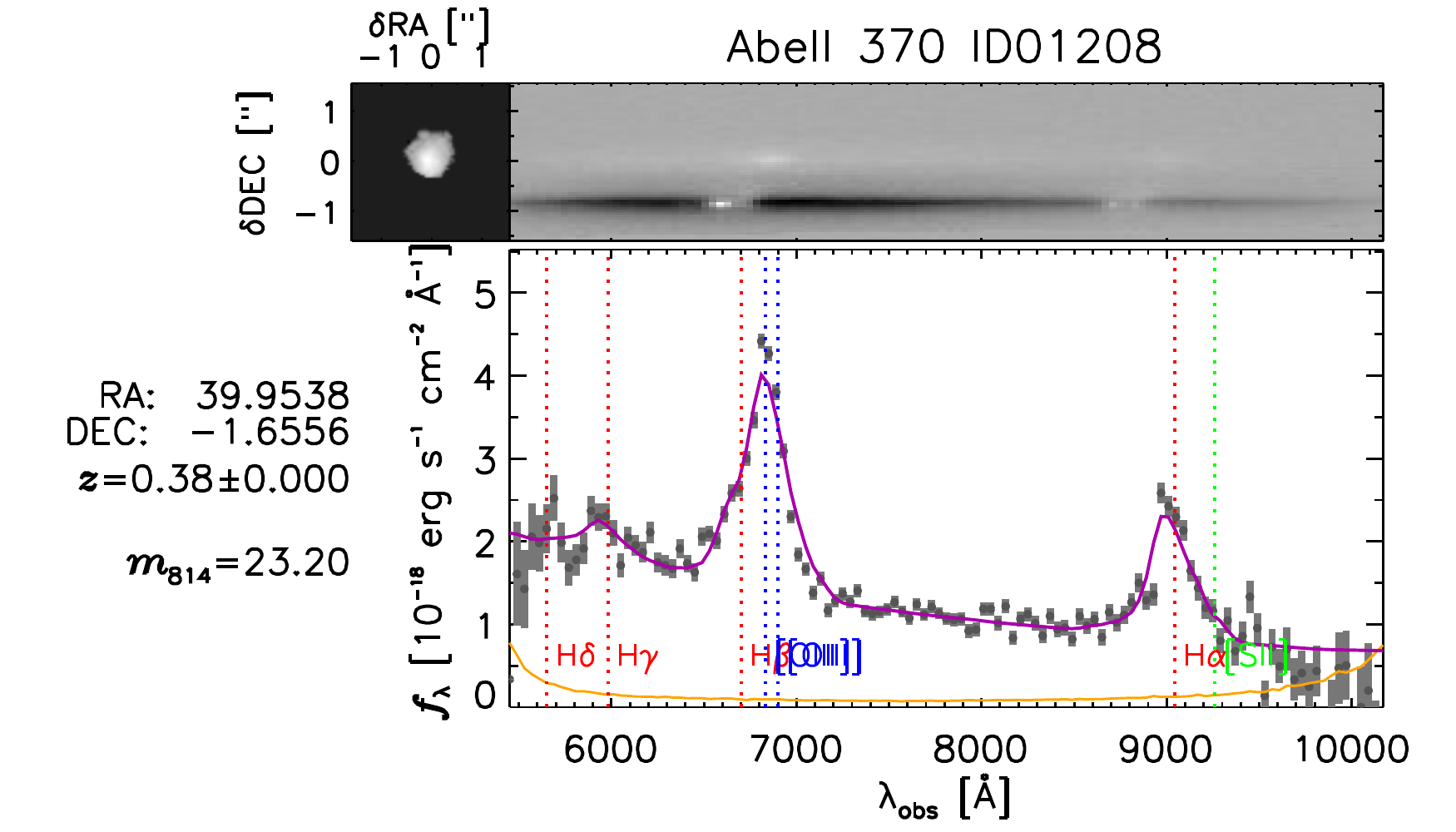}    
	\includegraphics[width=0.475\linewidth]{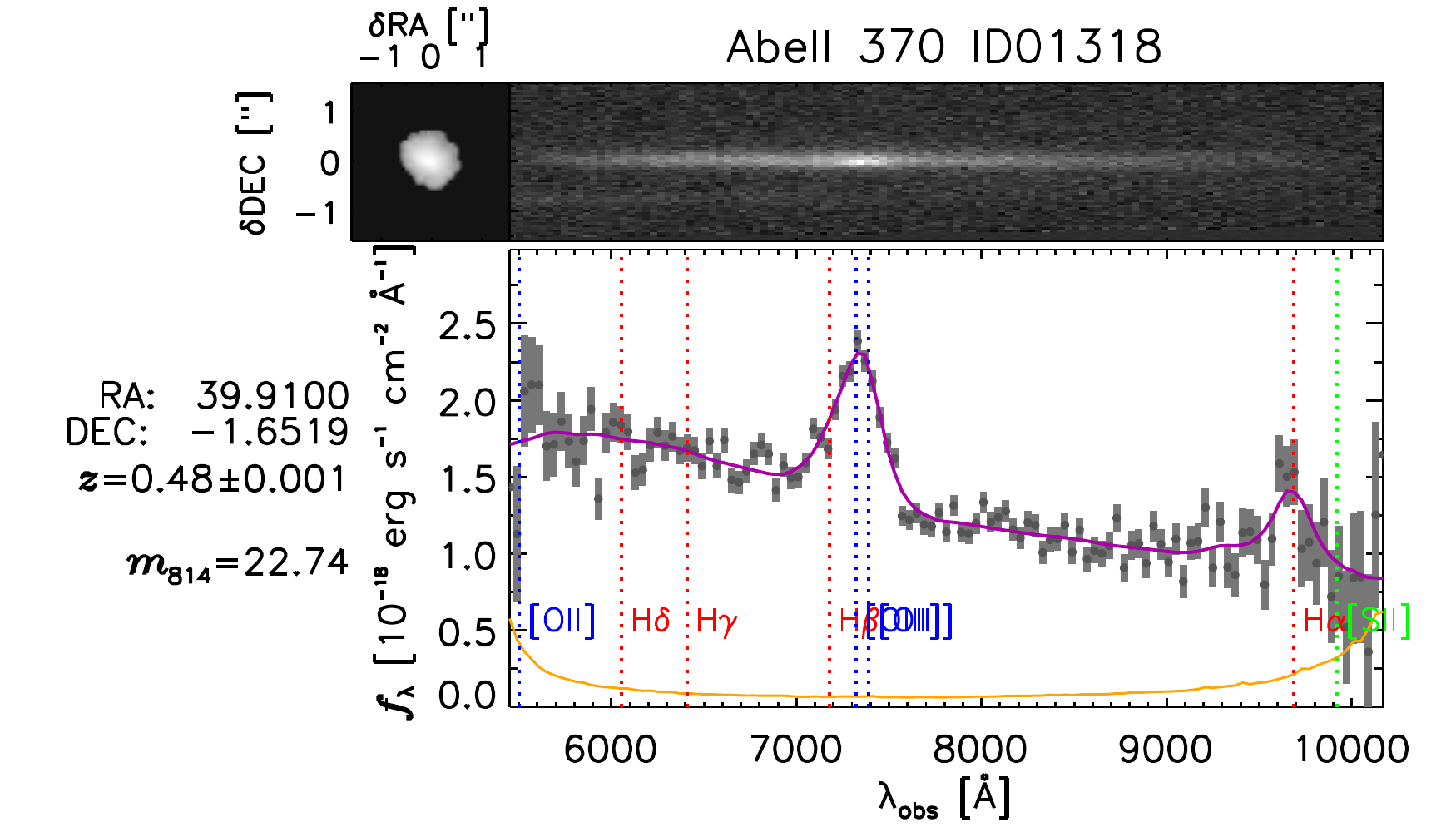}    
	\includegraphics[width=0.475\linewidth]{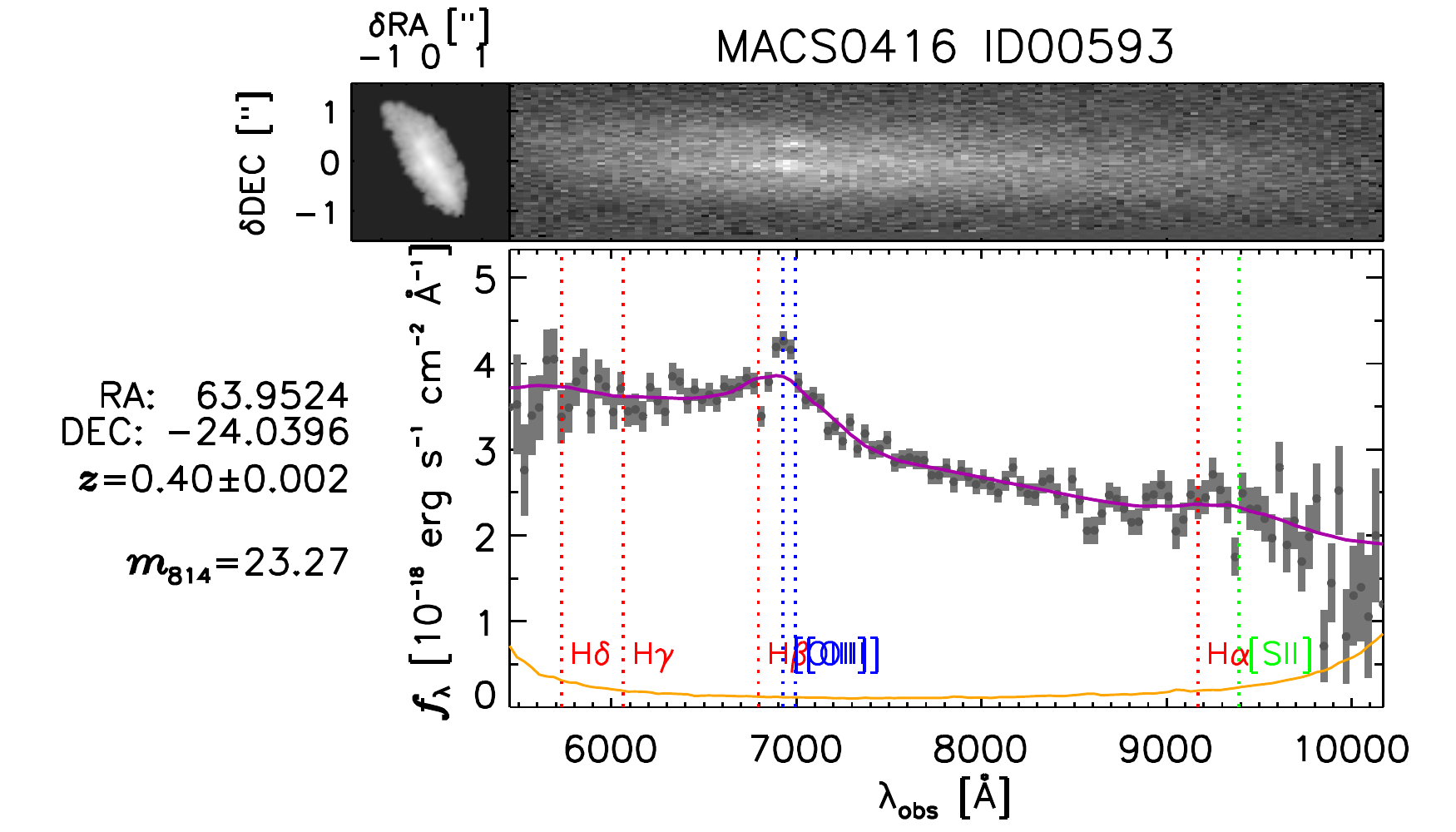}    
	\includegraphics[width=0.475\linewidth]{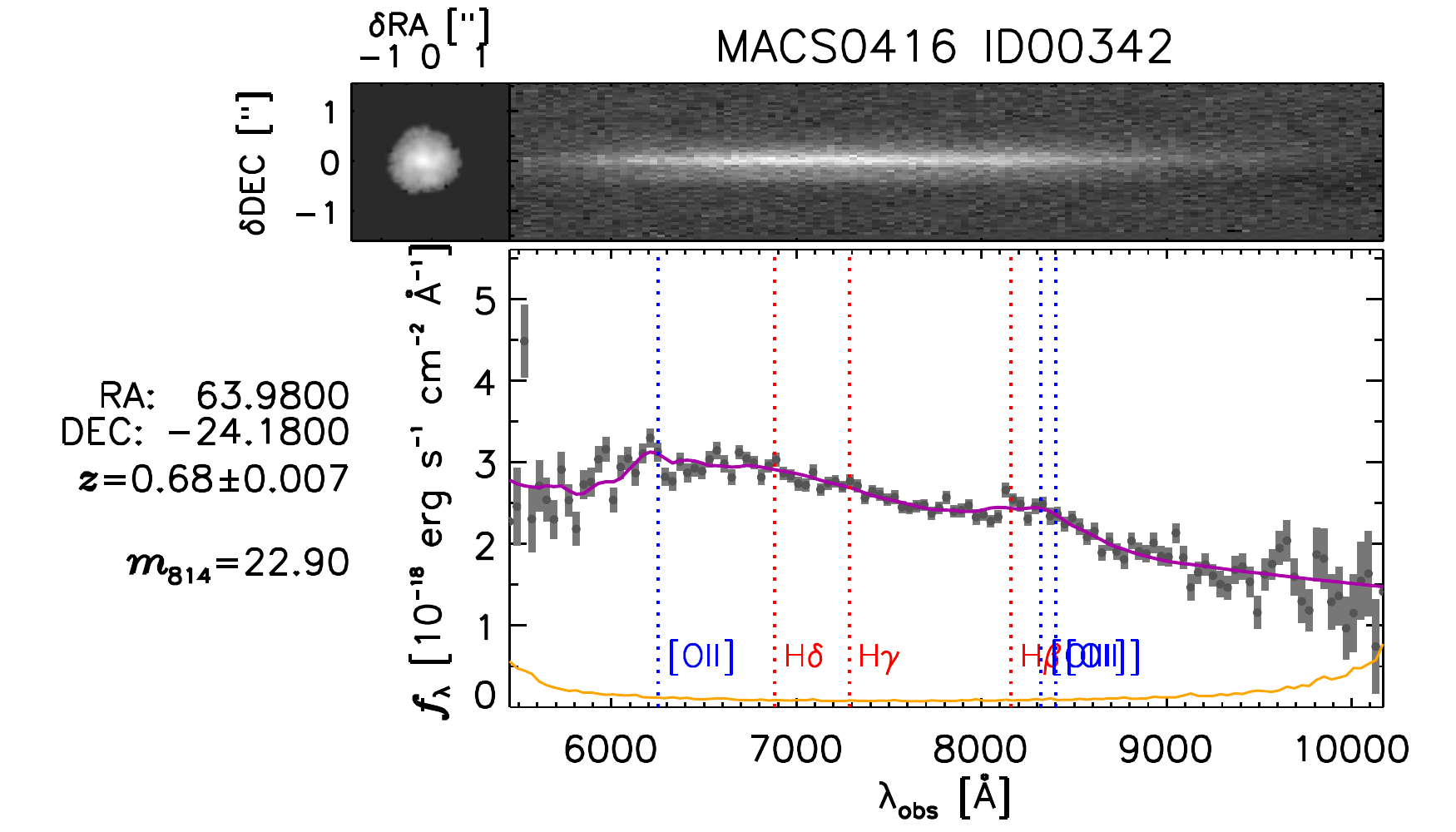}    
	\caption{Examples of the $\sim$300 bright, $S/N\geq5$ [\ion{O}{iii}] 
		emitters in the glass database. This figure continues on
		the following page.  This figure continues on
		the following page.}
\end{figure*}

\addtocounter{figure}{-1}

\begin{figure*}
	\includegraphics[width=0.475\linewidth]{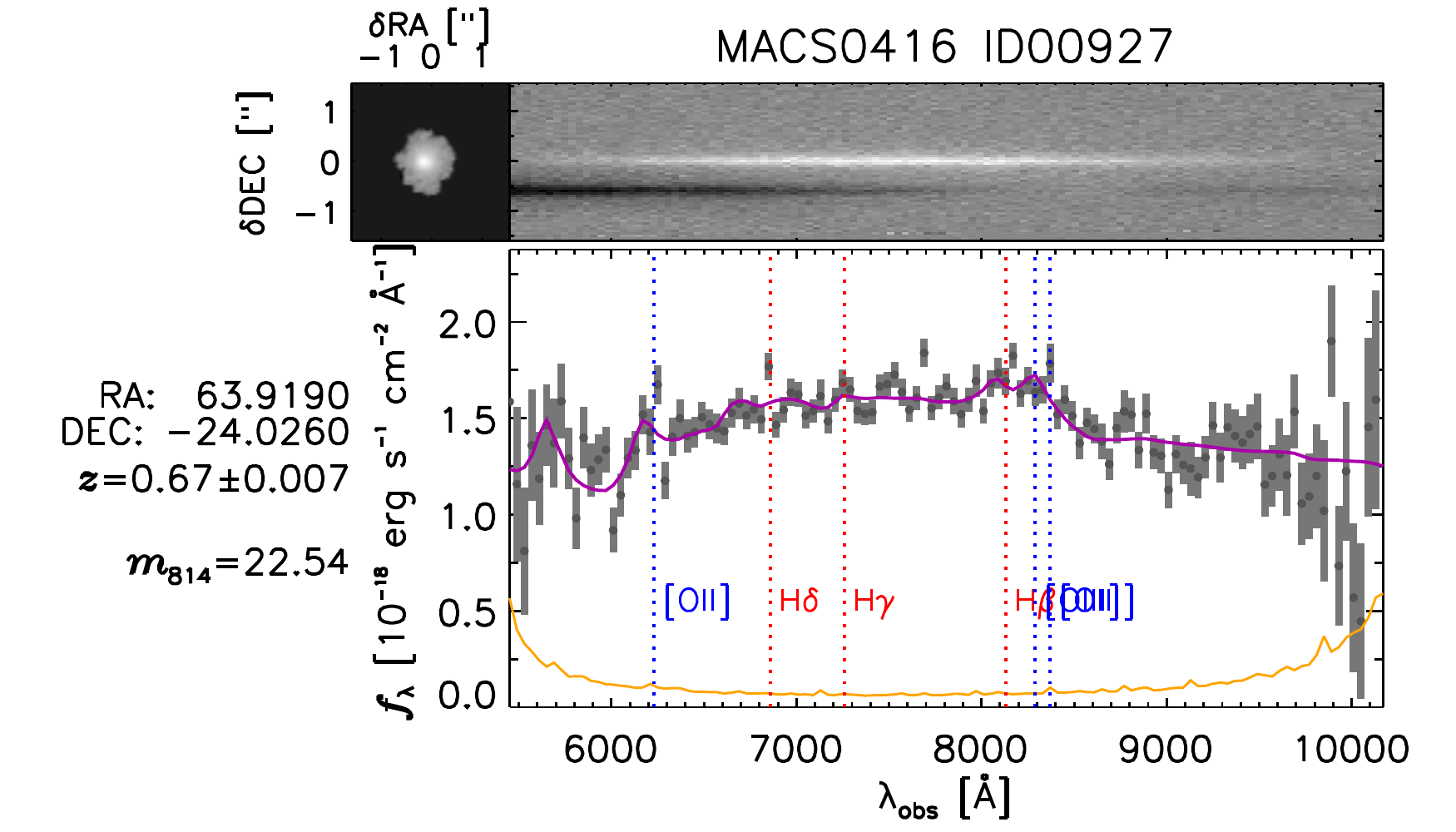}    
	\includegraphics[width=0.475\linewidth]{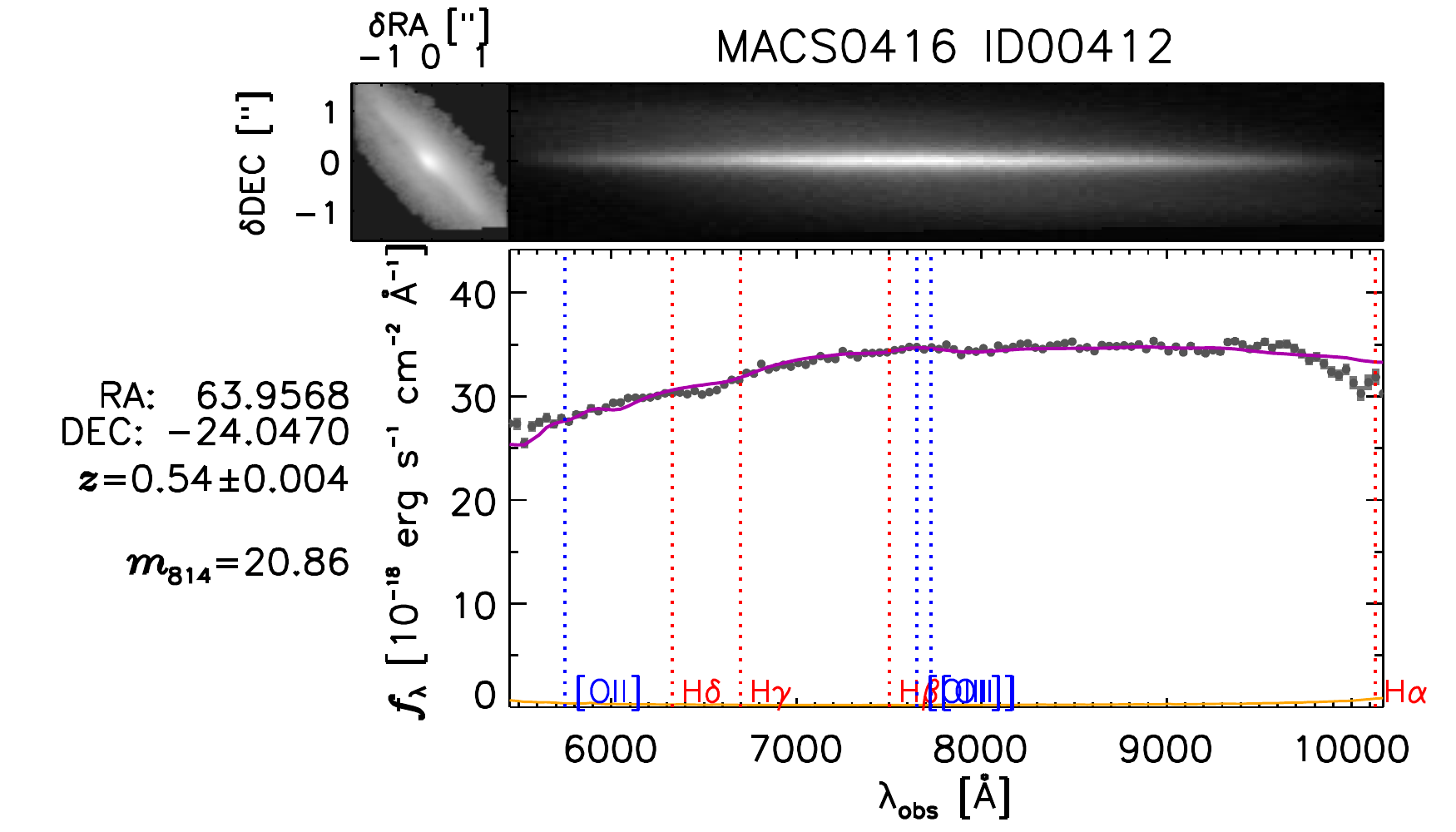}    
	\includegraphics[width=0.475\linewidth]{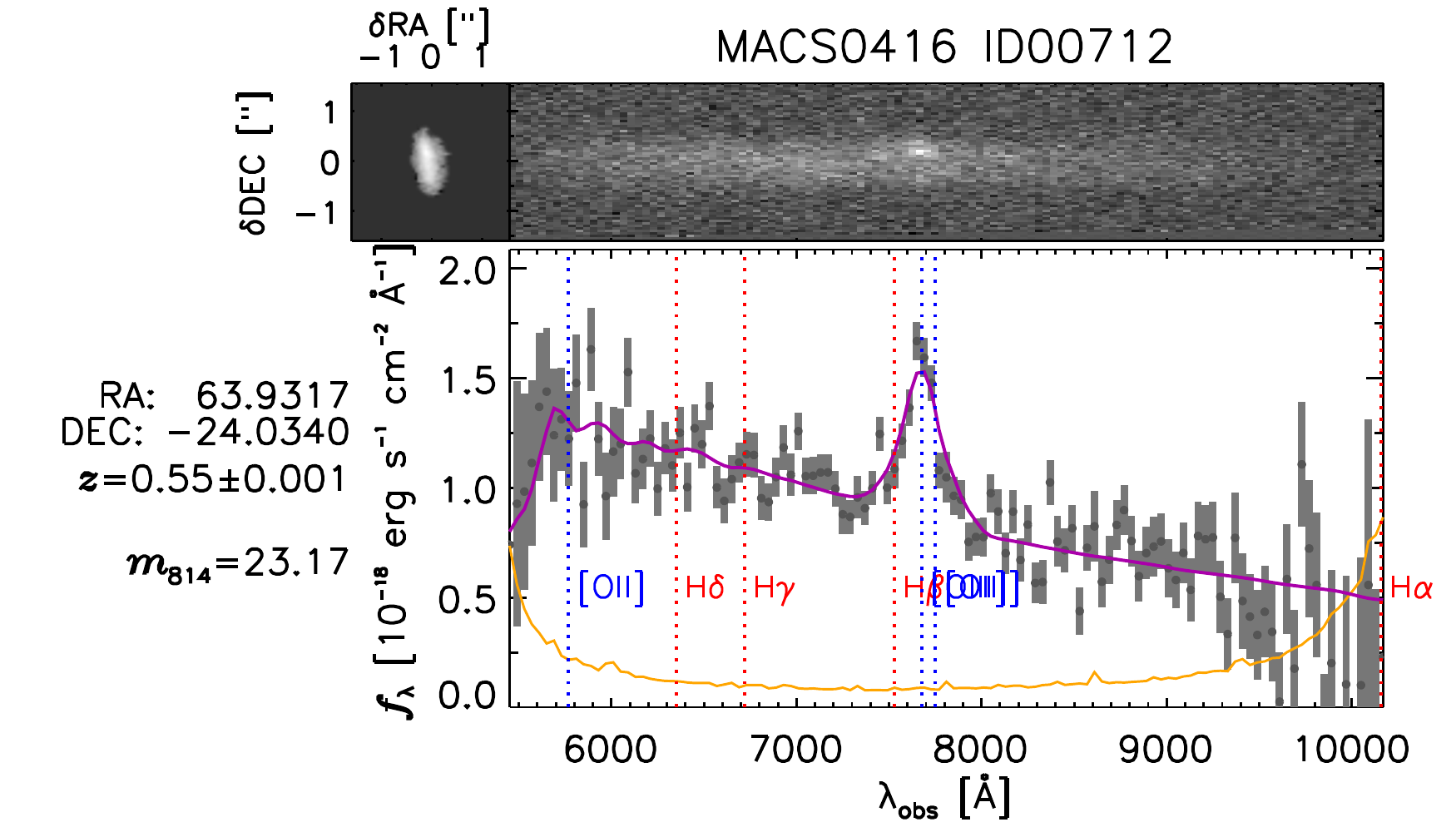}    
	\includegraphics[width=0.475\linewidth]{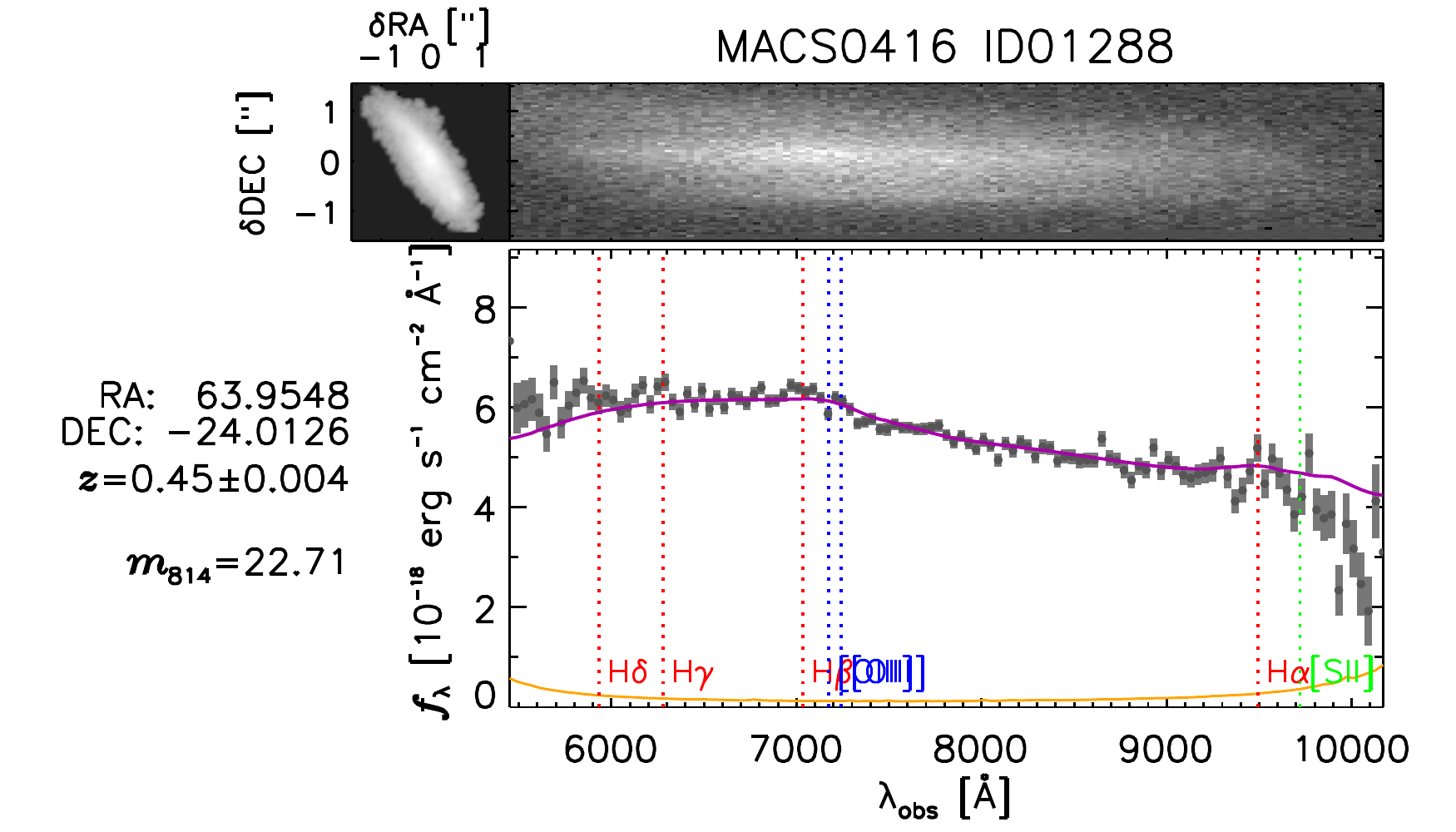}    
	\includegraphics[width=0.475\linewidth]{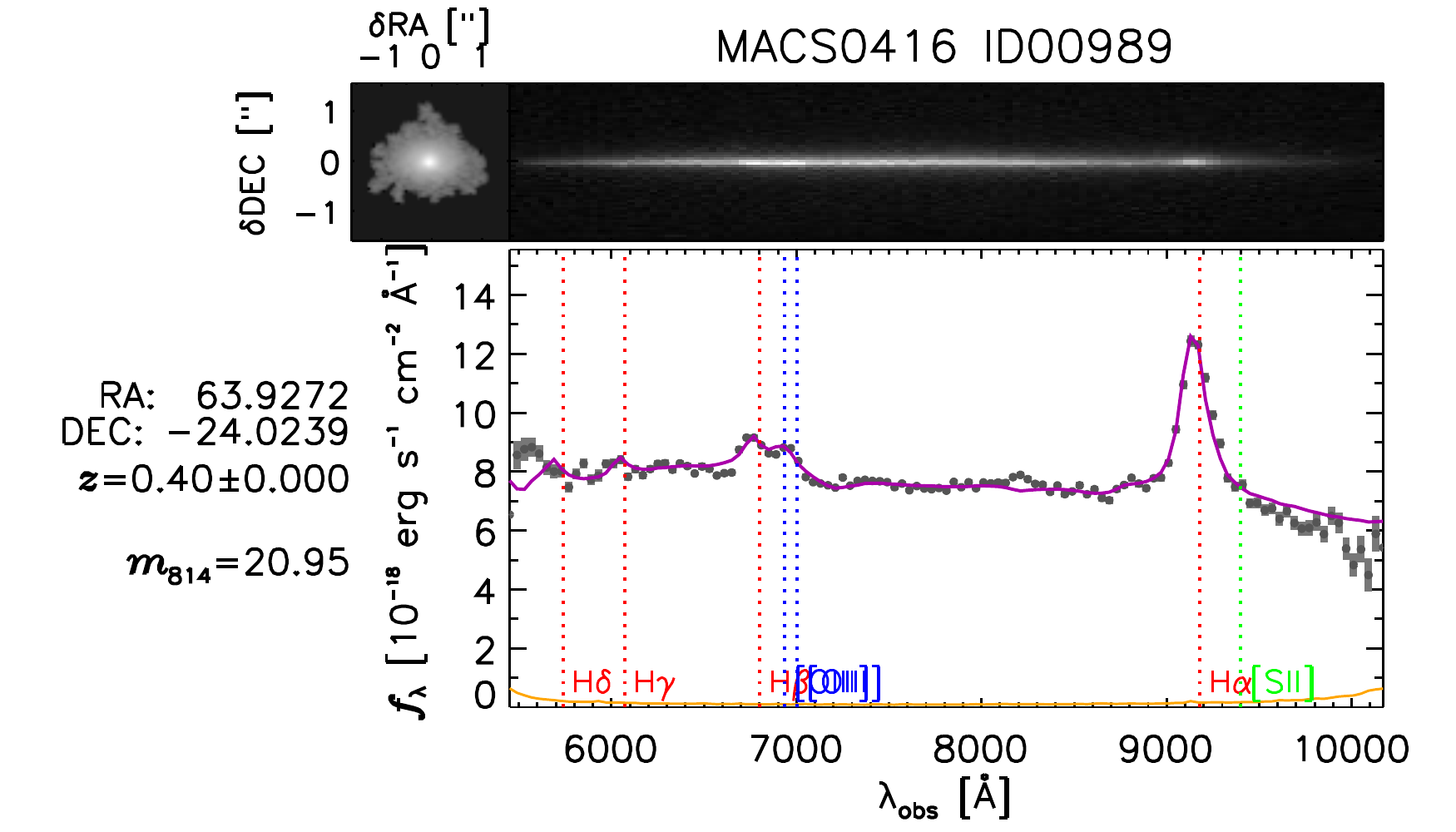}    
	\includegraphics[width=0.475\linewidth]{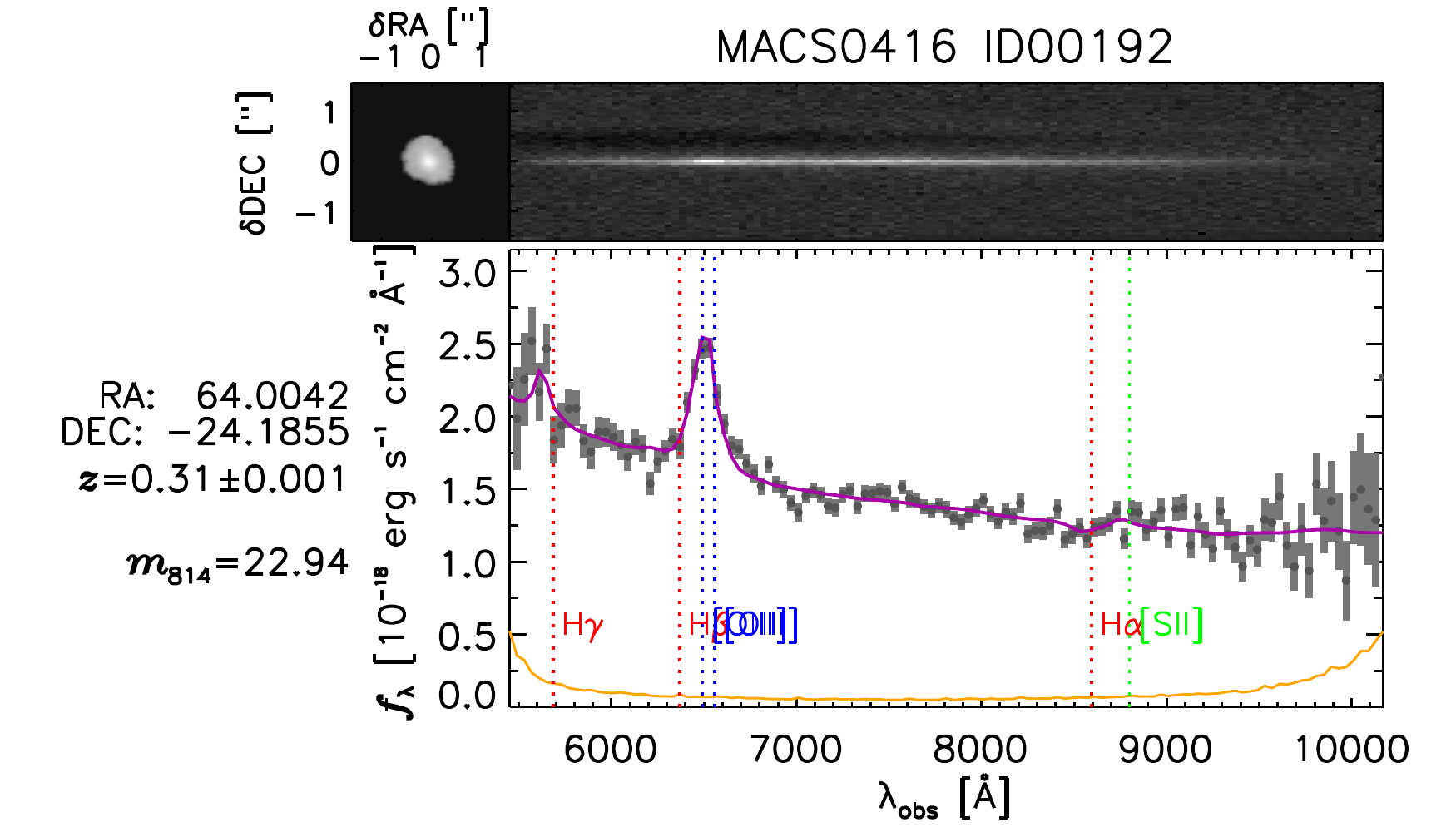}  
	\includegraphics[width=0.475\linewidth]{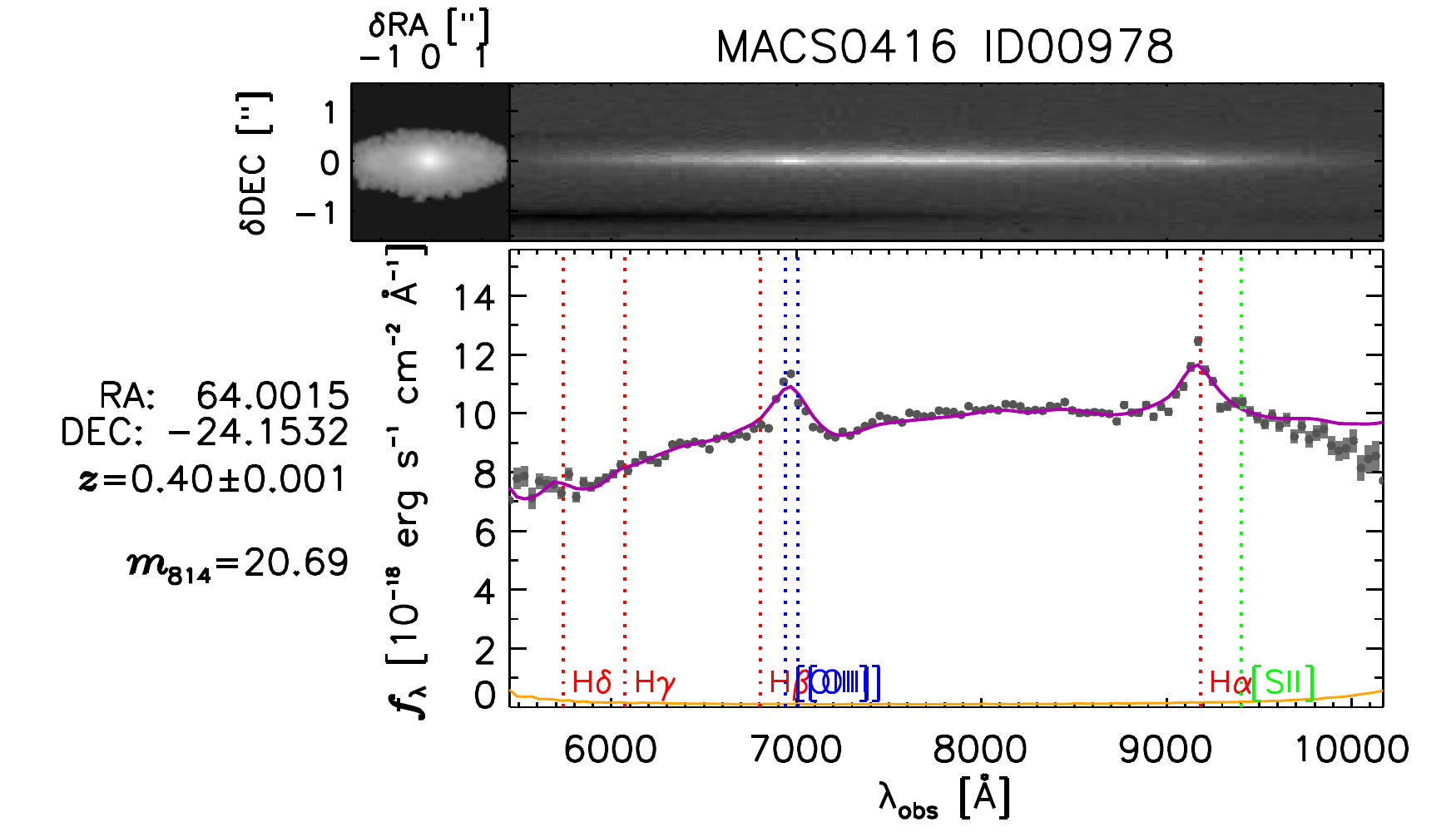}    
	\includegraphics[width=0.475\linewidth]{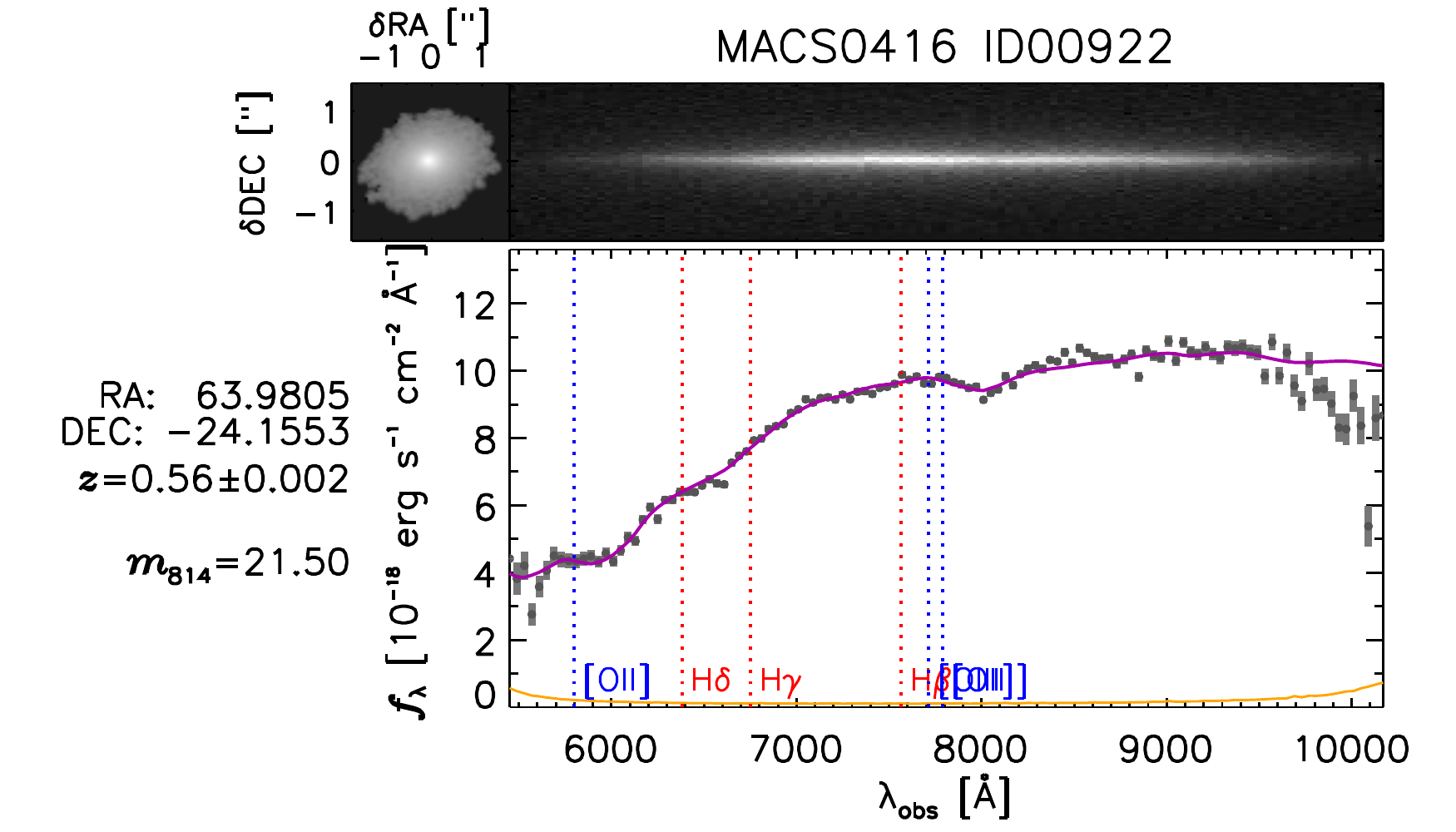}      
	\caption{Examples of the $\sim$300 bright, $S/N\geq5$ [\ion{O}{iii}] 
		emitters in the glass database.  This figure continues on
		the following page.}
\end{figure*}

\addtocounter{figure}{-1}

\begin{figure*}
	\includegraphics[width=0.475\linewidth]{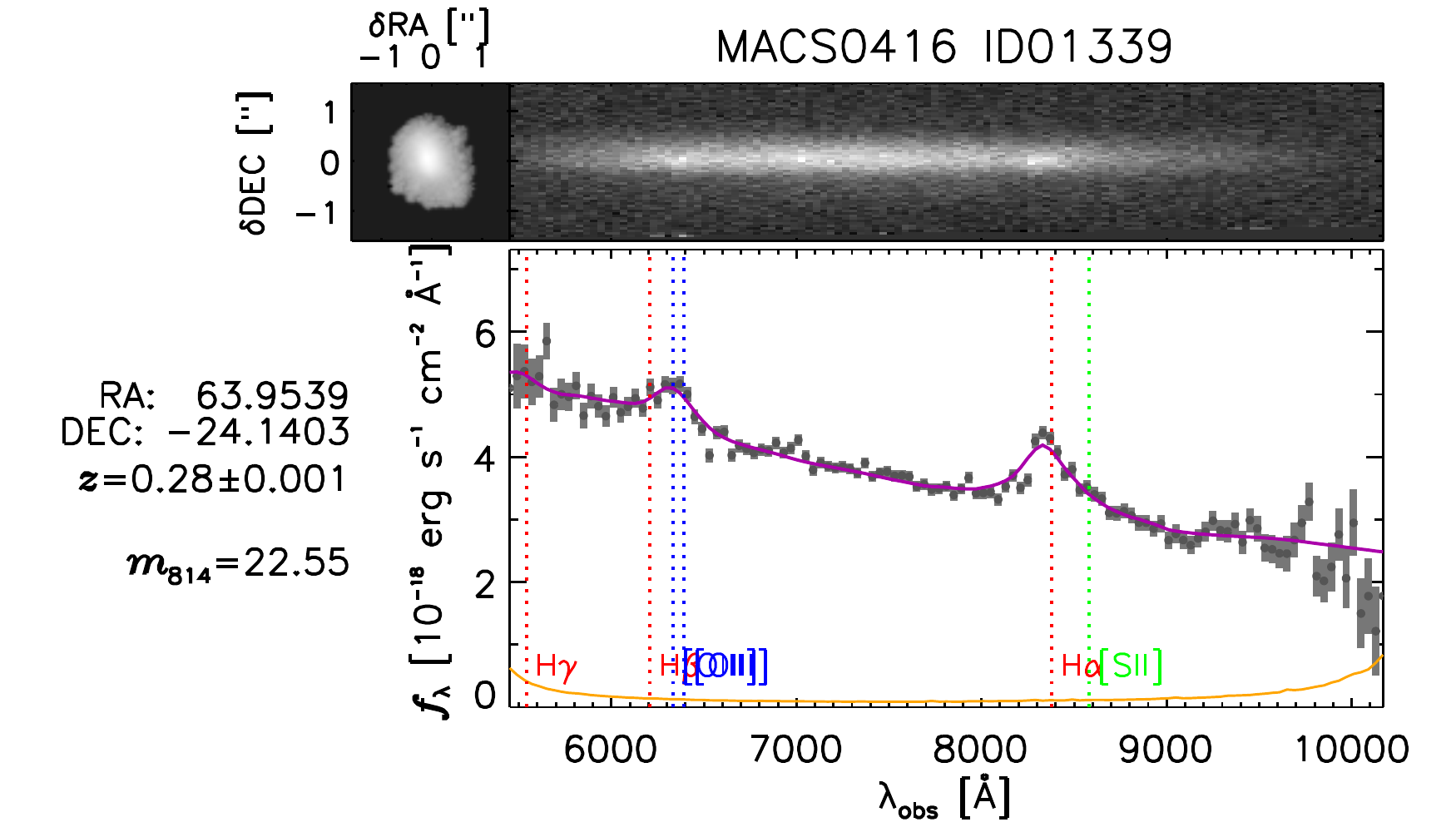}    
	\includegraphics[width=0.475\linewidth]{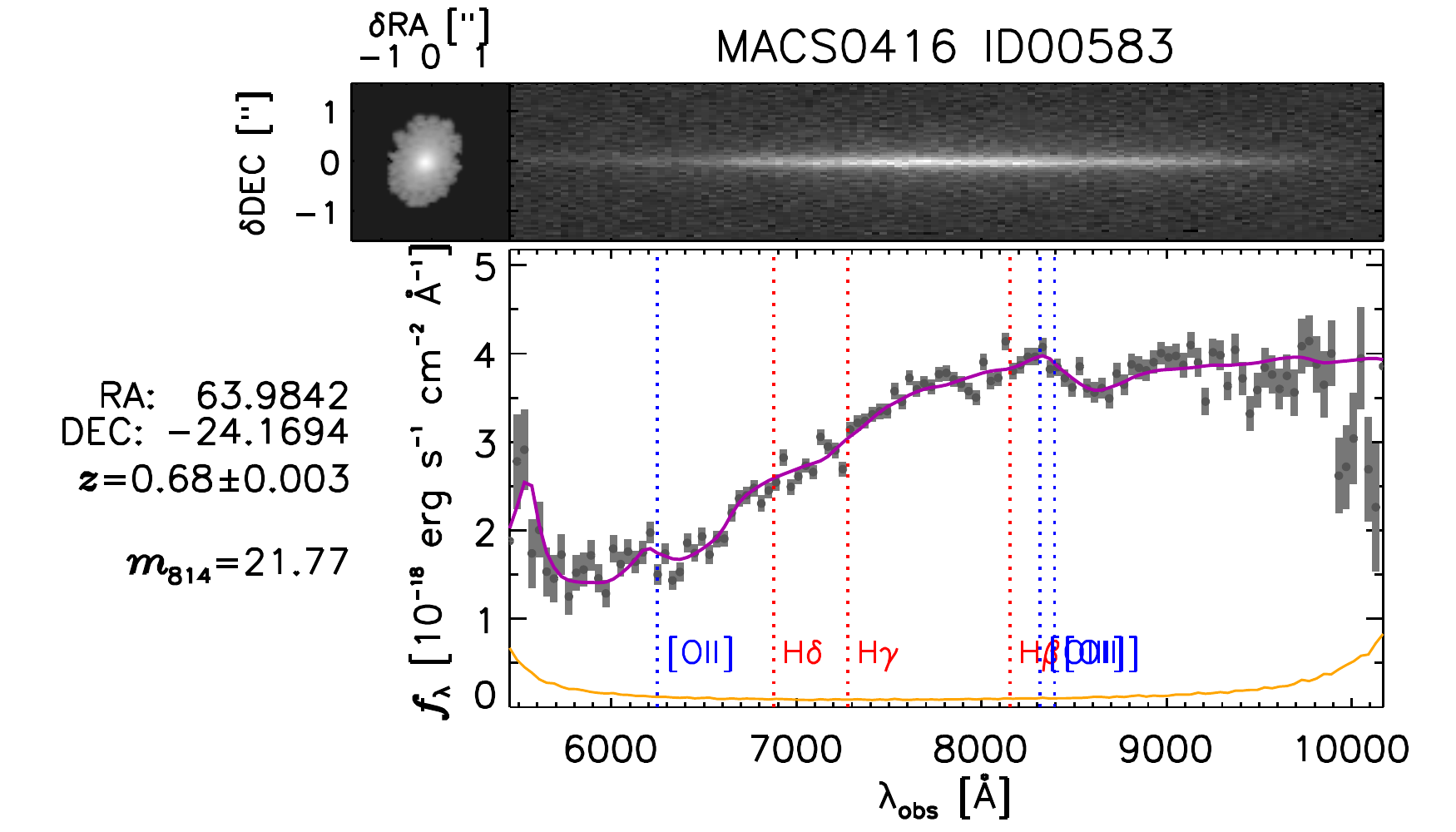}    
	\includegraphics[width=0.475\linewidth]{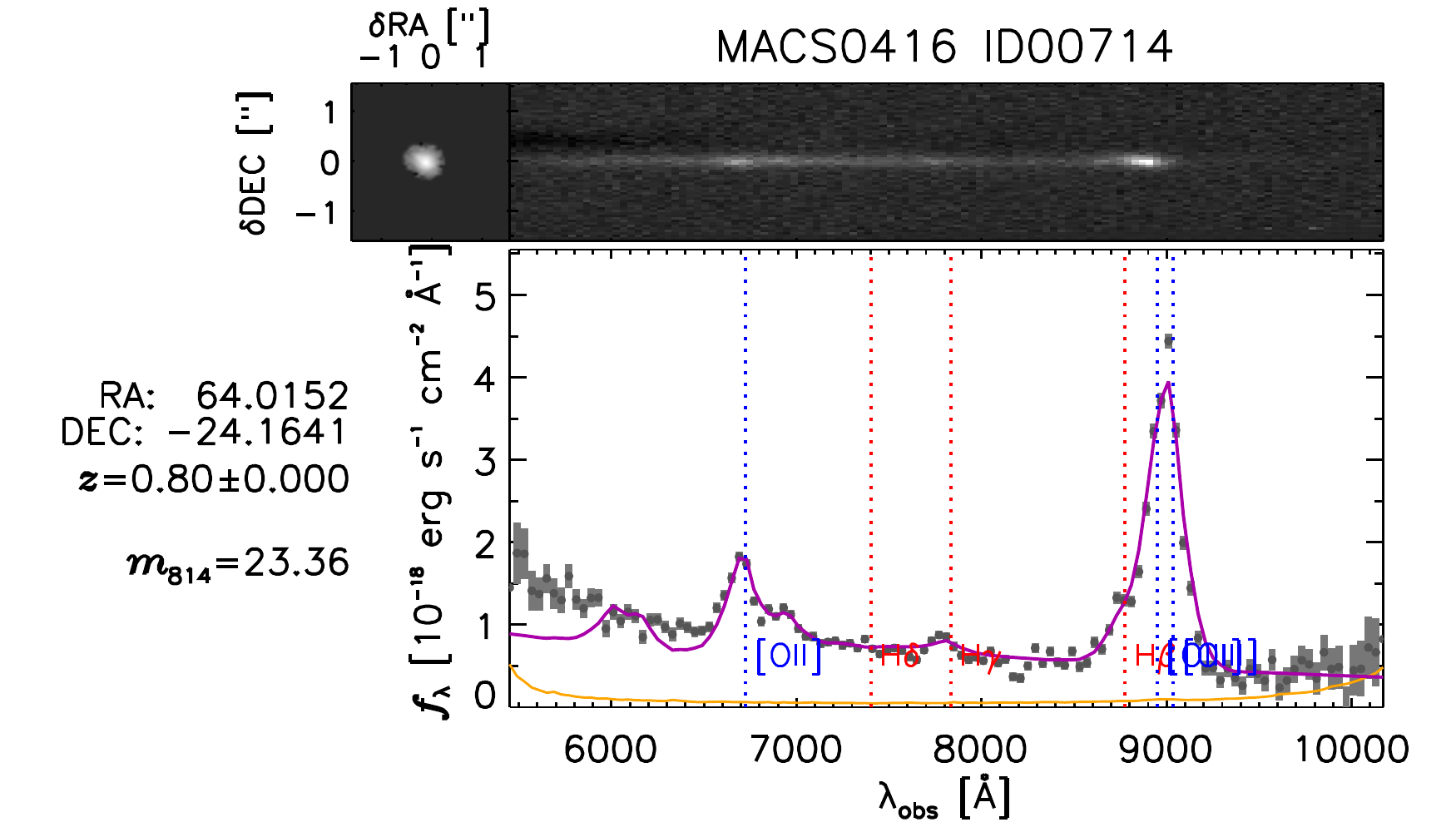}    
	\includegraphics[width=0.475\linewidth]{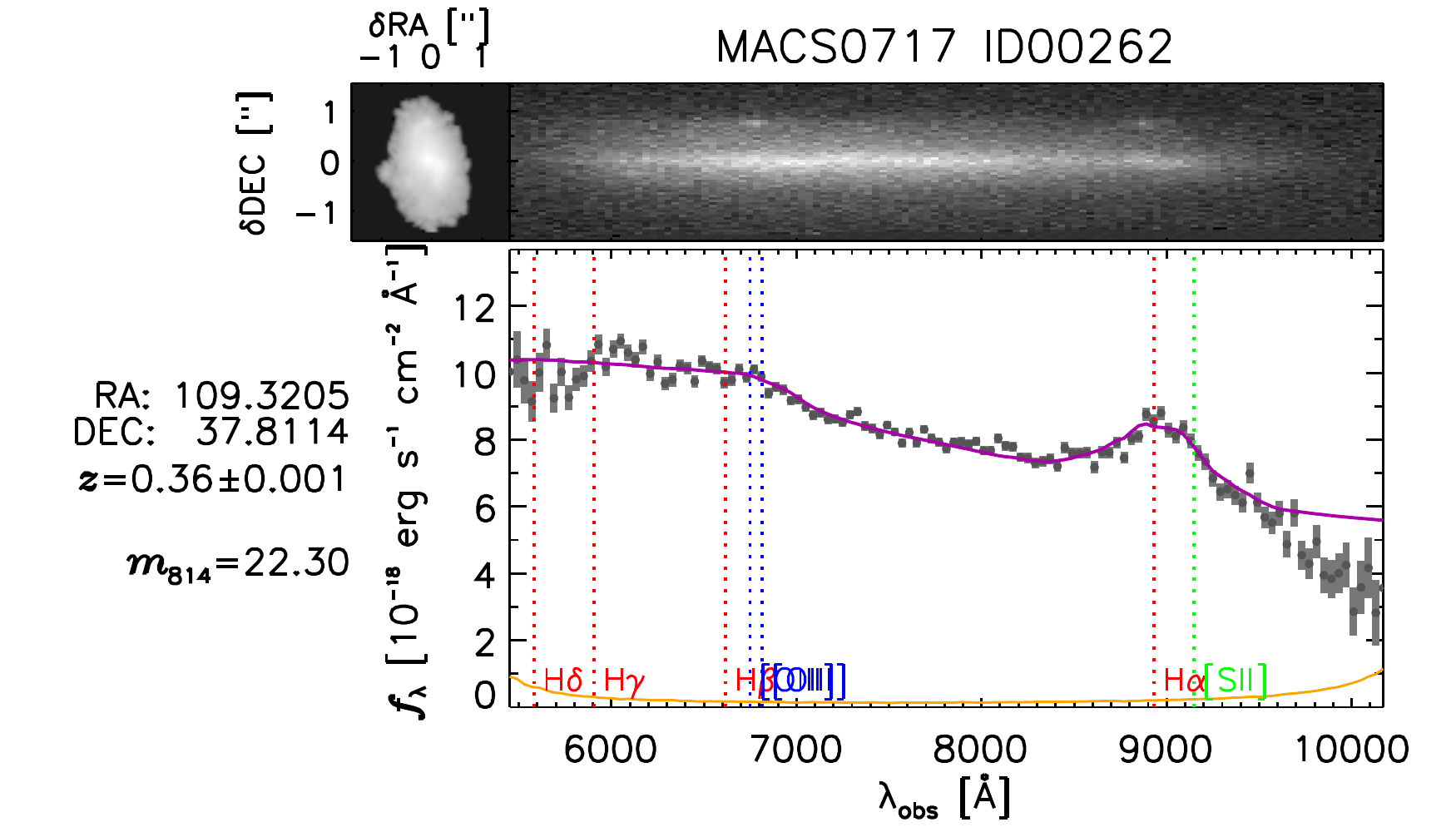}    
	\includegraphics[width=0.475\linewidth]{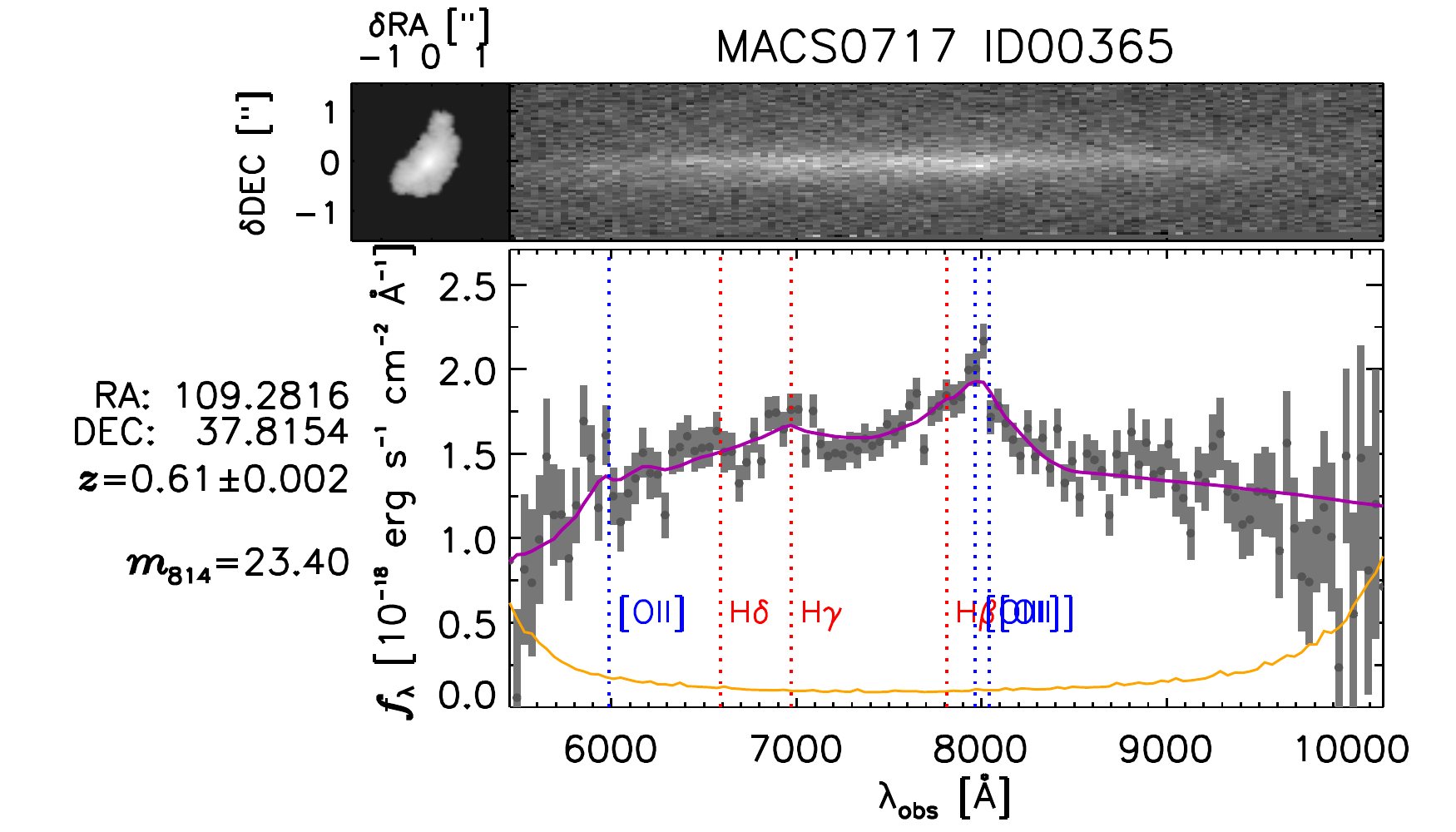}    
	\includegraphics[width=0.475\linewidth]{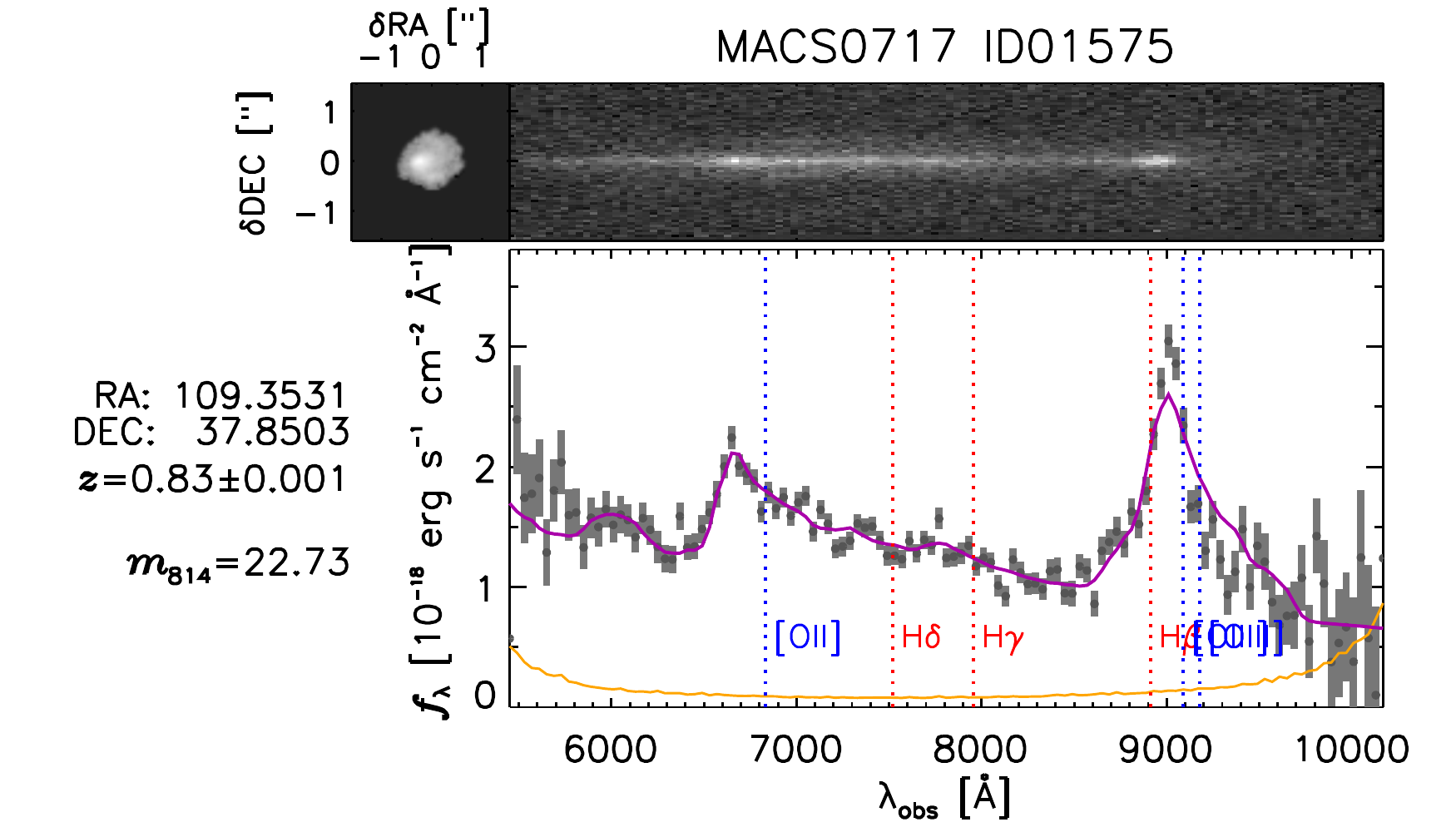}    
	\caption{Examples of the $\sim$300 bright, $S/N\geq5$ [\ion{O}{iii}] 
		emitters in the glass database.}
\end{figure*}

\begin{figure*}
	\includegraphics[width = 0.475\linewidth]{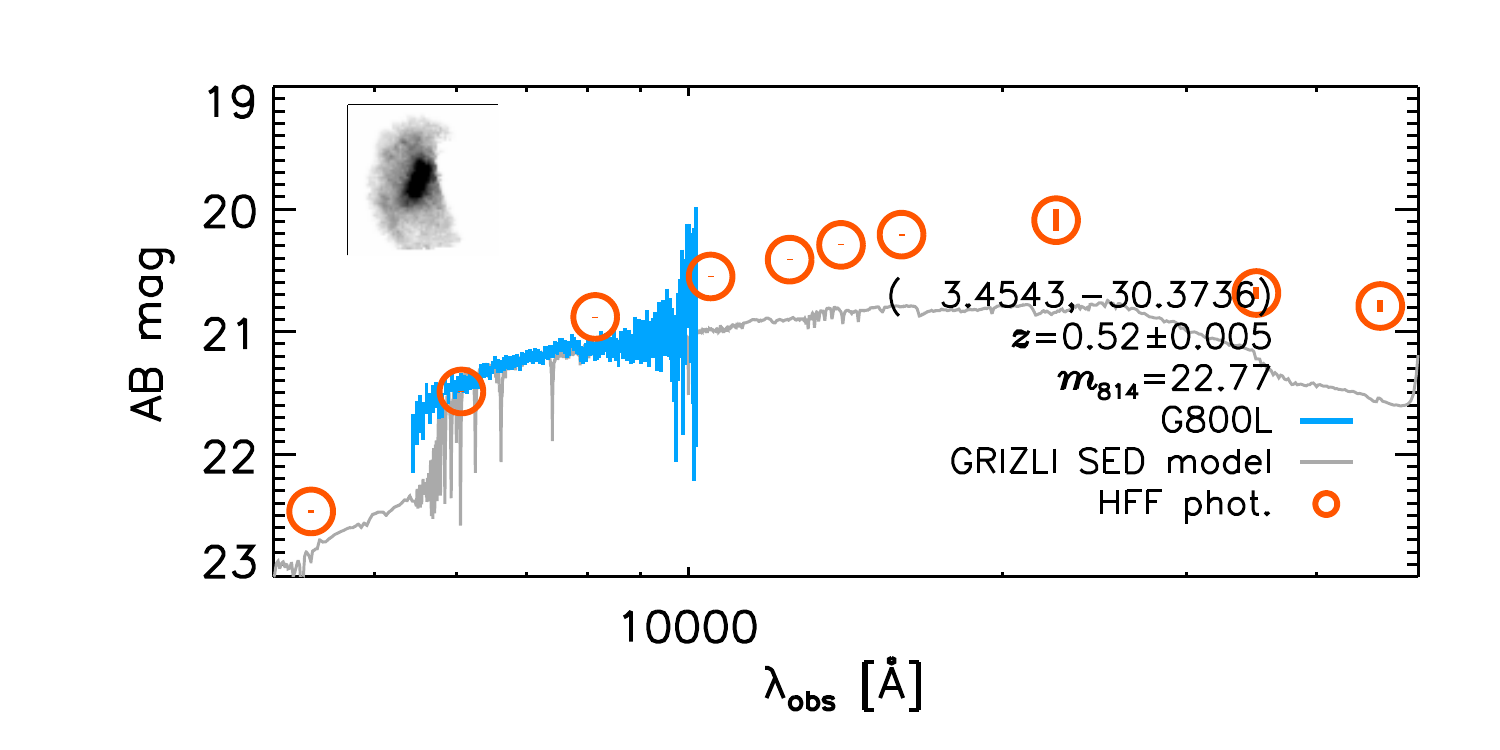}
	\includegraphics[width = 0.475\linewidth]{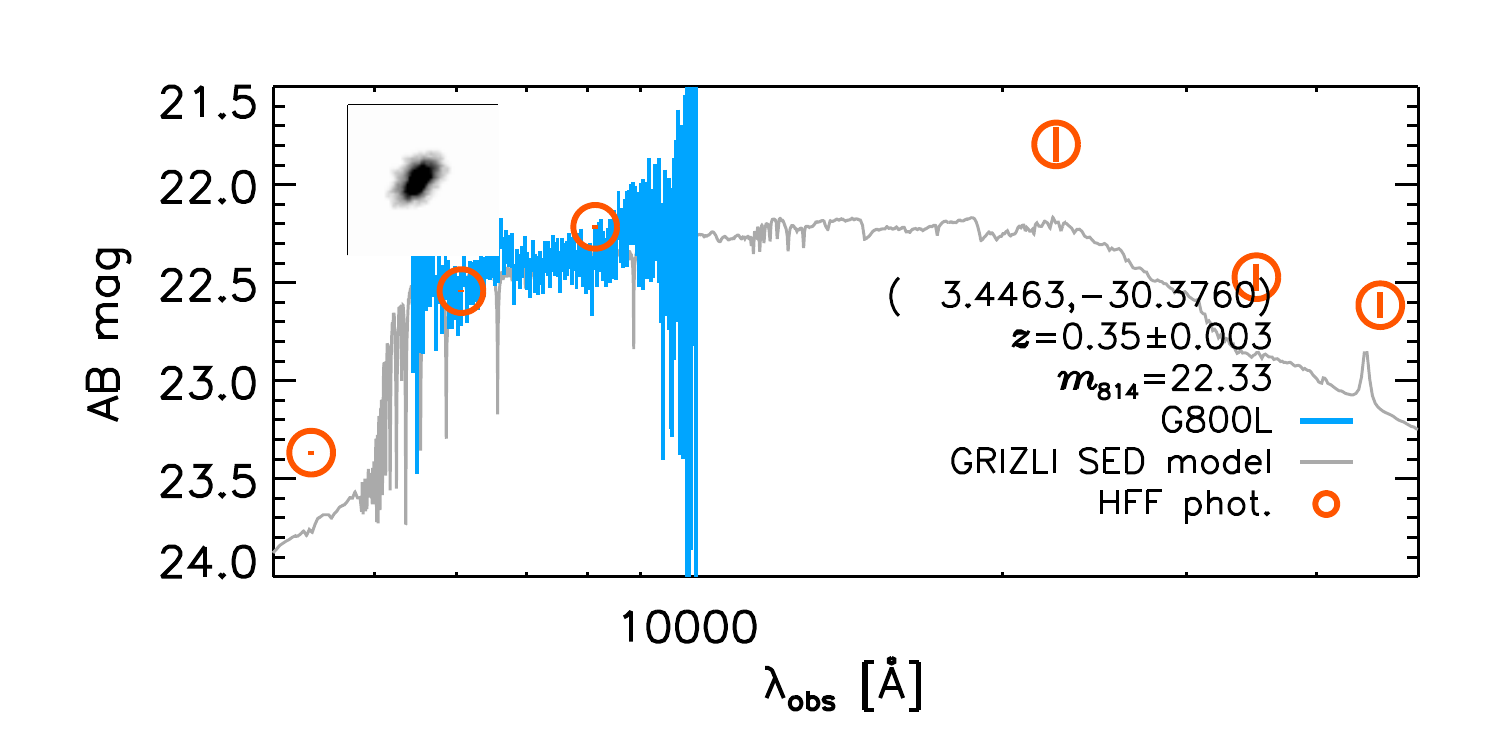}
	\includegraphics[width = 0.475\linewidth]{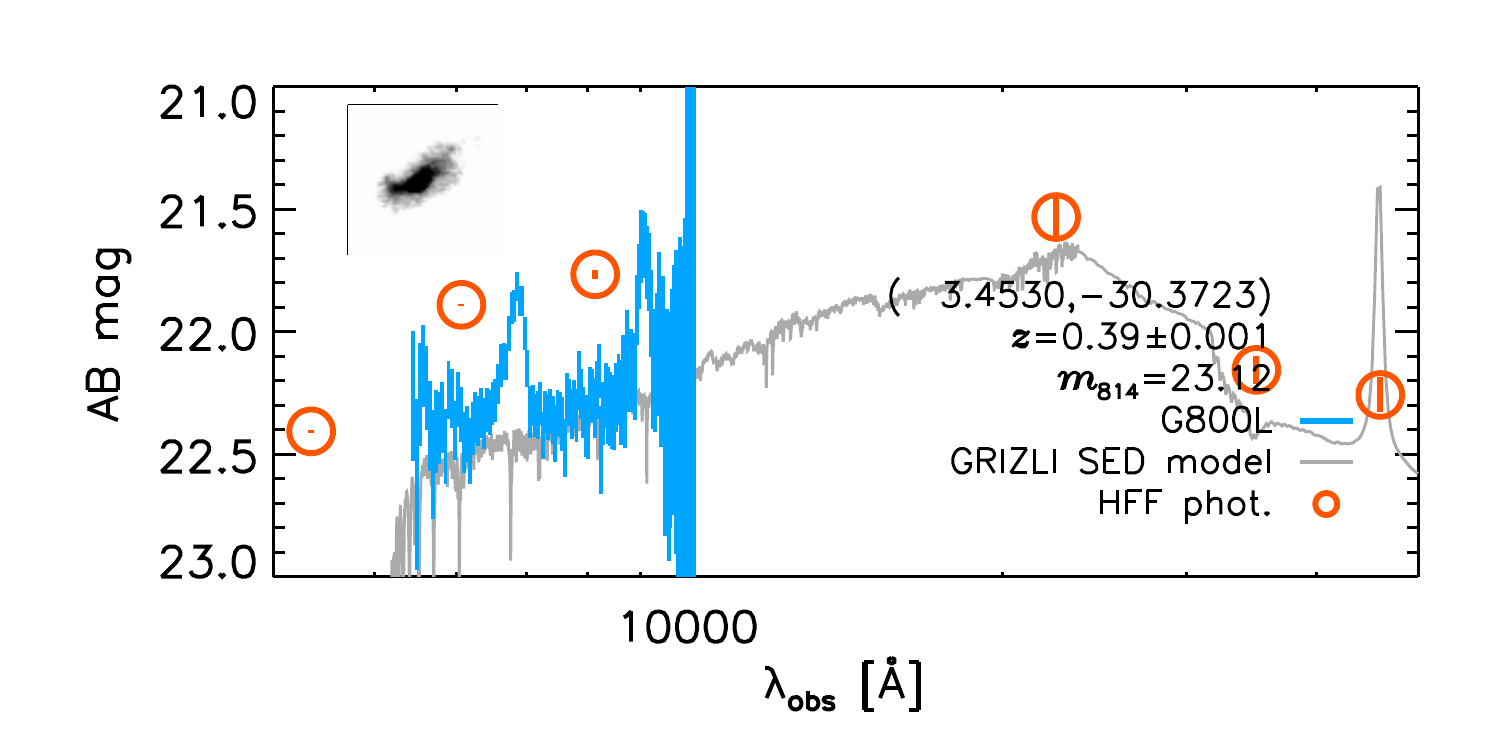}
	\includegraphics[width = 0.475\linewidth]{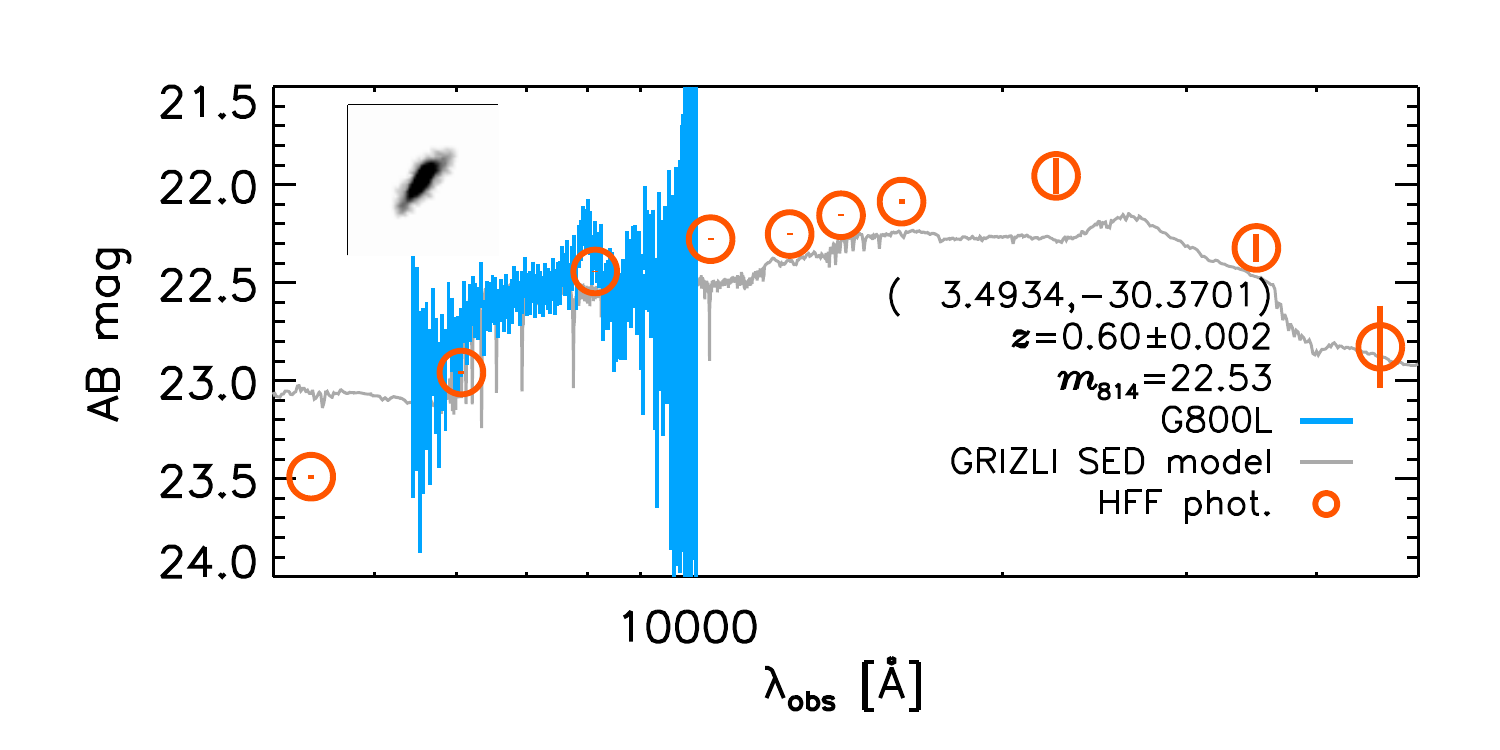}
	\includegraphics[width = 0.475\linewidth]{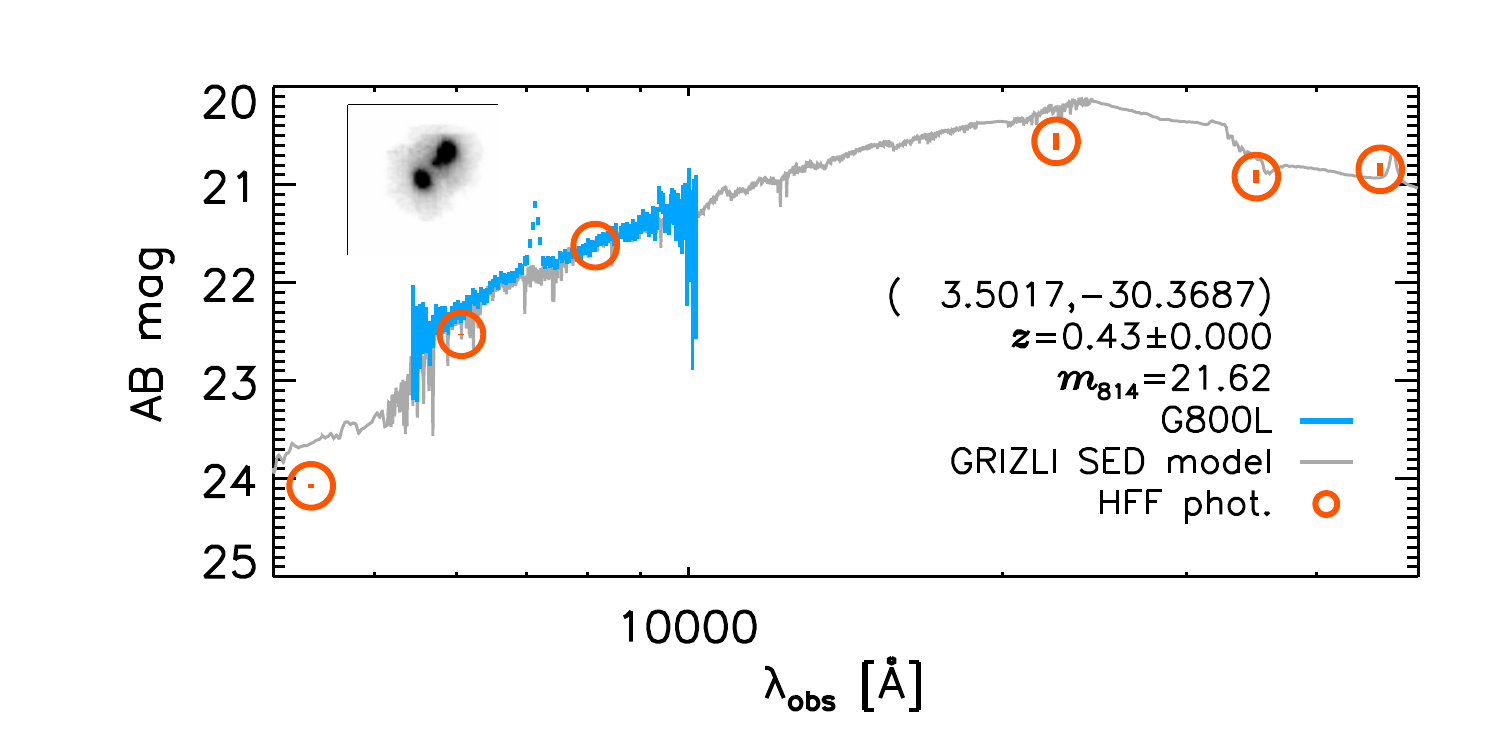}
	\includegraphics[width = 0.475\linewidth]{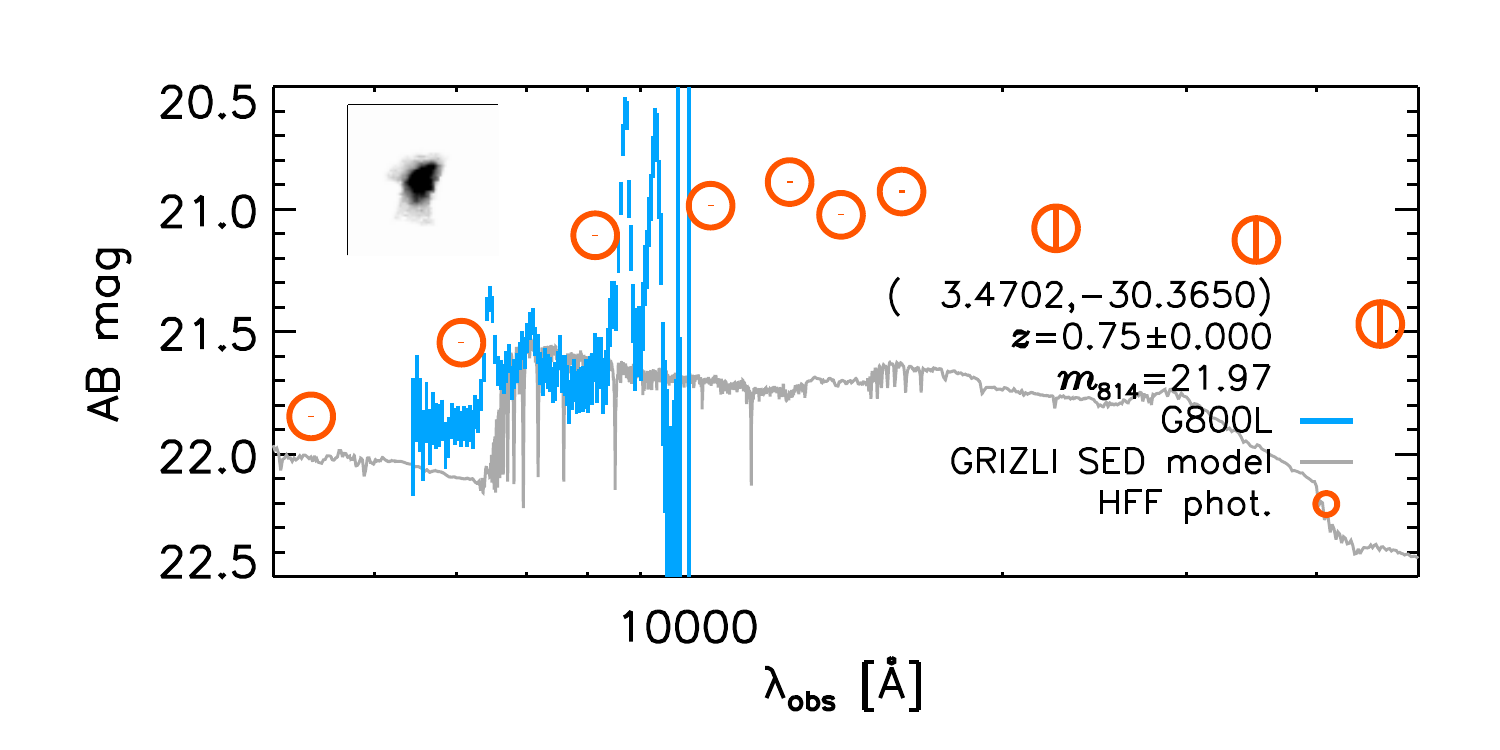}
	\includegraphics[width = 0.475\linewidth]{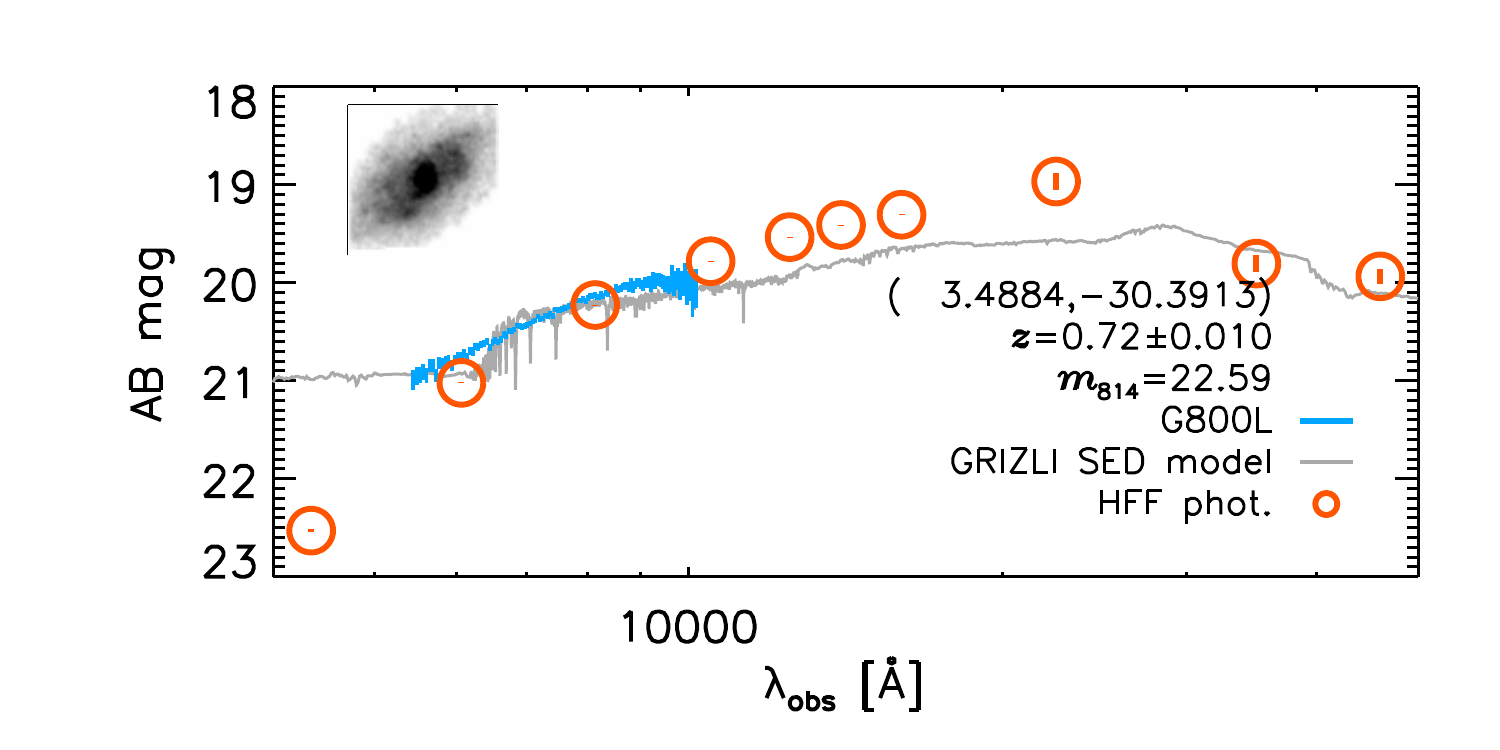}
	\includegraphics[width = 0.475\linewidth]{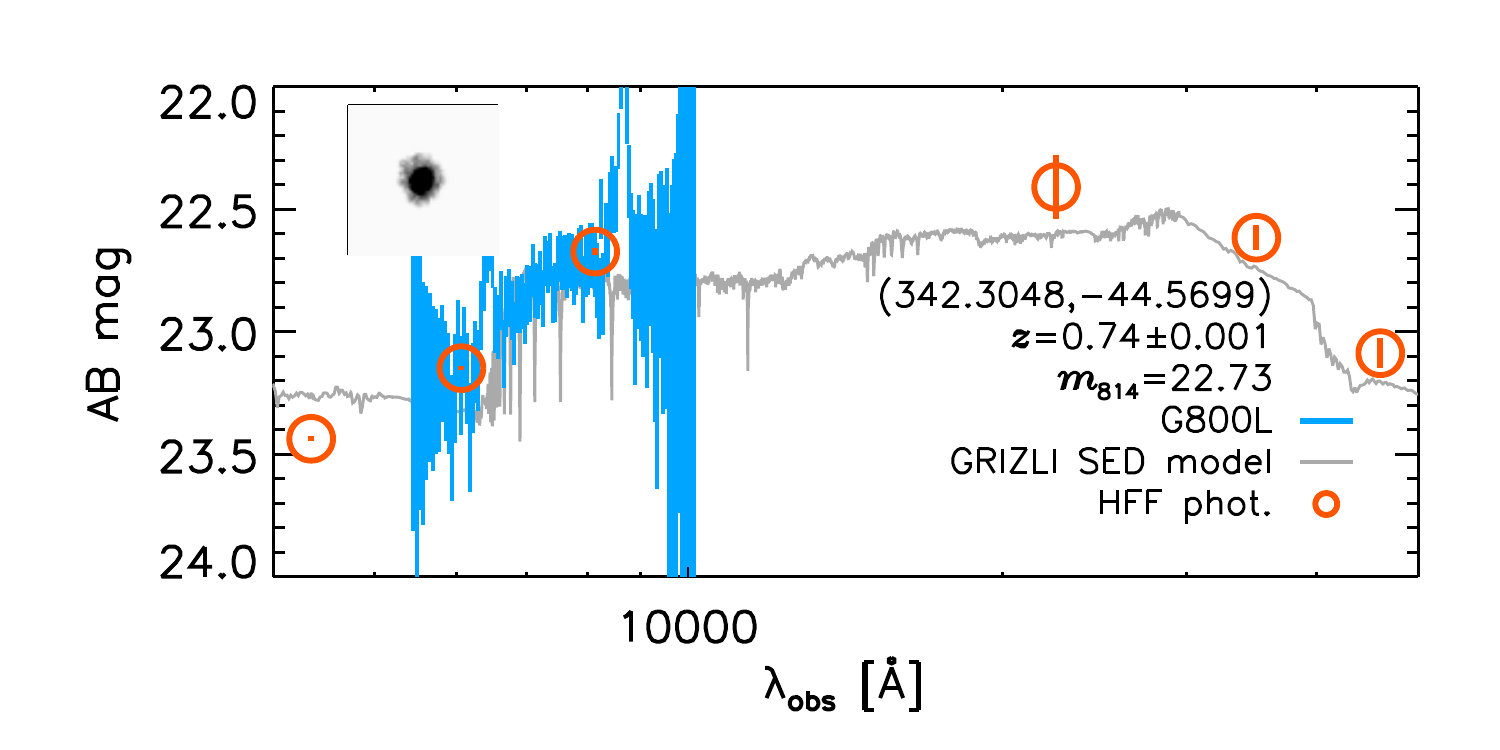}
\caption{GLASS G800L sources in the Abell 2744 and RXJ2248 parallel fields
		matched to HFF photometry from \citet{Shipley18}. This figure is continued on the
		following page.}
\label{fig:specPhotExamples2}
\end{figure*}

\addtocounter{figure}{-1}

\begin{figure*}
	\includegraphics[width = 0.475\linewidth]{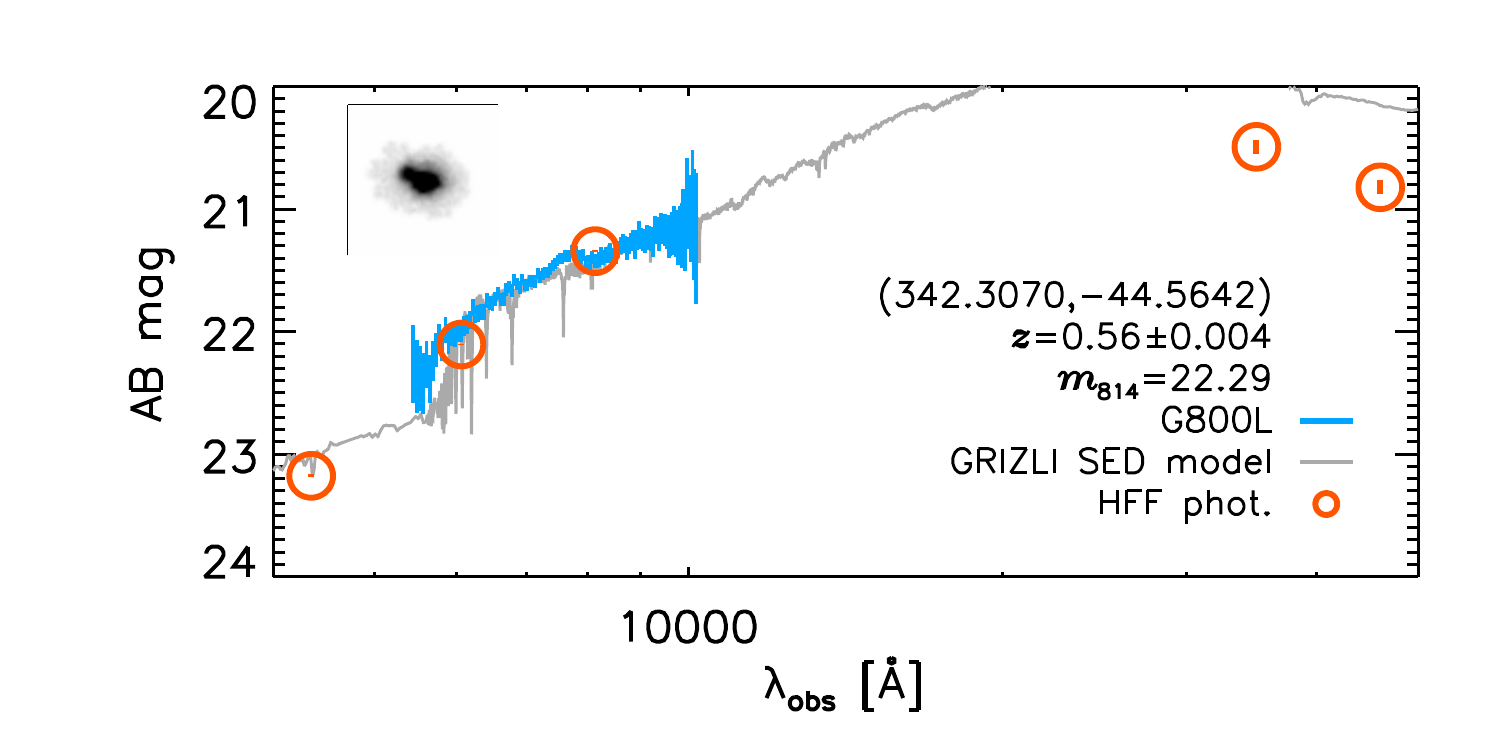}
	\includegraphics[width = 0.475\linewidth]{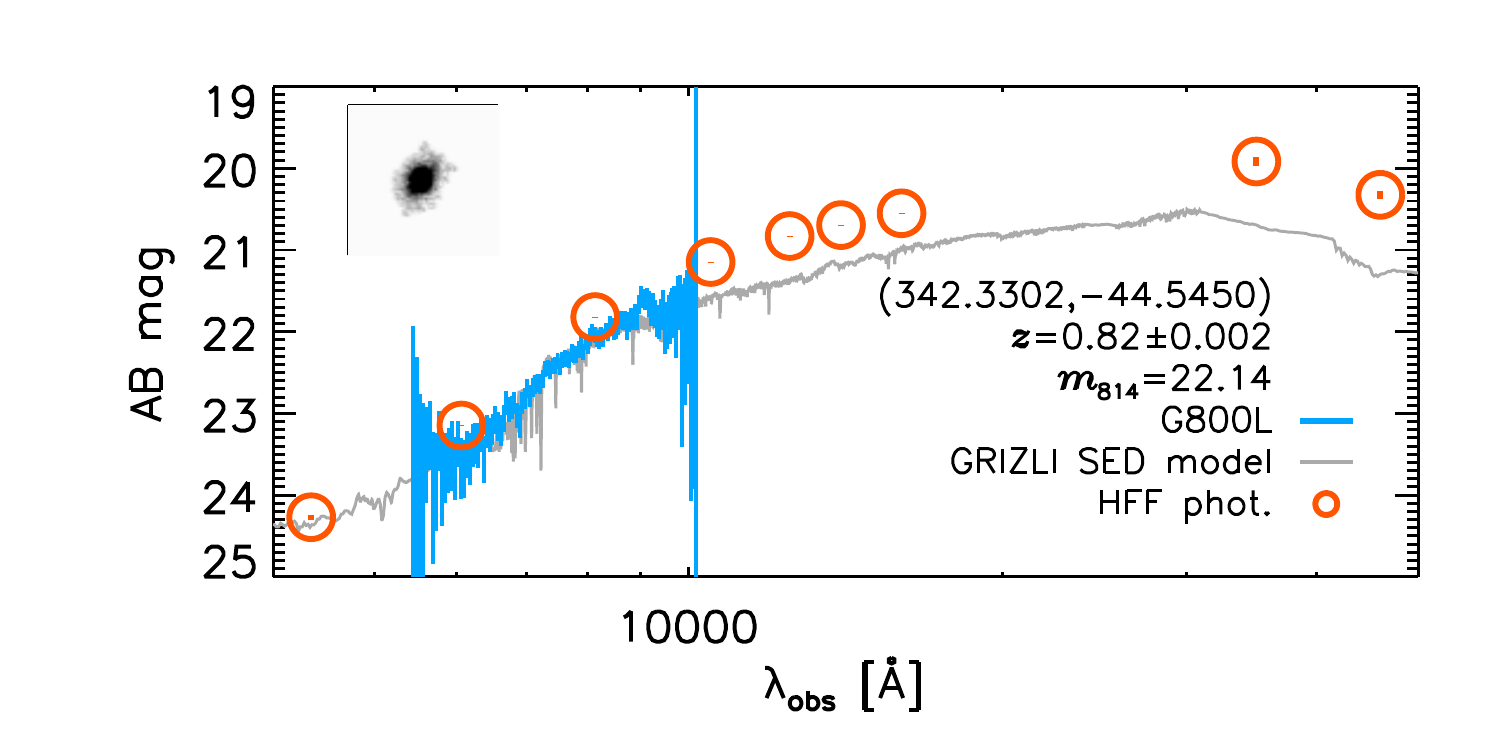}
	\includegraphics[width = 0.475\linewidth]{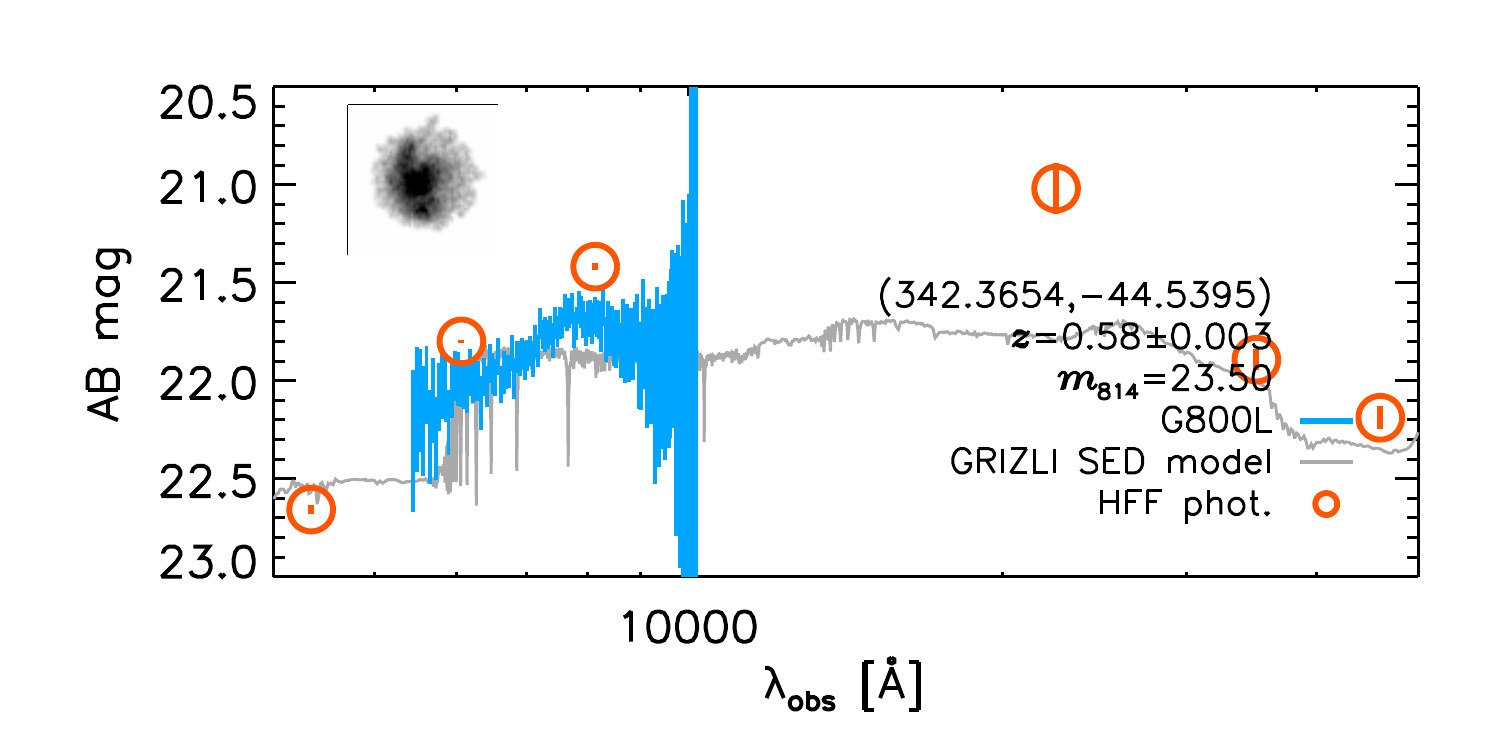}
	\includegraphics[width = 0.475\linewidth]{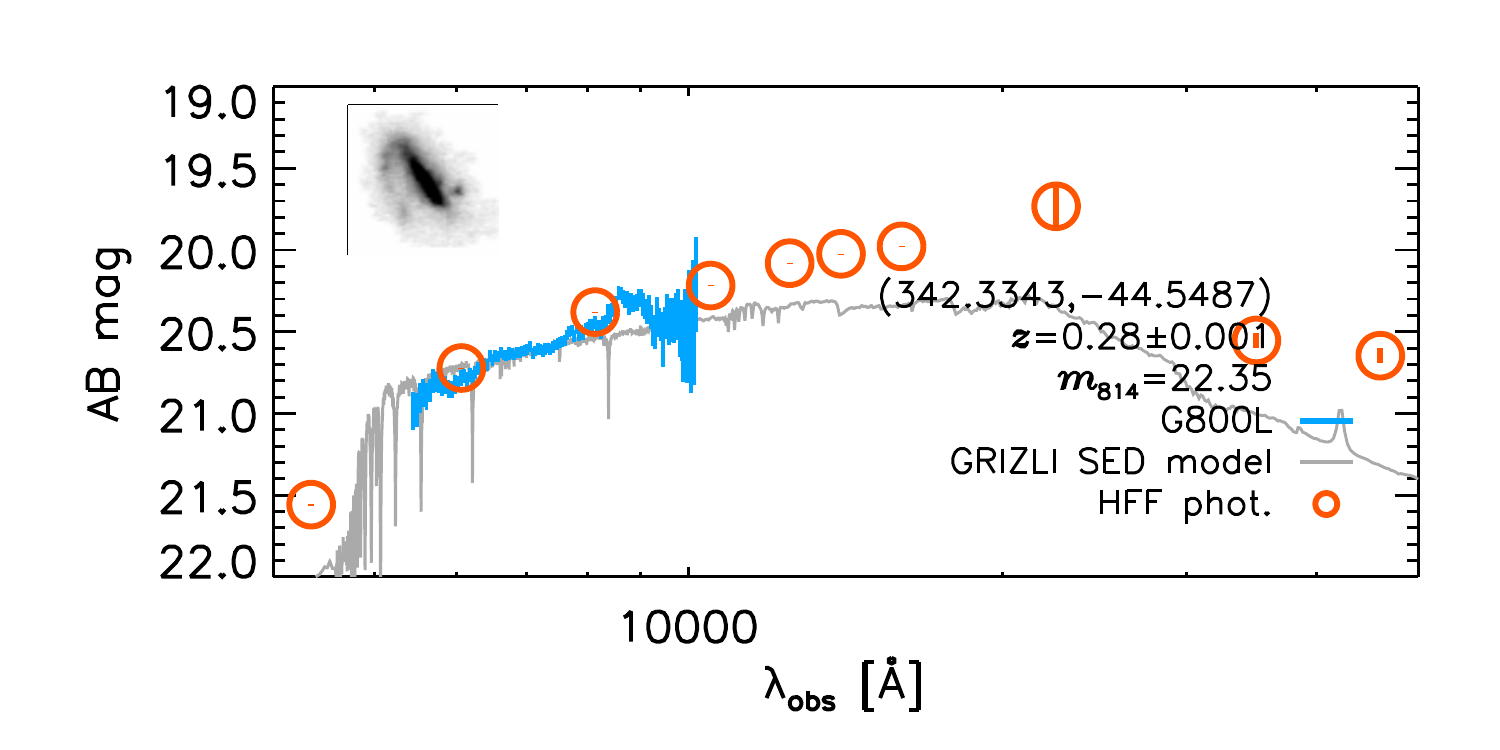}
	\includegraphics[width = 0.475\linewidth]{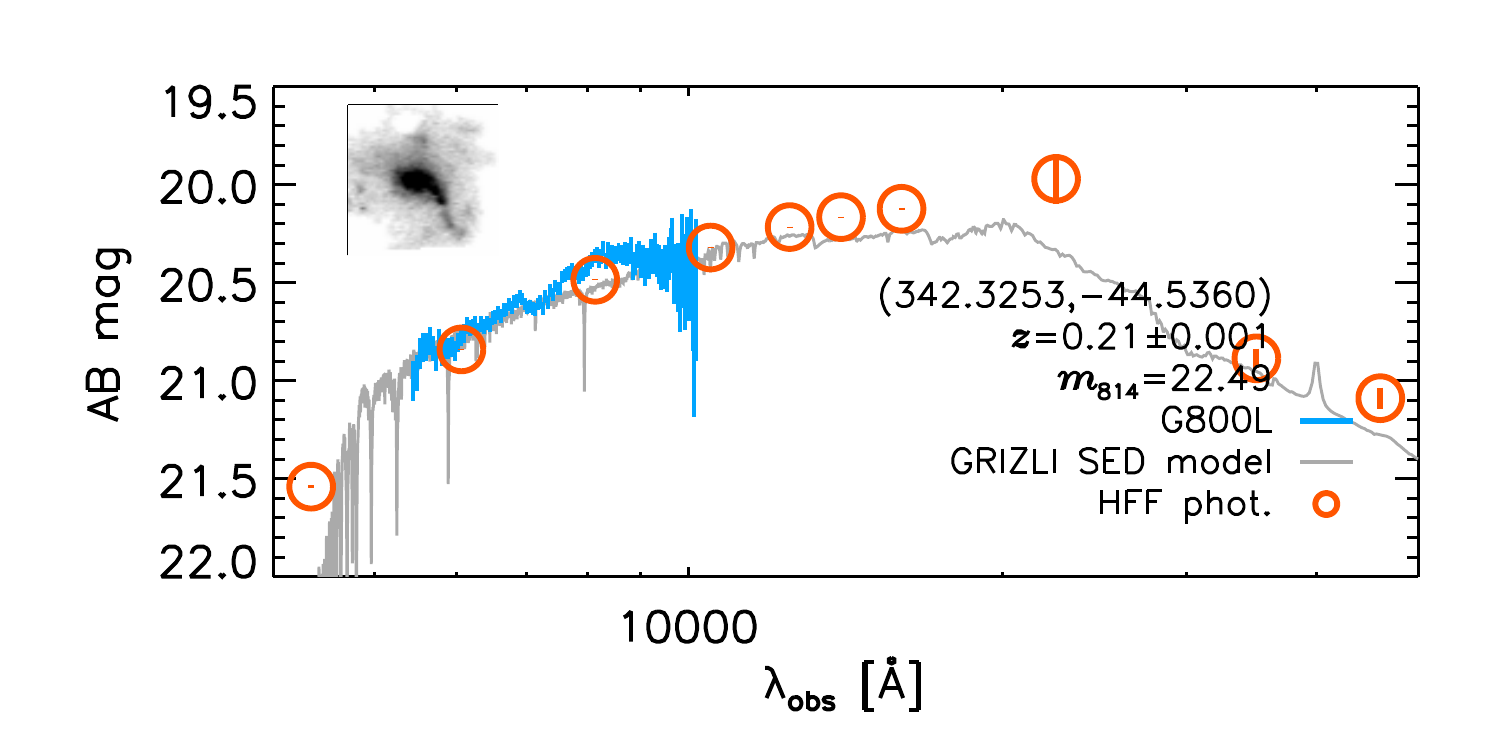}
	\includegraphics[width = 0.475\linewidth]{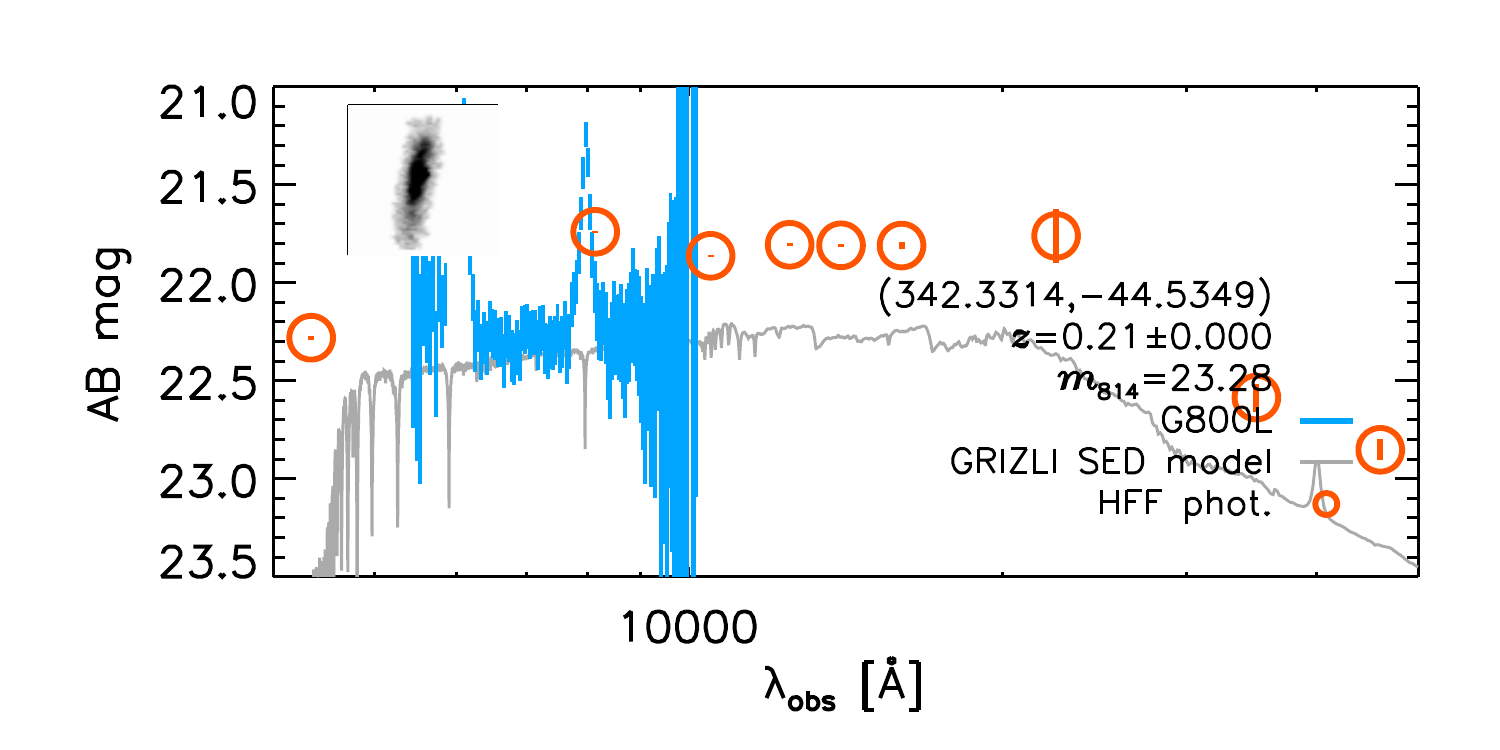}
	\includegraphics[width = 0.475\linewidth]{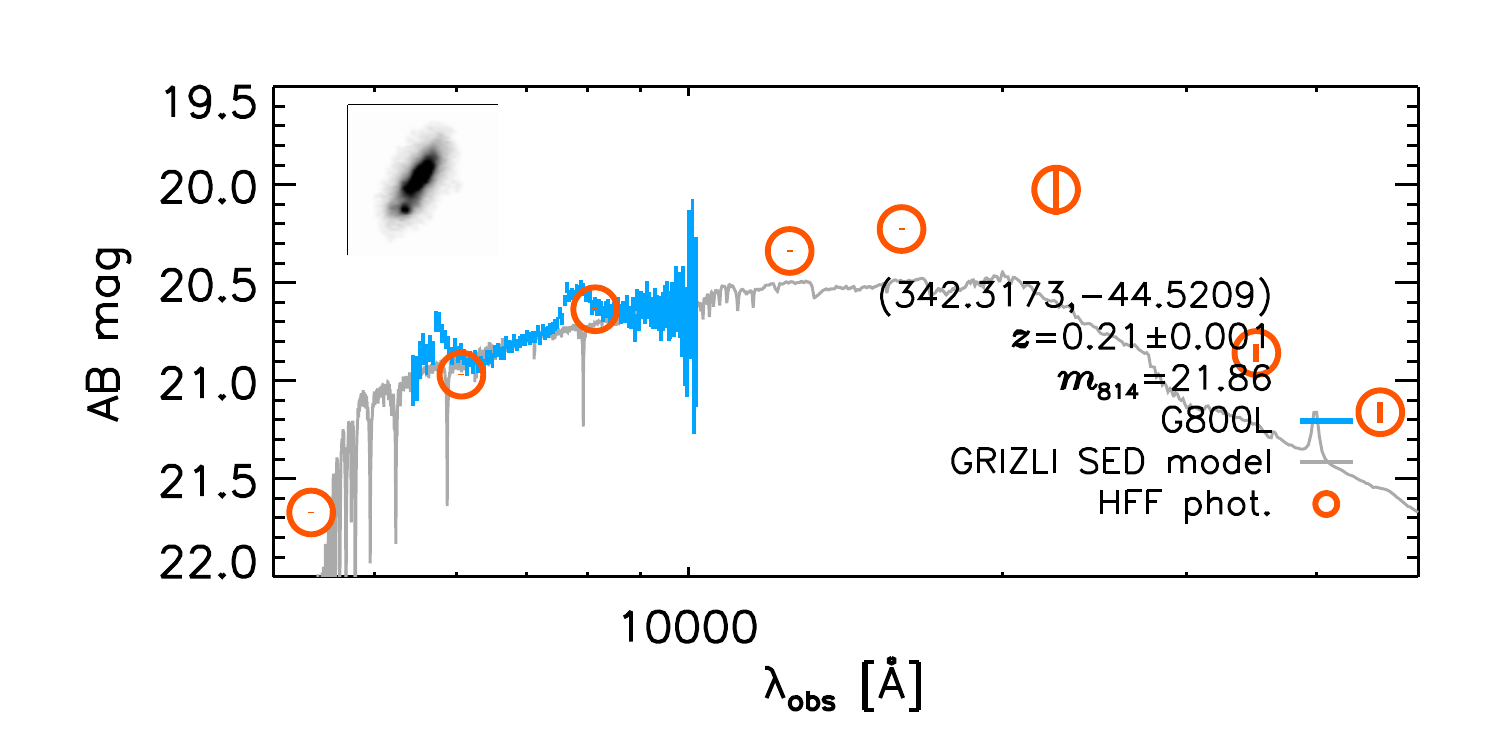}
	\includegraphics[width = 0.475\linewidth]{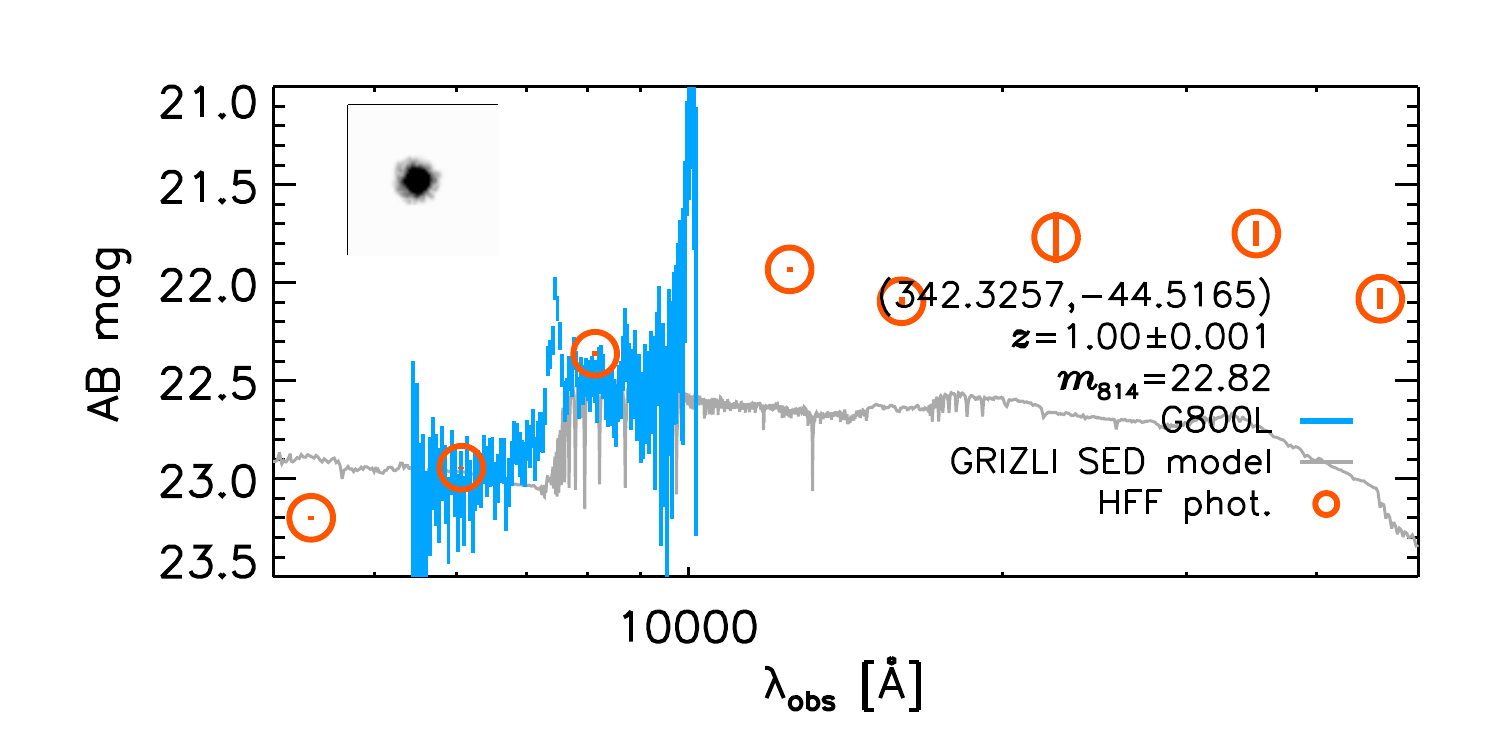}
	\caption{GLASS G800L sources in the Abell 2744 and RXJ2248 parallel fields
		matched to HFF photometry from \citet{Shipley18}. This figure is continued on the
		following page.}
\end{figure*}

\addtocounter{figure}{-1}

\begin{figure*}
	\includegraphics[width = 0.475\linewidth]{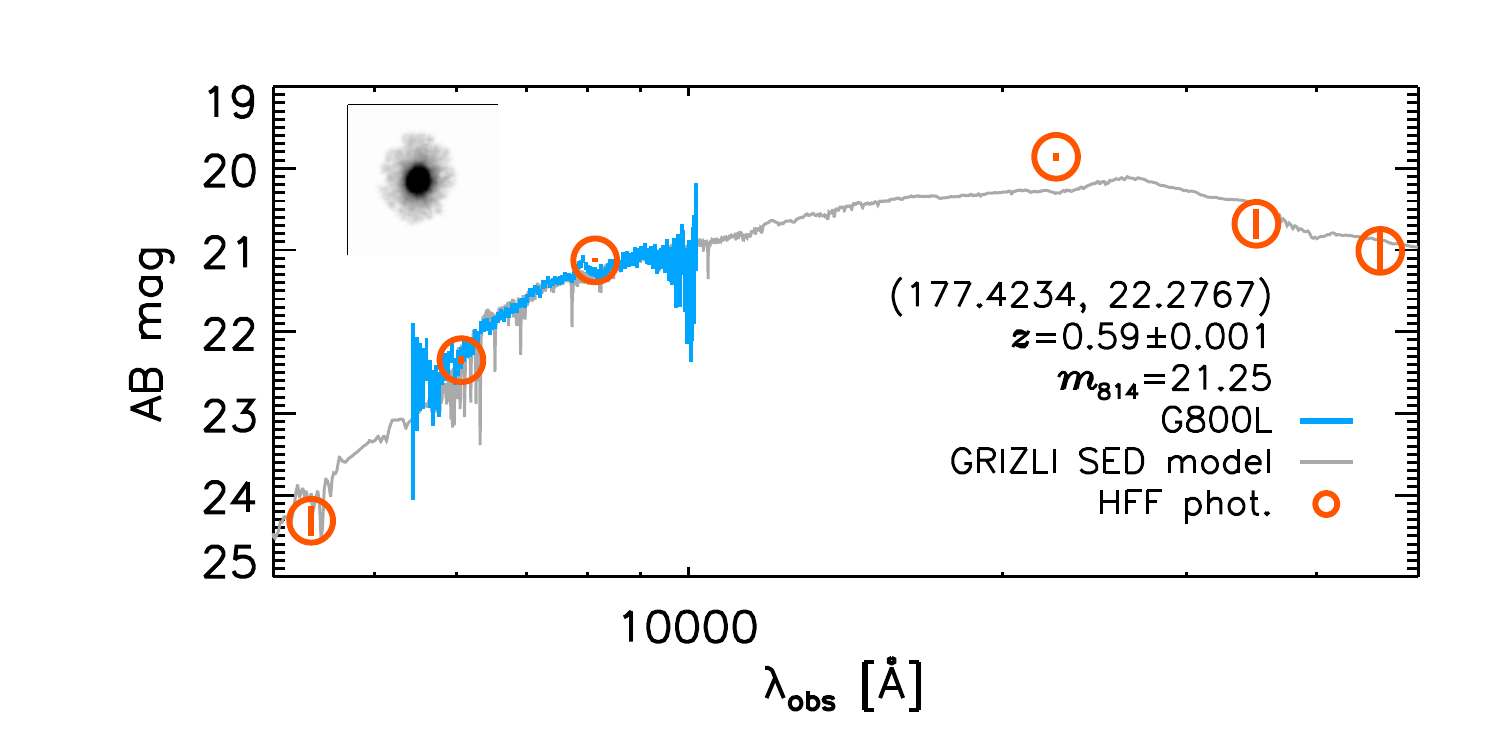}
	\includegraphics[width = 0.475\linewidth]{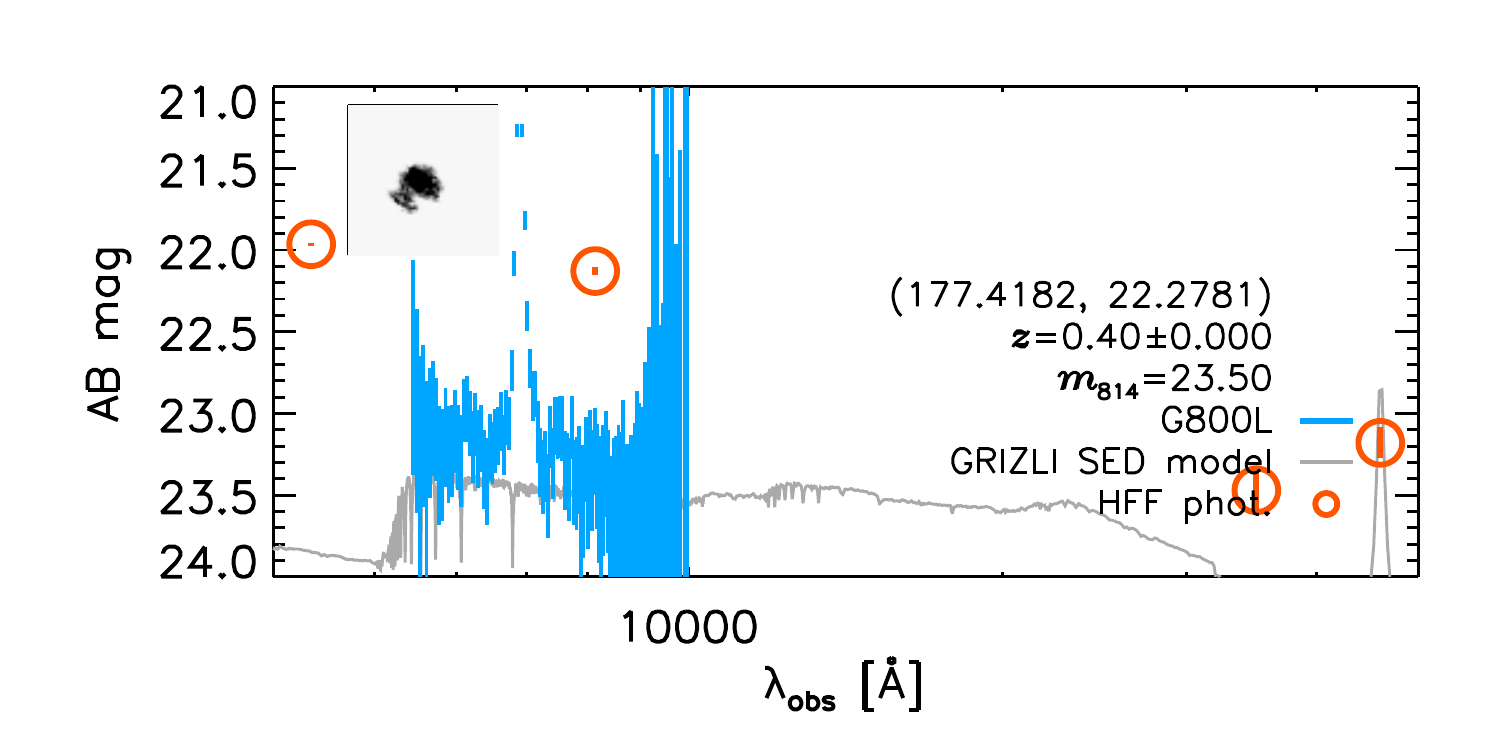}
	\includegraphics[width = 0.475\linewidth]{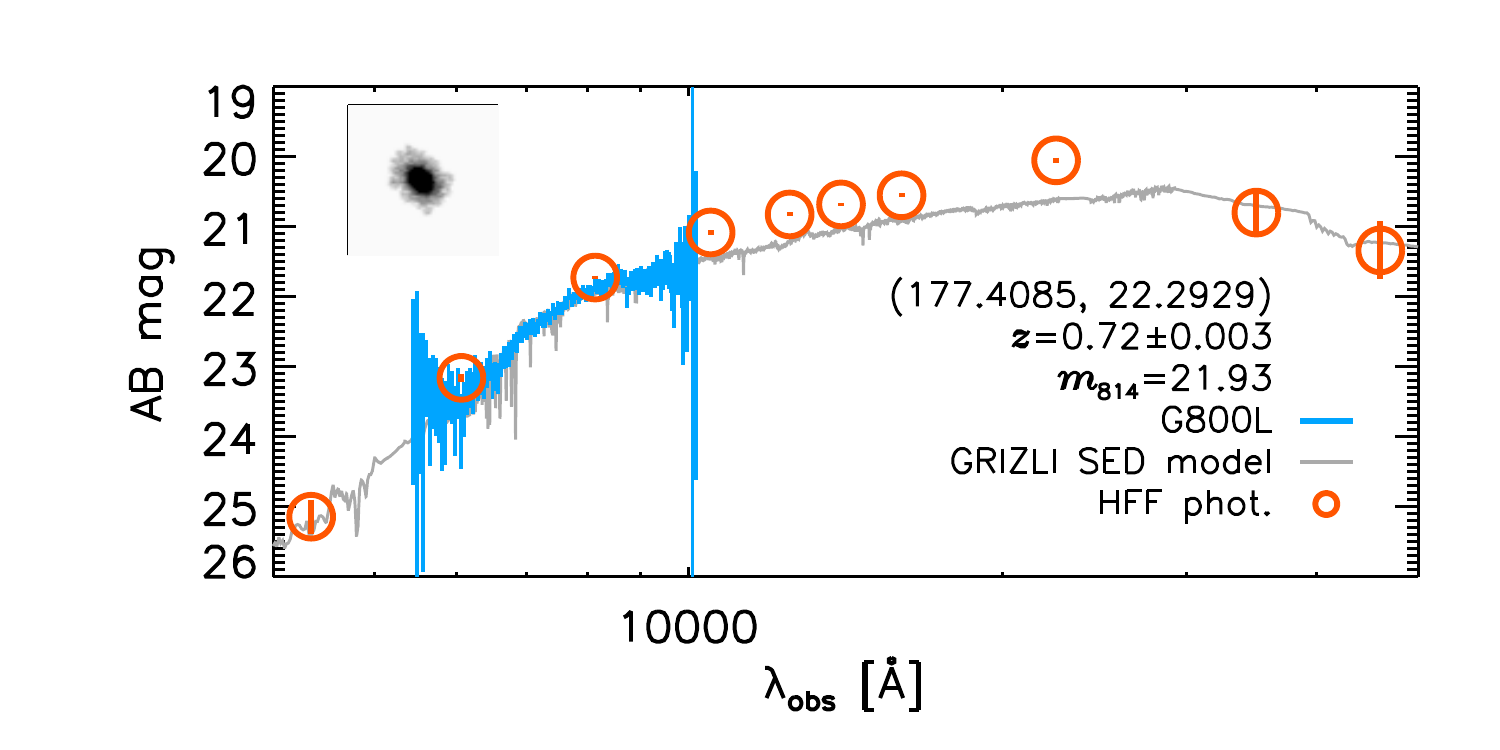}
	\includegraphics[width = 0.475\linewidth]{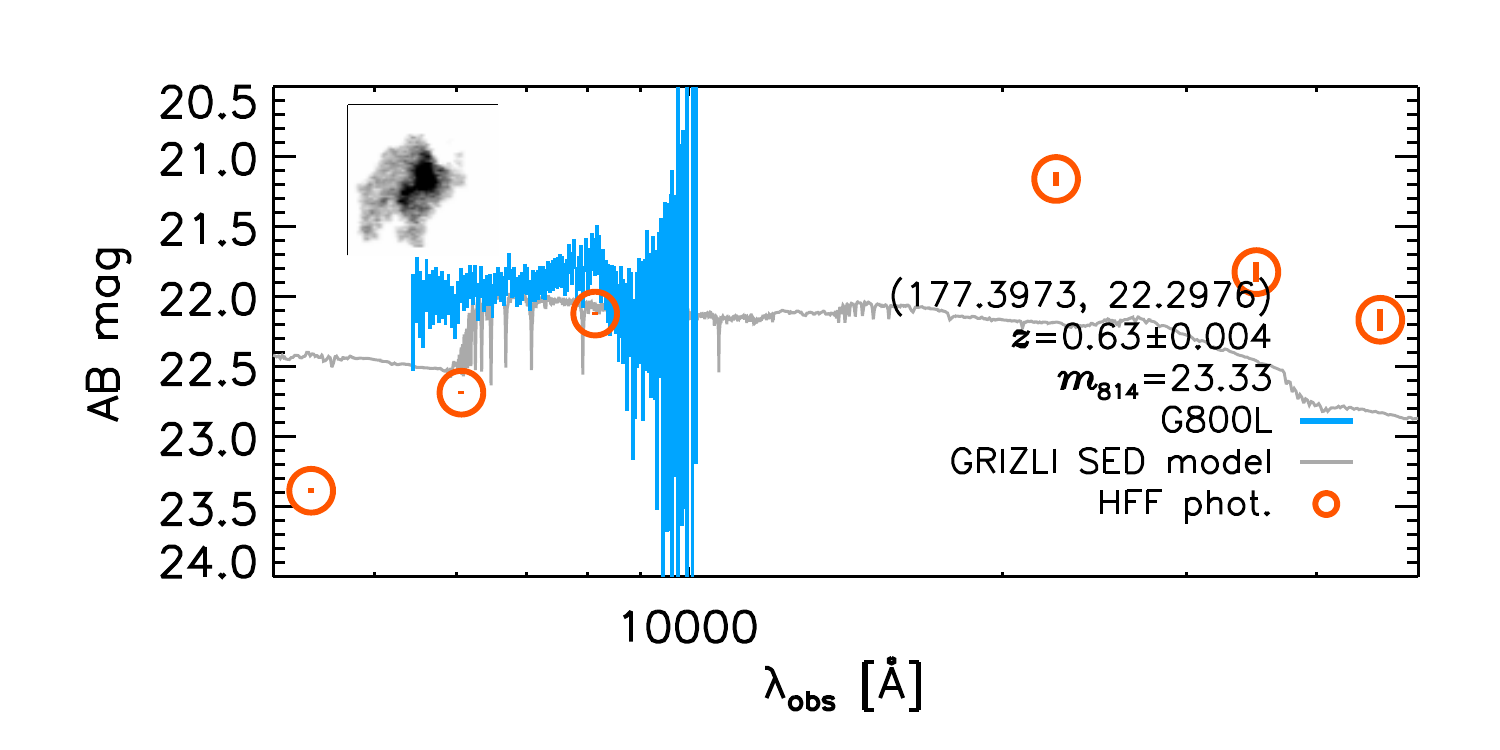}
	\includegraphics[width = 0.475\linewidth]{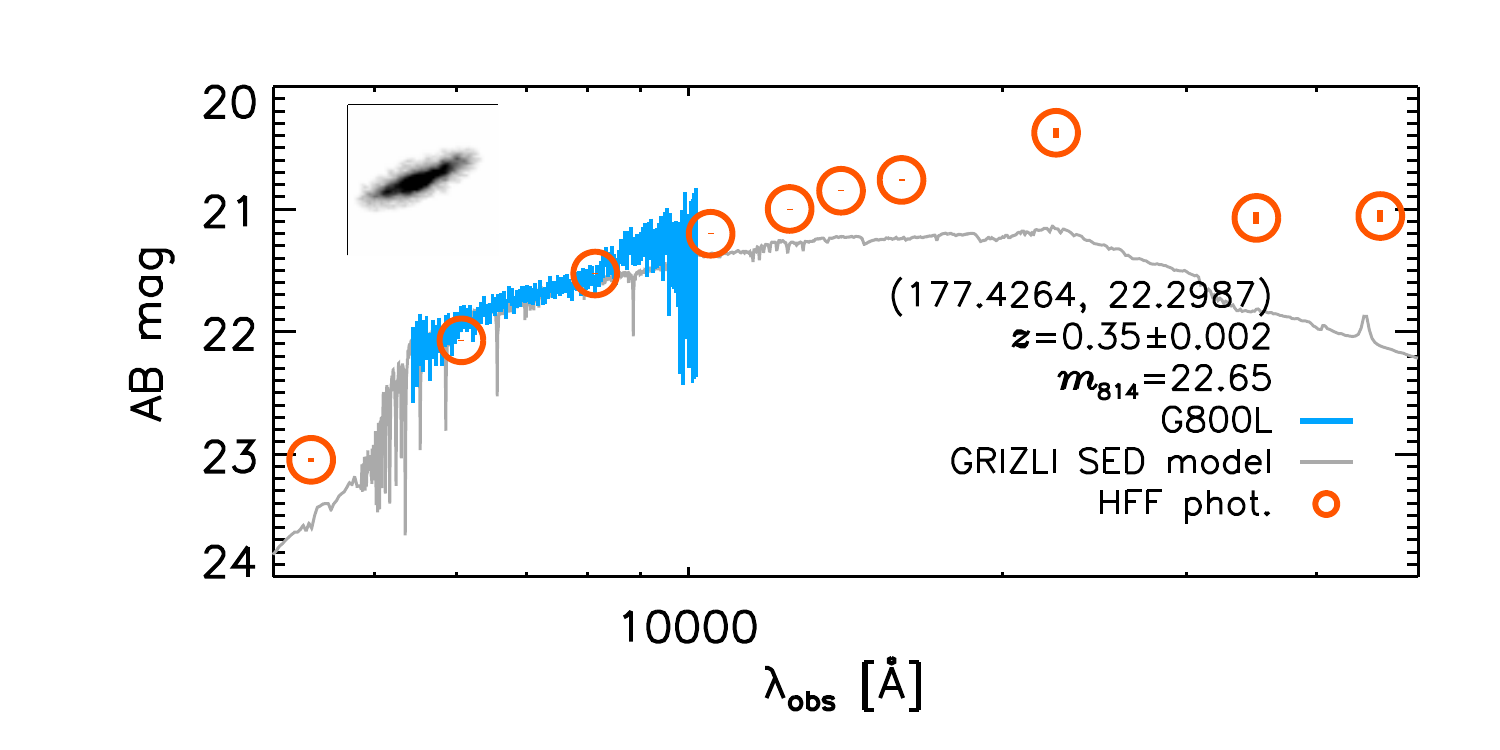}
	\includegraphics[width = 0.475\linewidth]{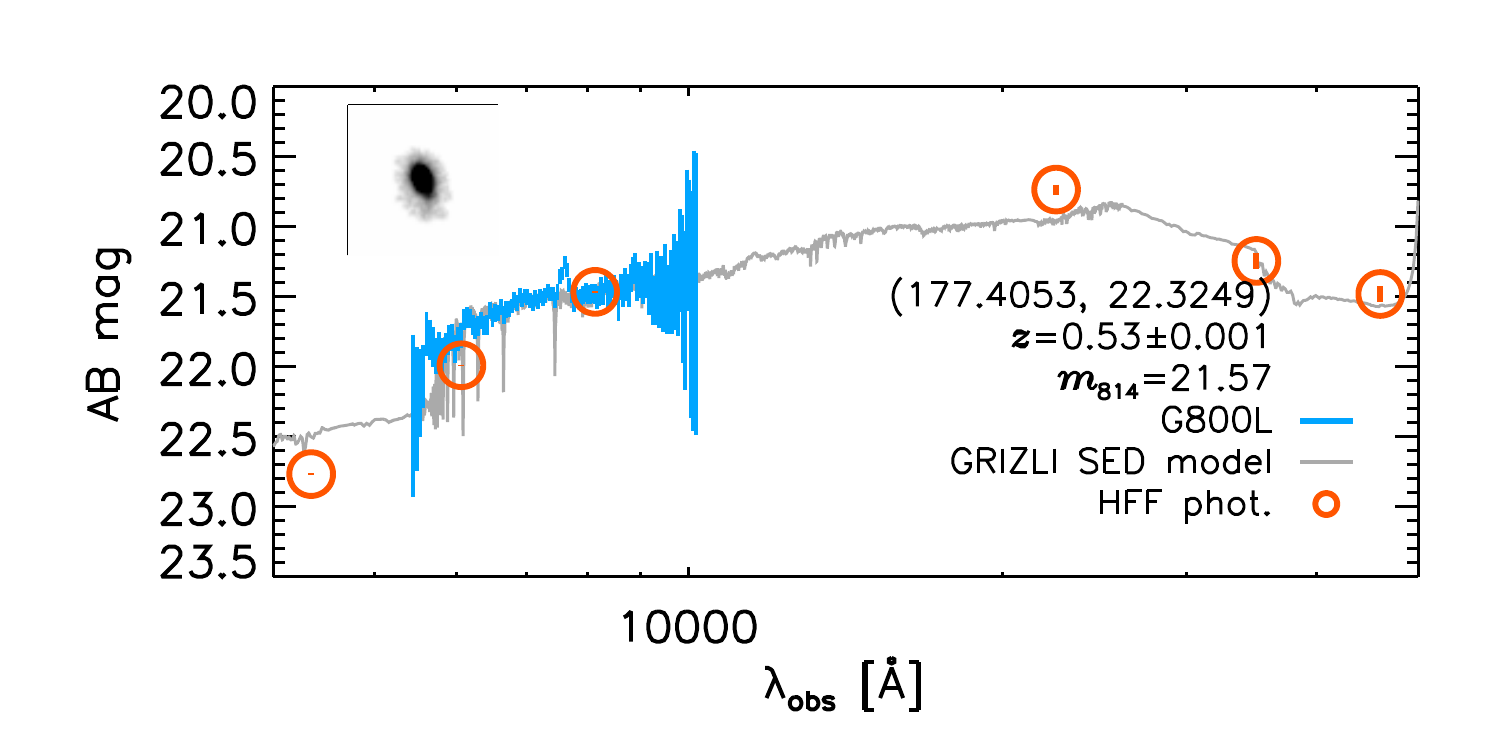}
	\includegraphics[width = 0.475\linewidth]{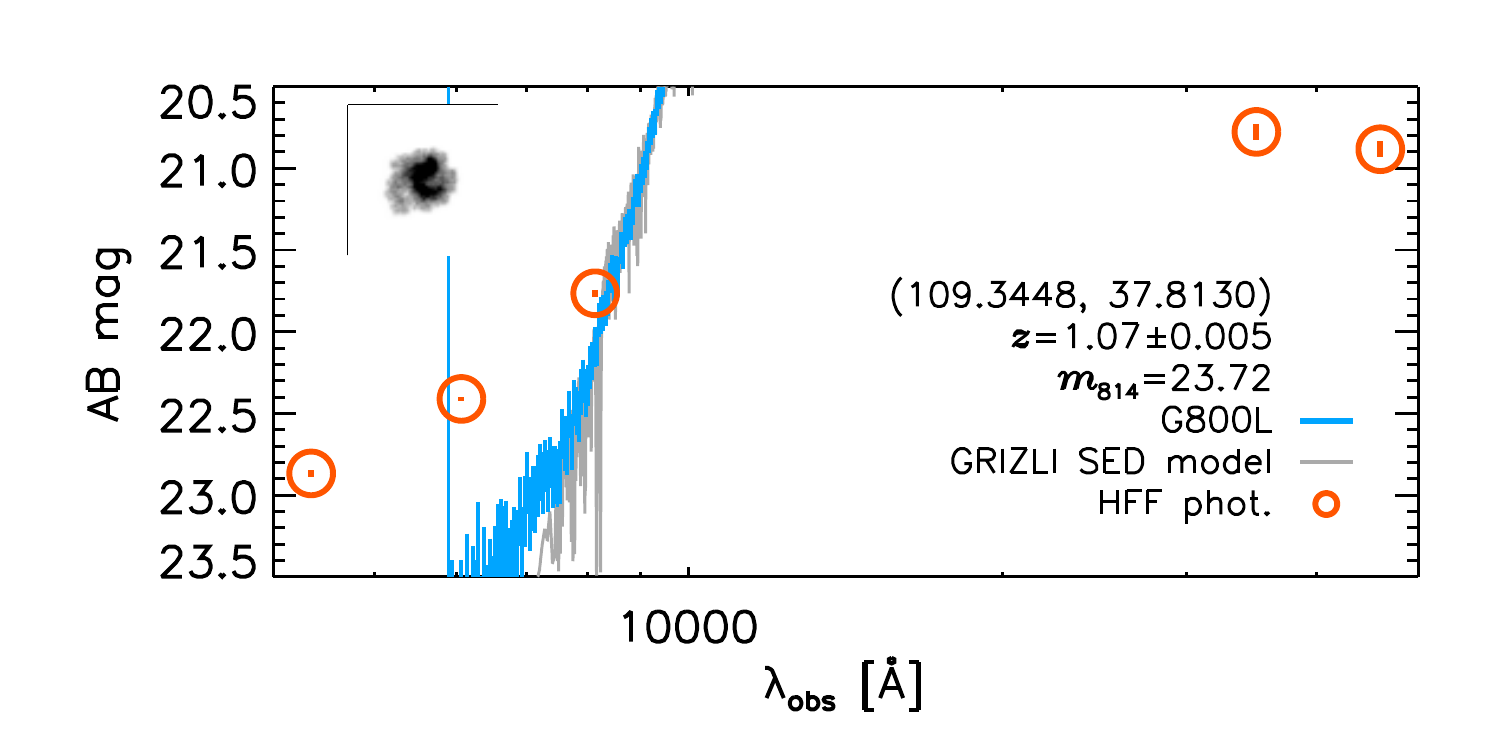}
	\includegraphics[width = 0.475\linewidth]{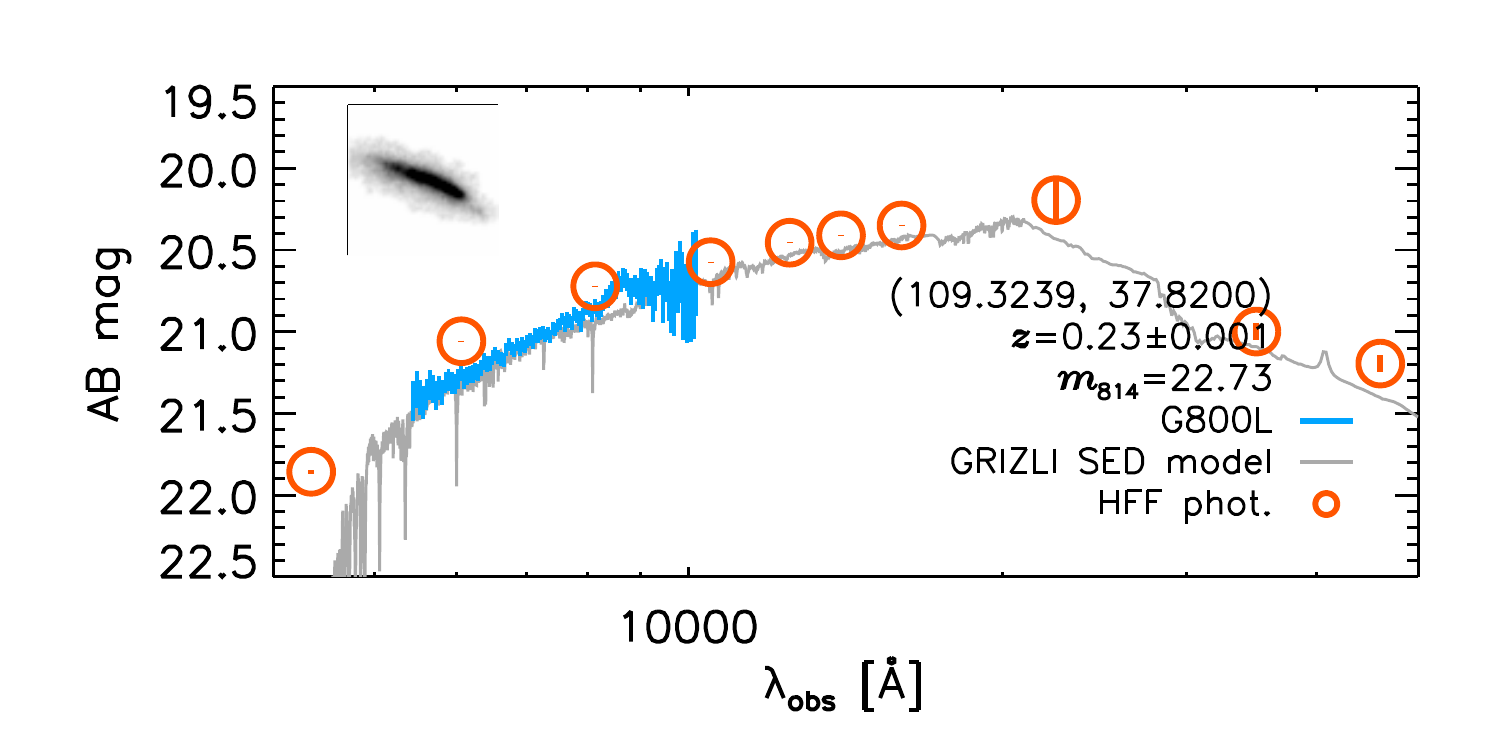}
	\caption{GLASS G800L sources in the MACS0717 and MACS1149 parallel fields 
		matched to HFF photometry from \citet{Shipley18}.}
\end{figure*}


\bsp	
\label{lastpage}
\end{document}